\documentclass[aps,twocolumn,showpacs]{revtex4}
\usepackage{graphicx}
\usepackage{bm}
\usepackage{multirow}
\usepackage{array}
\usepackage{graphicx}
\usepackage{amsmath}
\usepackage{mathrsfs}
\usepackage{subfigure}
\usepackage{amssymb}
\usepackage{caption}
\usepackage{float}
\usepackage{ulem}
\usepackage{color}
\usepackage{overpic}
\definecolor{dgreen}{cmyk}{1.,0.,1.,0.2}        
\definecolor{orange}{cmyk}{0.,0.353,1.,0.}    

\newcommand{\bra}[1]{\langle #1|}
\newcommand{\ket}[1]{|#1\rangle}

\newcommand{\di}{{\rm d}}

\newcommand{\tr}{{\rm tr}}

\newcommand{\be}{\begin{equation}}
\newcommand{\ee}{\end{equation}}                                                                               
\newcommand{\bea}{\begin{eqnarray}}
\newcommand{\eea}{\end{eqnarray}} 
\usepackage{slashed}
\usepackage{lipsum}
\usepackage{hyperref}
\hypersetup{hypertex=true,
	colorlinks=true,
	linkcolor=dgreen,
	anchorcolor=blue,
	citecolor=blue}

\begin{document}
 \bibliographystyle{IEEEtran}
\title{Recent progresses on QCD phases in a strong magnetic field — views from Nambu--Jona-Lasinio model}
\author{Gaoqing Cao}
\affiliation{School of Physics and Astronomy, Sun Yat-sen University, Zhuhai 519088, China.}

\begin{abstract}
In this review, we summarize recent progress on the possible phases of quantum chromodynamics (QCD) in the presence of a strong magnetic field, mainly from the views of the chiral effective Nambu--Jona-Lasinio model. Four kinds of phase transitions are explored in detail: chiral symmetry breaking and restoration, neutral pseudoscalar superfluidity, charged pion superfluidity and charged rho superconductivity. In particular, we revisit the unsolved problems of inverse magnetic catalysis effect and competition between the chiral density wave and solitonic modulation phases. It is shown that useful results can be obtained by adopting self-consistent schemes.
\end{abstract}

\maketitle

\tableofcontents

\section{Introduction}\label{sec:intro}
\subsection{A Brief history of magnetic field}\label{subsec:hist}
According to historical records, the ancient Chinese were the earliest people to discover and make use of magnets. The history can be traced back to at least the third century B.C.~\cite{Zhang1954}, when "Sinans",  devises like a compass with south-pointing handles, were invented to facilitate travels and divinations. The magnitude of magnetic field produced by natural magnets is of the order $5*10^3\,{\rm G}$. It wasn't until 1600 when British scientist W. Gilbert proposed that "the earth is a giant magnet" that people realized that the mechanism driving compasses is mainly based on the geomagnetic field naturally generated by the earth itself, even though its origin is still mysterious. In 1832, the great German mathematician and scientist C.F. Gauss developed the earliest experiment to measure the magnitude of the geomagnetic field and found the order to be of $0.5\,{\rm G}$~\cite{Chapman1940}. It is generally accepted that the distribution of the geomagnetic field around the earth is quite similar to the dipole field produced by a bar magnet and the geomagnetic north and south poles are roughly opposite to the geometrical ones of the earth. However, this is not always the case: the studies of natural remanent magnetization of rocks and marine sediments surprisingly indicated that the geomagnetic field has reversed $\sim200$ times in the past 100 million years with the last time occurring 0.7 million years ago~\cite{Campbell2003}.

In 1800s, the subject of electromagnetism was created and allowed rapid new developments due to the contributions of H.C. Oersted, A.M. Ampere, M. Faraday and J.C. Maxwell~\cite{Feynman2013}. In particular, the discovery of Ampere's law opened the door to man-made magnetic fields which could be much larger than natural ones. In 1831, by effectively increasing the number of windings in a coil, American electrician J. Henry developed a strong electromagnet that was able to generate a magnetic field with a magnitude around $3*10^4\,{\rm G}$~\cite{Henry1831}. In 1911, Dutch physicist H.K. Onnes discovered superconducting phenomena in hydrargyrum after cooling the temperature down to $4.2\,{\rm K}$. With the electric resistance reduced to almost zero, superconductors were the ideal means to generate large electric currents and thus large, stable magnetic fields according to Ampere's law. In modern times, this feature of superconductors has already played an important role in high energy particle accelerators worldwide. For example,  the Large Hadron Collider in CERN is equipped with $1600$ superconducting magnets to maintain the particles in a circle with a length of $26.7\,{\rm km}$ and the produced magnetic field is $7.7*10^4\,{\rm G}$~\cite{Bottura:2020vmt}. Currently, with the development of nuclear techniques, people pay a lot of attention to induce nuclear fusion reactions, just like those occurring in the interior of the Sun, in order to produce more power for manufacturing and living. The foremost installation to realize that scientific objective is the Tokamak~\cite{Wesson2011}, which was proposed to confine the nuclei by a huge magnetic field of the order $10^7\,{\rm G}$. So the prerequisite for these developments is first generating a huge magnetic field. The possibility for this has been tested again and again in laboratories; in 2018, Japanese scientists created a new record for magnetic fields -- $1.2*10^7\,{\rm G}$, though the lifetime is very short~\cite{Nakamura2018}.

Apart from all those mentioned above, there are three well-known sources of extremely strong magnetic field: magnetars, heavy ion collisions and the early universe, which can not be applied to daily life. In 1979, the first magnetar, ${\rm SGR}~1900+14$, was discovered by the $\gamma$-ray burst detector Konus of Union of Soviet Socialist Republics~\cite{Mazets:1979uu}, and subsequent detections gave the surface magnetic magnitude of $7\times 10^{14}\,{\rm G}$. Following that, more and more magnetars were discovered with the surface magnetic fields in the order $10^{14}-10^{15}\,{\rm G}$~\cite{Turolla:2015mwa}. Actually, if the neutron stars are born within a period of the order of $1\,{\rm ms}$ and convection is present, the magnetic field was proposed to be as large as $\sim10^{17}\,{\rm G}$~\cite{Thompson:1993hn}, which could affect the QCD transitions. Next, by following the big bang theory, a lot of properties of the early universe have been explored and the primordial magnetic field could be estimated to be of the order $10^{23}\,{\rm G}$ during the electroweak phase transition~\cite{Vachaspati:1991nm}. In fact, the fireballs left after heavy ion collisions (HICs) are quite like the matter in the QCD stage of the early universe, thus often called "little bang". In the peripheral collisions, the beams are so energetic and close to each other that extremely large magnetic field can be generated by the spectator currents -- model calculations predicted the magnitude to be of the order $10^{18}-10^{20}\,{\rm G}$~\cite{Skokov:2009qp,Voronyuk:2011jd,Bzdak:2011yy,Deng:2012pc}. We note that the lifetime of the magnetic field is also an important issue for the study of the magnetic effects on QCD matter. Naively, the magnetic field was expected to decay quickly around $1-2\,{\rm fm/c}$ in HICs, but it could survive as long as the lifetime of QGP once the interplay with QGP was taken into account~\cite{Tuchin:2013ie,Tuchin:2013apa,McLerran:2013hla,Guo:2019mgh,Yan:2021zjc}. The measurement of magnetic fields in HICs is technically easier and more controllable than tracing the primordial magnetic field from the current universe, thus we expect the magnetic fields in HICs would come out first in the future.

\begin{figure}[!htb]
	\centering
	\includegraphics[width=0.45\textwidth]{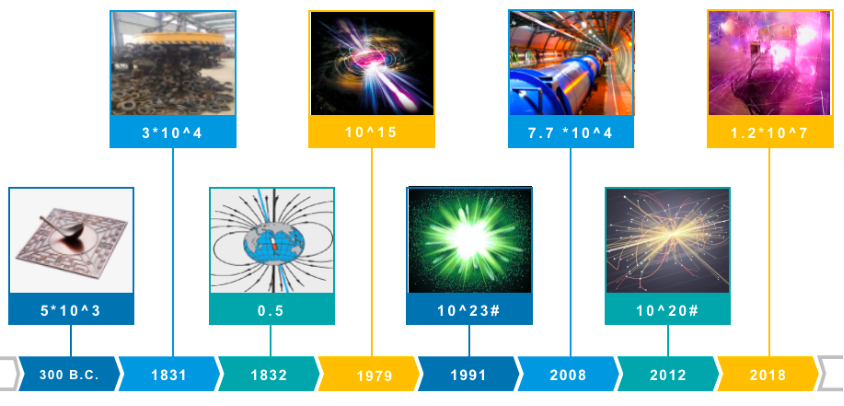}
	\caption{A brief history of magnetic field with the unit $"{\rm G}"$. The theoretical predictions are marked by "$\#$".}\label{history}
\end{figure}

Finally, it is instructive to summarize the brief history of magnetic field through an integrated diagram, see Fig.\ref{history}~\cite{baidu}. 

\subsection{Magnetic effect in QCD matter}\label{subsec:Meff}

In this section, we will only focus our attention on the studies of magnetic field effects on QCD matter, some of which will be discussed in more detail in the following sections. The earliest study of the effect of electromagnetic (EM) field on chiral symmetry should be attributed to S. P. Klevansky and R. H. Lemrner's work in 1989, although they mainly focused on the chiral symmetry restoration ($\chi$SR) effect induced by strong electric field~\cite{Klevansky:1989vi}. Later, from 1994 to 1996, V. P. Gusynin, V. A. Miransky, and I. A. Shovkovy published four important papers that well established the basic concepts of magnetic catalysis effect (MCE) on chiral symmetry breaking ($\chi$SB) and dimensional reduction in magnetic field under the framework of Nambu--Jona-Lasinio (NJL) model~\cite{Gusynin:1994re,Gusynin:1994va,Gusynin:1994xp,Gusynin:1995nb}. During this period, there were several parallel works about the QCD matter in the chromomagnetic field, which simply modeled gluon condensation. It was found that the non-Abelian background field would also catalysis chiral symmetry breaking but only effectively reduced the dimensions by $1$ compared to  $2$ in the Abelian case~\cite{Vshivtsev:1994si,Klimenko1994,Shovkovy:1995td}. 

Following that, in 21st century, the QCD phase diagrams in the presence of magnetic field were explored in greater detail.  In 2000, a brief study was carried out in chiral perturbation theory for finite temperature $T$ and magnetic field $B$~\cite{Agasian:2000hw}, and the pseudo-critical temperature $T_c$ for $\chi$SR was found to increase with magnetic field. About ten years later, not only this feature was well confirmed in the studies of PNJL model and its variants, but the $T_c$ for deconfinement was shown to be qualitatively the same in mean field approximation~\cite{Fukushima:2010fe,Gatto:2010qs,Gatto:2010pt}. The findings were summarized together with those from Quark-Meson model~\cite{Ruggieri:2013cya} in the review Ref.~\cite{Gatto:2012sp} and was supported by the study within the Polyakov linear sigma model~\cite{Tawfik:2014hwa,Tawfik:2015tga}. However, in the extended Polyakov linear sigma model~\cite{Mizher:2010zb} and Polyakov-Quark-Meson model~\cite{Li:2019nzj}, other researchers found  a splitting between $\chi$SR and deconfinment: the $T_c$ increases with $B$ for $\chi$SR but  decreases for deconfinement. In principle, these features are consistent with the spirits of the magnetic catalysis effect and asymptotic freedom in strong magnetic fields, but most recent lattice QCD (LQCD) simulations do not support such splitting up to quite large $B$~\cite{Bali:2011qj,Bali:2012zg}. It is not clear why the entanglement between chiral symmetry and confinement can be so strong.

Also in 2000, magnetic oscillations were discovered in dense and cold quark matter for chiral condensate. This is actually a demonstration of  de Haas–van Alphen (dHvA) effect in the relativistic case~\cite{Ebert:1999ht}, see Ref.~\cite{Chatterjee:2011ry,Andersen:2011ip,Avancini:2012ee,Costa:2013zca,Grunfeld:2014qfa,Ruggieri:2014bqa,Ferreira:2017wtx,Shao:2019hen} for further explorations. Even after turning on the diquark interaction channels, the dHvA effect penetrates into the color superconductor phase at larger baryon chemical potential $\mu_{\rm B}$ with the diquark condensate or $\mu_{\rm B}-B$ phase boundary oscillating~\cite{Noronha:2007wg,Fukushima:2007fc,Fayazbakhsh:2010bh,Cao:2015xja}. Actually, the case with large $\mu_{\rm B}$ is quite complicated: besides the color superconductivity, inhomogeneous phases are also possible around the first-order phase boundary between the homogeneous $\chi$SB and $\chi$SR phases. In Ref.~\cite{Frolov:2010wn}, the inhomogeneous phase of chiral density wave (CDW) was explored for finite $\mu_{\rm B}$ and $B$ and it was found to be always favored over $\chi$SB. The reason is that the wave vector of CDW induces an effective chemical potential to particles and antiparticles at the lowest Landau level (LLL), which then prefers to reduce the original $\mu_{\rm B}$ in order to minimize the thermodynamic potential. As expected, there is still a trail of dHvA effect in the features of order parameters and phase boundaries~\cite{Frolov:2010wn}. The work had been extended by considering another kind of inhomogeneous chiral condensate -- the solitonic modulation~\cite{Cao:2016fby}, and it turned out that the transitions remain second order up to large enough $B$ and the critical $\mu_{\rm B}$'s oscillate with $B$ at zero temperature. Later, the study with Ginzburg-Landau approximation seemed to favor the CDW phase over the solitonic modulation phase~\cite{Abuki:2016zpv,Abuki:2018iqp}.

In 2010, by increasing only the magnetic field in QCD vacuum, the ground state was found to transit to a phase with rho condensations, both neutral and charged~\cite{Chernodub:2010qx,Chernodub:2011mc}. The main argument was that the mass of charged rho mesons $\rho^\pm$ with typical spins would decrease to zero at some critical $B$, which then indicates the instability of QCD vacuum against the charged rho condensations with vortical structures~\cite{Chernodub:2010qx}. The situation is quite like that of charged vector bosons $W^\pm$ at stronger magnetic field and had been checked in two-flavor NJL model~\cite{Chernodub:2011mc,Frasca:2013kka,Liu:2014uwa}. However, the LQCD simulations from different groups~\cite{Hidaka:2012mz,Luschevskaya:2016epp,Bali:2017ian}, where there is no sign problem for the case of pure magnetic field, didn't support the vanishing of $\rho^\pm$ mass at all. The point is that rho mesons are composed of quarks and antiquarks, and the fundamental QCD theory does not allow vector condensations according to the Vafa-Witten theorem~\cite{Vafa:1983tf,Vafa:1984xg,Hidaka:2012mz}. In 2019, the contradiction between the conclusions of the two-flavor NJL model and LQCD was well discussed by extending the study to three-flavor NJL model~\cite{Cao:2019res}, and the underlying reason was attributed to mass splitting between $u$ and $d$ quarks in the external magnetic field. At this point, it is worth mentioning that the possibility of $\rho^\pm$ superconductivity can be shown to be much more realistic by introducing a large rotation parallel to $B$~\cite{Cao:2020pmm}.

In 2012, two serial papers of the Regensburg group on the QCD phase transition at finite temperature and magnetic field, Refs.~\cite{Bali:2011qj,Bali:2012zg}, had triggered great interests all over the world in the features of QCD matter in strong magnetic fields.  In these LQCD simulations, though the MCE was well verified at zero temperature, the pseudo-critical temperature was unexpectedly found to decrease with $B$, thus called "inverse magnetic catalysis effect" (IMCE). Incidentally, they discovered one year later that the features of gluon action density are similar to those of chiral condensate in strong magnetic field~\cite{Bali:2013esa}, which then binds $\chi$SB and confinement even more tightly. After the watershed year of 2012, the number of works exploring the effects of magnetic field grew rapidly.  First of all, the following years saw a lot of attempts to explain the IMCE, which mainly fit into three categories: inhibition from collective excitations~\cite{Fukushima:2012kc,Mao:2016fha,Mao:2017tcf}, chiral chemical potential effect~\cite{Fukushima:2010fe,Chao:2013qpa,Yu:2014sla,Yu:2014xoa,Cao:2014uva} and running coupling~\cite{Kojo:2012js,Ferrer:2014qka,Ferreira:2014kpa,Fraga:2013ova,Andersen:2014oaa,Farias:2014eca,Ayala:2014gwa,Farias:2016gmy,Mueller:2015fka,Endrodi:2019whh}. The scheme with running coupling has seen a much wider market, even though the concrete realizations of the IMCE were diverse. Of course, the concept of asymptotic freedom is well known in QCD, but it is still challenging to consistently work out the $B$ dependence of the running coupling.  We also note that there were some related works about the magnetic, baryon, and electric charge susceptibilities of QCD matter in strong magnetic field~\cite{Frasca:2011zn,Bali:2012jv,Bonati:2013vba,Kamikado:2014bua,Bali:2020bcn,Fu:2013ica,Fukushima:2016vix,Bhattacharyya:2015pra,Ferreira:2018pux,Ding:2021cwv}.

Other than chiral symmetry breaking and restoration, there are other topics of interest. Firstly, as charged pion superfluidity is an important concept in the explorations of QCD phase diagram, the possibility was checked in the presence of strong magnetic field~\cite{Cao:2015xja}. Usually, the mass of charged pions increases with $B$~\cite{Wang:2017vtn,Liu:2018zag,Mao:2018dqe,Coppola:2019uyr,Li:2020hlp,Kojo:2021gvm}, so charged pion superfluidity is more difficult in magnetic field which is consistent with the Meissner effect. However, in a rotational system, the effective isospin chemical potential, induced by the interplay between parallel magnetic field and rotation, seemed to favor the superfluidity~\cite{Liu:2017spl,Cao:2019ctl}, though some ambiguity with the Schwinger phase might change that conclusion~\cite{Chen:2019tcp}. Secondly, the uniform neutral pion ($\pi^0$) condensation was rediscovered in parallel electromagnetic (PEM) field due to the chiral anomaly effect~\cite{Cao:2015cka}, which was later found to induce other pseudoscalar condensations with $s$ quark components~\cite{Cao:2020pjq}. Before that, the domain walls of neutral pseudoscalars were proposed to exist in some windows of finite $\mu_{\rm B}$ and $B$~\cite{Son:2007ny}. Indeed, these results could be related to each other if one alternatively understands the temporal potential for electric field as a coordinate dependent chemical potential. We note that the possibility of charged pion superfluidity was also discussed in PEM at finite temperature and isospin chemical potential~\cite{Chao:2018ejd}.

\begin{figure}[!htb]
	\centering
	\includegraphics[width=0.45\textwidth]{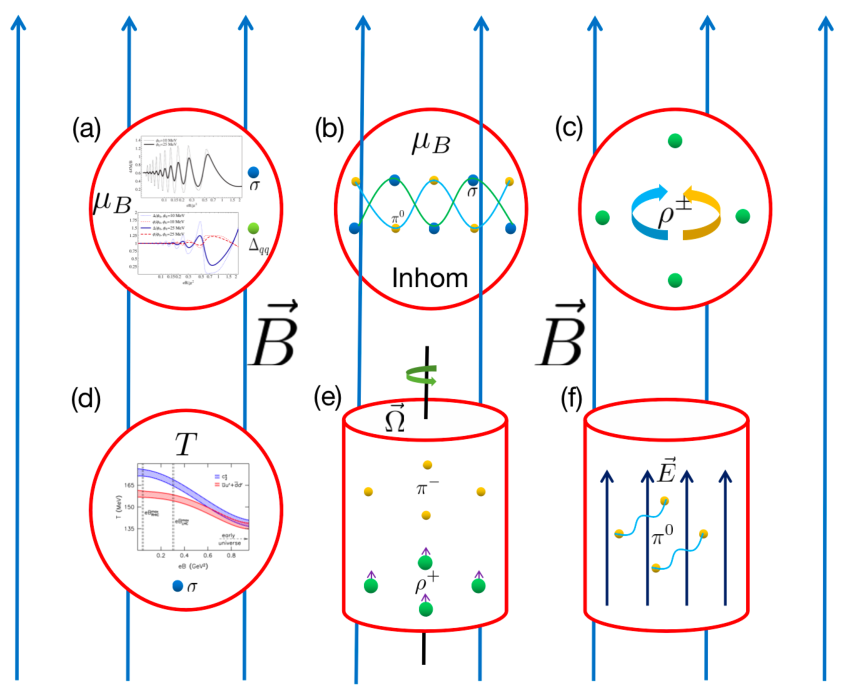}
	\caption{A summary of the possible phases and nontrivial features in the presence of strong magnetic field and additional circumstances: (a) de Haas–van Alphen effect in both chiral ($\sigma$) and diquark ($\Delta_{qq}$) condensates with the insertions from Ref.~\cite{Noronha:2007wg}; (b) inhomogeneous phases, such as chiral density wave, solitonic modulation and domain wall of $\pi^0$; (c) type-II vacuum superconductivity; (d) inverse magnetic catalysis effect  with the insertion from Ref.~\cite{Bali:2011qj}; (e) type-II charged pion superfluidity and rho superconductivity; (f) neutral pseudoscalar superfluidity, such as that of $\pi^0$. The colored bullets correspond to different types of particles which condensate in the systems for specific setups, see the notations in the plots (a-f).}\label{phases_B}
\end{figure}

 For the purpose of visualization, all the proposed phases and nontrivial features in the presence of strong magnetic field are summarized in Fig.\ref{phases_B}, which will be elucidated in more detail in the following separated sections. By recalling the previous subsection, it would be pragmatical to point out where the relevant environments could be found: large $\mu_B$ and $B$ in the interior of magnetars, pure strong $B$ in early Universe and large parallel $\Omega$, $E$ and $B$ in peripheral heavy ion collisions, though the last two cases are always accompanied by high temperatures. Concerning these topics, there are recently two relevant reviews on the magnetic effects, see Refs.~\cite{Miransky:2015ava,Andersen:2014xxa}; and several studies within holographic QCD are meant to facilitate our understanding of QCD matter from higher viewpoints, see Refs.~\cite{Rougemont:2015oea,Zayakin:2008cy,Preis:2010cq,Callebaut:2011uc,Preis:2012fh,Li:2016gtz,Gursoy:2017wzz,Bu:2018trt}.

Finally, it should be mentioned that there is another very important branch of research on the magnetic effect in QCD matter, the macroscopic anomalous transport phenomena. In 2008, D. E. Kharzeev, L. D. McLerran, H. J. Warringa and K. Fukushima proposed and studied the "chiral magnetic effect" (CME)~\cite{Kharzeev2008,Fukushima2008}, that is, a non-dissipative electric current could be induced along the external magnetic field when nonvanishing topological charges are generated due to the chiral anomaly. This is a completely novel notion and implies that quantum anomaly can be detected directly in experiments. In 2016, the CME was first observed in a QED system, the Weyl semi-metal system ZrTe5~\cite{Li2016}, where physicists set up an external PEM to induce chiral imbalance in the system. However, the realization of CME is much harder in QCD system as one has to appeal to the less controllable relativistic HICs in order to generate a system of approximately massless quarks. Recently, there was an important breakthrough in the BES II experiment of STAR group~\cite{STAR:2020crk}, where they seemed to detect a CME-driven charge separation in the Au-Au collisions. More conclusive results will have to be waited until the outcomes of isobar experiments in the near future. Furthermore, there are cousins of CME in the family of macroscopic anomalous transport where the external magnetic field is presented together with various circumstances, to name some: chiral separation effect~\cite{Son:2004tq,Metlitski:2005pr}, chiral magnetic wave~\cite{Kharzeev:2010gd}, magnetovorticity~\cite{Hattori:2016njk}, and chiral electric effect~\cite{Cao:2021jjy,Yamamoto:2021gts}. The interested readers could refer to the reviews Refs.~\cite{Liao:2014ava,Kharzeev:2015znc,Huang:2015oca,Kharzeev:2020jxw} for more details.

\subsection{Theoretical basises for quantum EM effect}\label{subsec:SRF}

It will be very helpful to briefly introduce the sophisticated mathematic tools for theoretical studies of quantum EM effects: the well-known Schwinger formalism~\cite{Schwinger:1951nm} and the more intuitive Ritus formalism~\cite{Ritus:1972ky,Ritus:1978cj}. In 1951, American physicist J.S. Schwinger formally worked out the fermion propagator directly in arbitrary configuration of external EM field by introducing proper time integral. Within his beautiful framework, Schwinger subtly transferred the problem of deriving the eigenvalue of conjugate momentum operator to that of the expectation value in quantum field theory~\cite{Schwinger:1951nm} -- that is the reason why the auxialiary parameter was called "proper time". In 1972, Russian physicist V.I. Ritus managed to derive the Green's functions of electron and photon in a plane-wave EM background field~\cite{Ritus:1972ky} by adopting the well-known formal eigenfunctions for the Dirac equation~\cite{Wolkow:1935zz,Nikishov:1964zza}. In 1978, he elegantly solved the Dirac equation in a constant and parallel EM field and reproduced Schwinger's result after substituting the eigenfunctions into the basic definition of the fermion propagator~\cite{Ritus:1978cj}. As we know, particle propagators are the basic computing elements in quantum field theory -- armed with those and interaction vertices, all physical quantities can be evaluated in principle. Actually, Schwinger formalism is more powerful as he presented the propagator in such a compact form that regularization and higher-loop calculations are much more convenient, see the first part of Ref.~\cite{Cao:2014uva} for example. But Ritus formalism is simpler especially when we are only interested in the thermodynamic potential of the system, where only the eigenvalues are required from the Dirac equation. Furthermore, Ritus formalism provides a much general procedure to construct particle propagators when magnetic field is present together with extra parameters, see the second part of Ref.~\cite{Cao:2014uva} for example.

\subsubsection{Schwinger formalism}
In the following, we roughly expound the Schwinger formalism for charged fermions first~\cite{Schwinger:1951nm}. In Euclidean space, the Dirac equation in the presence of external EM field is given by
\begin{eqnarray}
\left[\gamma^\mu\Pi_\mu(x)+m\right]\psi(x)\equiv\left[\gamma\cdot\Pi(x)+m\right]\psi(x)=0,\label{DE}
\end{eqnarray}
where $\Pi_\mu\equiv-i\partial_\mu-qA_\mu$ is the conjugate momentum with $q$ the charge and $A_\mu$ the vector potential. Then, according to the definition of the Green's function: $G(x,x')= i\langle{\cal{T}}\psi(x)\bar{\psi}(x')\rangle$, the corresponding equation of motion (EOM) is 
\begin{eqnarray}
\left[\gamma\cdot\Pi(x)+m\right]G(x,x')=\delta^4(x-x'),
\end{eqnarray}
or in matrix form with space-time indices:
\begin{eqnarray}
\left[\gamma\cdot\Pi+m\right]G=1.
\end{eqnarray}
So, the Green's function is solely determined by $\Pi_\mu$ and can formally expressed as:
 \begin{eqnarray}
G&=&\left[\gamma\cdot\Pi+m\right]^{-1}=\left[-\gamma\cdot\Pi+m\right]\left[-(\gamma\cdot\Pi)^2+m^2\right]^{-1}\nonumber\\
&=&i\left[-\gamma\cdot\Pi+m\right]\int_0^\infty {\rm d}s~e^{-i\left[-(\gamma\cdot\Pi)^2+m^2-i\,\epsilon\right]s},\label{Green's}
\end{eqnarray}
where $\epsilon\rightarrow 0^+$ guarantees the convergence of the proper time integral. Now, regardless of the trivial part $m^2-i\,\epsilon$, we can immediately identify the integrand $e^{-i\left[-(\gamma\cdot\Pi)^2\right]s}$ as a partition function with $s$ the time and 
$$H\equiv-(\gamma\cdot\Pi)^2=\Pi^2-{1\over 2}qF\!:\!\sigma$$
the Hamiltonian. Here, $F\!:\!\sigma\equiv F^{\mu\nu}\sigma_{\mu\nu}$ with $\sigma_{\mu\nu}={i\over 2}[\gamma_\mu,\gamma_\nu]$ and the EM field strength tensor $F_{\mu\nu}\equiv\partial_\mu A_\nu-\partial_\nu A_\mu$.

To derive the fermion propagator in coordinate space, we are interested in the transition amplitude of the Green's function Eq.\eqref{Green's} between two different space-times, that is, $\langle x|G|x'\rangle$ or alternatively $\langle x|e^{-iHs}|x'\rangle$. By adopting Heisenberg's picture, we regard $x_\mu(s)$ and $\Pi_\mu(s)$ as the expectation values at the instant $s$, which are given by
$$x_\mu(s)=\bra{0}e^{iHs}x_\mu e^{-iHs}\ket{0},\ \Pi_\mu(s)=\bra{0}e^{iHs}\Pi_\mu e^{-iHs}\ket{0}.$$
Then, the corresponding EOMs follow directly as
\bea
\!\!\!\!{\di x_\mu(s)\over \di s}\!\!\!&=&\!\!\!-i[x_\mu,H]=2\Pi_\mu(s),\\ 
\!\!\!\!{\di \Pi_\mu(s)\over \di s}\!\!\!&=&\!\!\!-i[\Pi_\mu,H]=2qF_{\mu\nu}\Pi^\nu\!-\!iq{\partial F_{\mu\nu}\over \partial x_\nu}\!+\!{q\over 2}{\partial \sigma\!:\!F\over \partial x_\mu},
\eea
where average over the eigenstate should also be understood for the Poisson brackets. This review mainly focuses on the case of constant EM field, then the EOMs become simply
\bea
{\di x(s)\over \di s}=2\Pi(s), \ {\di \Pi(s)\over \di s}=2qF\Pi(s)\label{EOMs}
\eea
in matrix form, which can be immediately solved to give
\bea
\Pi(s)&=&e^{2qFs}\Pi(0),\nonumber\\
 x(s)-x(0)&=&(e^{2qFs}-1)(qF)^{-1}\Pi(0).\label{Pis}
\eea
For future use, we can reversely express $\Pi(0)$ as functions of the space-time coordinates $x(s)-x(0)$:
\bea
\Pi(0)&=&qF(e^{2qFs}-1)^{-1}[x(s)-x(0)]\nonumber\\
&=&e^{-qFs}{\cal P}[x(s)-x(0)],\label{Pi0}\\ 
&&{\cal P}={1\over 2}qF[\sinh(qFs)]^{-1}.\nonumber
\eea

We are now in a position to calculate the key quantity for the Green's function, the transition amplitude between two different coordinates:
$$\langle{x'(s)}\ket{x''(0)}\equiv\bra{x'}e^{-iHs}\ket{x''}$$
with the differential equation 
\bea
i{\di \langle{x'(s)}\ket{x''(0)}\over \di s}=\bra{x'(s)}H\ket{x''(0)}.\label{diff}
\eea
Recalling the explicit form of the Hamiltonian, we have
\bea
H\!\!&=&\!\!{\cal{T}}~[x(s)-x(0)]{\cal P}^2[x(s)-x(0)]-{i\over 2}\tr\ qF\coth(qFs)\nonumber\\
&&-{1\over 2}qF:\sigma;
\eea
thus Eq.\eqref{diff} becomes explicitly
\bea
i{\di \langle{x'(s)}\ket{x''(0)}\over \langle{x'(s)}\ket{x''(0)}}\!\!\!&=&\!\!\!\left[\Delta x{\cal P}^2\Delta x\!-\!{i\over 2}\tr\ qF\coth(qFs)\right.\nonumber\\
&&\left.-{1\over 2}qF:\sigma\right]\di s
\eea
with $\Delta x=x'-x''$ ,which can be solved to give
\bea
\langle{x'(s)}\ket{x''(0)}&=&{C(x',x'')\over s^2}e^{{i\over4}\Delta x\, qF\coth(qFs)\Delta x}\nonumber\\
&&e^{{1\over2}\tr\ln(2{\cal P}s)+{i\over 2}qF:\sigma s}.\label{xsx0}
\eea
Here, the $s$-independent coefficient $C(x',x'')$ can be fixed by applying the differential equations for $\Pi_\mu$:
\bea
\!\!\!\!\!\!\!\!\left[-i\partial_\mu'\!-\!qA_\mu(x')\right]\!\langle x'(s) |x''(0)\rangle\!\!\!&=&\!\!\!\langle x'(s)|\Pi_\mu(s)| x''(0) \rangle,\\
\!\!\!\!\!\!\!\!\left[i\partial_\mu''\!-\!qA_\mu(x'')\right]\!\langle x'(s) |x''(0)\rangle\!\!\!&=&\!\!\!\langle x'(s)|\Pi_\mu(0)| x''(0) \rangle.
\eea
Substituting the explicit forms of $\Pi(s)$, $\Pi(0)$ and $\langle{x'(s)}\ket{x''(0)}$ shown in Eqs.\eqref{Pis}, \eqref{Pi0} and \eqref{xsx0}, we find the differential equations for $C(x',x'')$ as:
$$[-i\partial_\mu'-qA_\mu(x')]C(x',x'')\!=\![i\partial_\mu''-qA_\mu(x'')]C(x',x'')=0.$$
Thus, $C(x',x'')$ is gauge dependent with the solution:
\bea
C(x',x'')=-{i\over (4\pi)^2}\,e^{i\int_{x''}^{x'}qA\cdot \di x},
\eea
where the prefactor was fixed according to the normalization condition $\langle{x'(s)}\ket{x''(0)}_{s\rightarrow0}=\delta^4(x'-x'')$.

Finally, inserting all the relevant results into Eq.\eqref{Green's}, we find the explicit form of the Green's function as
\begin{widetext}
 \begin{eqnarray}
G(x',x'')={e^{i\int_{x''}^{x'}qA\cdot \di x}\over (4\pi)^2}\int_0^\infty {{\rm d}s\over s^2}\left[m-{1\over2}qF(\coth(qFs)-1)\Delta x\cdot\gamma\right]e^{-i\,m^2s+{i\over4}\Delta x\,qF\coth(qFs)\Delta x+{1\over2}\tr\ln(2{\cal P}s)+{i\over 2}qF:\sigma s}.
\end{eqnarray}
More explicitly, the last but one exponential had been evaluated as~\cite{Schwinger:1951nm}:
\bea
e^{{1\over2}\tr\ln(2{\cal P}s)}&=&-{q^2({\bf E\cdot B})s^2\over\Im \cos\left[q\sqrt{({\bf B}+i\, {\bf E})^2}s\right]},
\eea 
and the last one can be presented as
\bea
e^{{i\over 2}qF:\sigma s}&=&\sum_{n=0}^\infty{1\over 2n!}\left({i\over 2}qF:\sigma s\right)^{2n}+{i\over 2}qF:\sigma s\sum_{n=0}^\infty{1\over (2n+1)!}\left({i\over 2}qF:\sigma s\right)^{2n}\nonumber\\
&=&\sum_{n=0}^\infty{1\over 2n!}\left(iq({\bf B}-i\gamma^5{\bf E})s\right)^{2n}+{i\over 2}qF:\sigma s\sum_{n=0}^\infty{1\over (2n+1)!}\left(iq({\bf B}-i\gamma^5{\bf E})s\right)^{2n}\label{Lexp}
\eea
\end{widetext}
by utilizing the property $\left({1\over 2}F:\sigma \right)^2=({\bf B}-i\gamma^5{\bf E})^2$~\cite{Schwinger:1951nm}. It is very convenient to apply Eq.\eqref{Lexp} to crossed EM field with $({\bf B}-i\gamma^5{\bf E})^2=B^2-E^2\equiv I_1^2$, we get
\bea
e^{{i\over 2}qF:\sigma s}=\cos\left(qI_1s\right)+{i\,F:\sigma\over 2I_1}\sin\left(qI_1s\right).
\eea
For parallel EM field, the electric and magnetic parts in $F:\sigma$ commute with each other, thus the last term can be rewritten as
$$e^{{i\over 2}qF:\sigma s}=e^{{i\over 2}q(F:\sigma)_E s}e^{{i\over 2}q(F:\sigma)_B s},$$
which is just the product of the pure electric and pure magnetic results.

\subsubsection{Ritus formalism}\label{Ritus}
Next, we introduce the Ritus formalism by only focusing on the case of parallel EM field. For the fields along $z$ direction, the vector potential can be chosen as $A_\mu=(0,0,Bx_1,-Et)$ without loss of generality. And then there are four independent operators that commute with each other:
\bea
-i\partial_2,-i\partial_3,(-i\gamma\cdot\Pi)^2,\Pi_1^2+\Pi_2^2-qB\sigma_{12}\label{Opes}
\eea
instead of the four energy momenta in the case without external field. So, their eigenvalues $P=(p_2,p_3,p^2,2lqB)$ ($l\in \mathbb{N}^+$) can be simultaneously used to characterize the eigenstates. Taking into account the spinor nature of the eigenfuntions, the operators in Eq.\eqref{Opes} also commute with the spin operator ${1\over 2}\Sigma_3\equiv{1\over2}\sigma_{12}$ and chirality operator $\gamma^5$, whose eigenvalues can be $\tau=\pm {1\over 2}$ and $h=\pm1$. Hence, in the more convenient Weyl representation, the eigenfunctions were found to be of the forms~\cite{Berestetskii1971,Akhiezer1965}
\bea
\phi_{p\tau h}={e^{i{\pi\over4}\lambda}\Gamma(-\lambda)e^{i\,(p_2x_2+p_3x_3)}\over (2\pi|E/B|)^{1\over4}(n!)^{1\over2}}D_n(\bar{x}_1)D_\lambda(\bar{t})v_{\tau h}.
\eea
Here, the vector $v_{\tau h}$, given by different columns of a $4\times4$ identity matrix, are the eigen bispinors of spin and chirality operators; and $D_n(\bar{x}_1)$ and $D_\lambda(\bar{t})$ are parabolic cylinder functions with the indices and variables given by
\bea\label{nl}
\left\{\!\!\!\!\!\!\begin{array}{ll}
&n\!=\!l+\tau\mathcal{S}(qB)-{1\over 2},\ \ \ \ \ \ \ \ \bar{x}_1\!=\!(2 |qB|)^{1\over2}\left(x_1\!-\!{p_2\over qB}\right),\\
&\lambda\!=\!{p^2\!-\!|2kqB|\over -i|2qE|}\!-\!h\tau\mathcal{S}(qE)\!-\!{1\over 2}, \bar{t}\!=\!e^{i{\pi\over4}}\!(2 |qE|)^{1\over2}\!\!\left(\!t\!+\!{p_3\over qB}\!\right).
\end{array}\right.
\eea
Note that $\mathcal{S}(x)$ is the sign function in Eq.\eqref{nl} and the following. Then, for given $p$, we can define an eigenfunction matrix for the particle with positive frequency:
$$\Phi_p^+=(\phi_{p++},\phi_{p-+},\phi_{p+-},\phi_{p--})$$
and the negative-frequency one for antiparticle is just $\Phi_p^-=\Phi_p^+|_{\bar{t}\rightarrow-\bar{t}}$.

Now, the magic is that if we apply the Dirac operator $\gamma\cdot\Pi$ to $\Phi_p^\pm$, we would find
\bea
\gamma\cdot\Pi\, \Phi_p^\pm=\Phi_p^\pm i\,\gamma\cdot\tilde{p}^\pm,
\eea
where the effective energy momenta $\tilde{p}^\pm$ are respectively:
\bea
&&\tilde{p}^t_1=0,\tilde{p}^t_2=-\mathcal{S}(qB)|2nqB|^{1\over2},\nonumber\\
&&\tilde{p}^t_0=t{2|nqB|-p^2+2|qE|\over2(2|qE|)^{1/2}},\nonumber\\
&&\tilde{p}^t_3=t{2|nqB|-p^2-2|qE|\over2\mathcal{S}(qE)(2|qE|)^{1/2}}.
\eea
Then, by defining the eigenfunctions of the Dirac operator as $\psi_p^\pm\equiv \Phi_p^\pm u^\pm_p$, the Dirac equation Eq.\eqref{DE} simply reduces to
$$(i\,\gamma\cdot\tilde{p}^\pm+m)u^\pm_p=0,$$
which is just the EOM for "free" fermions and determines the eigenenergy $p_0$ through the equation $(\tilde{p}^t)^2+m^2=0$. At this step, one should keep in mind that $\tilde{p}^t_0\rightarrow-i\,t p_0$ and $\tilde{p}^t_3\rightarrow tp_3$ in the absence of electric field, thus $p_0$ is well defined and can be easily solved. Such scheme can also be used in the case with finite temperature or chemical potential and is enough to obtain the corresponding thermodynamic potential. 

On the $\Phi_p^\pm$ basis, the fermion propagator is free, that is, $G_p^\pm=i/(i\,\gamma\cdot\tilde{p}^\pm+m)$. Then the propagator in coordinate space can be derived directly by taking a Fourier-like transformation:
\bea
G(x,x')\!\!&=&\!\!\int\di^4p\sum_{t=\pm}\psi_p^t(x)G_p^t\gamma^0\psi_p^{t\dagger}(x')\gamma^0\nonumber\\
\!\!&=&\!\!i\!\int\!\!\di^4p\!\sum_{t=\pm}\!\! \Phi_p^t(x){m\!-\!i\gamma\cdot\tilde{p}^t\over p^2+m^2}\gamma^0\Phi_p^{t\dagger}(x')\gamma^0
\eea
with $\int\di^4p=\sum_{n=0}^\infty\int \di p^2\di p_2\di p_3$ for brevity. We're glad to find that the propagator can be divided into components with different Landau levels. It would be much simpler to take the LLL approximation for effective calculations, when $|qB|^{1/2}$ is much larger than other energy scales of the system, such as $|qE|^{1/2}, m, T$ and $\mu$.

\subsection{Organizations and conventions}

The review is organized as follows. In Sec.\ref{sec:chiral}, we explore the patterns of chiral symmetry breaking and restoration under the interplays between magnetic field and other physical parameters, such as temperature, baryon chemical potential and angular velocity. Specially, in Sec.\ref{ssubsec:IMCE}, the inverse magnetic catalysis effect at finite $T$ is naturally explained by adopting the $B$-dependent running coupling extracted from the LQCD data for neutral pion mass; and the inhomogeneous phases are compared with each other by utilizing a consistent regularization scheme at finite $\mu_{\rm B}$ in Sec.\ref{subsec:inhom}. In Sec.\ref{sec:pseudo}, neutral pseudoscalar condensations are discussed in parallel electromagnetic field, particularly in-in and in-out formalisms are compared in both two- and three-flavor cases. Due to the differentiation between $u$ and $d$ quark eigenstates in external magnetic field, mainly the Ginzburg-Landau approximation is implemented to study the charged pion superfluidity in Secs.\ref{sec:cpi} and charged rho superconductivity in Sec.\ref{sec:crho}, respectively. To provide a more complete scope for the studies of QCD phases, we briefly cover the topic about color superconductivity in a rotated magnetic field in Sec.\ref{sec:CS}. Finally, we conclude and give some perspectives in Sec.\ref{sec:summ}.

Conventions: for brevity, all the physical variables are assumed to be positive if not specified in the context, such as $e, B, E, T, \mu_{\rm B},\mu_{\rm I}$ and $\Omega$; the multiplication of electric charge and EM field is prior to other arithmetics, which means $1/eB=1/(eB)$. To facilitate reading, all the abbreviations are listed here in alphabetical order:\\ {\bf 2SC}\ \ \ \quad -- two-flavor color superconductor,\\ ${\bf B}\parallel\boldsymbol{\Omega}$\ \ \; -- parallel magnetic field and rotation,\\ {\bf CDW}\ \ \, -- chiral density wave,\\ {\bf CFL}\ \ \ \;\, -- color-flavor locking,\\ {\bf ChPT}\ \; -- chiral perturbation theory,\\ {\bf CME}\ \ \ \ -- chiral magnetic effect,\\ {\bf dHvA}\ \ \ -- de Haas–van Alphen,\\ {\bf EOM}\ \ \ \ -- equation of motion,\\ {\bf EM}\ \ \ \,\quad -- electromagnetic,\\ {\bf EV}\ \ \ \,\;\quad -- expectation value,\\ {\bf GL}\ \ \ \,\;\quad -- Ginzburg-Landau,\\ {\bf GMOR} -- Gell-Mann-Oakes- Renner,\\ {\bf GN}\ \ \;\;\quad -- Gross-Neveu,\\ {\bf HIC}\ \ \,\quad -- heavy ion collision,\\ {\bf IMCE}\ \; -- inverse magnetic catalysis effect,\\ {\bf LLL}\quad\ \; -- lowest Landau level,\\ {\bf LQCD}\ \,  -- lattice QCD,\\ {\bf MCE}\ \ \;\, -- magnetic catalysis effect,\\ {\bf MFA}\ \ \;\;  -- mean field approximation,\\ {\bf NJL}\ \ \ \ \; -- Nambu–Jona-Lasinio,\\ {\bf OAM}\ \ \ \ -- orbital angular momentum,\\ {\bf PEM}\ \ \ \, -- parallel electromagnetic,\\ {\bf QCD}\ \ \ \, -- quantum chromodynamics,\\ {\bf RPA}\ \ \ \; -- random phase approximation,\\ {\bf SM}\ \ \ \ \quad -- solitonic modulation,\\ {\bf SPP}\ \ \ \ \, -- Schwinger pair production,\\{\bf VS}\,\ \ \ \ \quad -- vacuum superconductor,\\ {\bf WZW}\ \;  -- Wess-Zumino-Witten,\\$\boldsymbol{\chi}${\bf SB}\ \ \ \ \, -- chiral symmetry breaking,\\ $\boldsymbol{\chi}${\bf SR}\ \ \ \ \, -- chiral symmetry restoration.

\section{Chiral symmetry breaking and restoration}\label{sec:chiral}
\subsection{Inverse magnetic catalysis at finite temperature}\label{subsec:IMC}

\subsubsection{Magnetic catalysis effect -- the baseline}\label{ssubsec:MCE}

As mentioned in the introduction, the studies of strong magnetic field effect in NJL model established the well accepted concept of magnetic catalysis effect on chiral symmetry breaking in QCD vacuum~\cite{Gusynin:1994xp,Gusynin:1995nb}. The MCE was then taken as the baseline for nontrivial explorations of systems in more complicated circumstances besides the magnetic field. So, first of all, we'd like to take the two-flavor NJL model for example to demonstrate how and in what sense the MCE was developed. Since the Lagrangian of NJL model maintains the same symmetries as the basic QCD theory~\cite{Nambu:1961fr,Klevansky:1992qe}, it is very useful and convincing to study the symmetry related properties of QCD system.

In the presence of a constant magnetic field, the Lagrangian density of NJL model can be modified from the original one~\cite{Nambu:1961fr,Klevansky:1989vi} to
\begin{eqnarray}
{\cal L}_{\rm NJL}=\bar\psi(i\slashed{D}-m_0)\psi+G[(\bar\psi\psi)^2+(\bar\psi i\gamma_5{\tau}\psi)^2]\label{LNJL2}
\end{eqnarray}
by adding vector potential $A_\mu=(0,0,-Bx,0)$ to the derivative thus defining a covariant one: $D_\mu=\partial_\mu-iQA_\mu$~\cite{Gusynin:1994xp,Gusynin:1995nb}. By neglecting possible feedback to the effective coupling constant $G$, this is a minimal coupling way to introduce a constant magnetic field along $z$-direction. Note that $\psi=(u,d)^T$ represents the two-flavor quark field, $m_0$ is the current mass of quarks and $Q={\rm diag}(q_u,q_d)$ is the electric charge matrix in Eq.\eqref{LNJL2}. 

To explore the feature of chiral symmetry breaking and restoration in magnetic field, it is intuitive enough to derive the thermodynamic potential in mean field approximation (MFA) level. For that purpose, we take Hubbard-Stratonovich transformation with the auxiliary fields $\sigma=-2G\bar\psi\psi$ and ${\boldsymbol \pi}=-2G\bar\psi i\gamma_5{\boldsymbol \tau}\psi$ and then get a bosonized action~\cite{Gusynin:1994xp,Gusynin:1995nb}:
\begin{eqnarray}\label{action}
{S}_b=\int{d^4x}{\sigma^2\!+\!{\boldsymbol \pi}^2\over 4G}-{\rm Tr}\ln\left[i{\slashed D}\!-\!m_0\!-\!\sigma\!-\!i\gamma_5{\boldsymbol \tau\cdot\pi}\right]
\end{eqnarray}
after integrating out the quark degrees of freedom. Eventually, the thermodynamic potential can be formally obtained by setting $\sigma=m-m_0$ and ${\boldsymbol \pi}=0$ in Eq.\eqref{action}:
\begin{eqnarray}
V(m)={(m-m_0)^2\over 4G}-{{\rm Tr}\ln\left[i{\slashed D}-m\right]\over V_{\rm 3+1}}.\label{Vm}
\end{eqnarray}
From finite temperature field theory~\cite{Kapusta2006}, in order to derive the explicit form of $V(m)$, it is enough to solve the eigenenergies from the EOM Eq.\eqref{DE}. According to the Ritus formalism, one can immediately determine the eigenenergies as
$$E_{\rm fn}(p_3)=\sqrt{2n|q_{\rm f}B|+p_3^2+m^2},$$
see the discussions in Sec.\ref{Ritus}. Then, the thermodynamic potential follows directly as
\begin{widetext}
\begin{eqnarray}
V_{\rm BT}(m)={(m-m_0)^2\over 4G}-2N_c\sum_{\rm f=u,d}{|q_{\rm f}B|\over2\pi}\sum_{n=0}^\infty\alpha_{\rm n}\int_{-\infty}^\infty{\di p_3\over2\pi} \left[E_{\rm fn}(p_3)+2T\ln\left(1+e^{-E_{\rm fn}(p_3)/T}\right)\right]\label{Omeg_BT}
\end{eqnarray}
for finite temperature, where $\alpha_{\rm n}=1-\delta_{n0}/2$ and ${|q_{\rm f}B|\over2\pi}$ are the degeneracy factors for higher Landau levels. 

The thermodynamic potential is divergent and needs regularization. Usually, soft-cutoff schemes were directly applied to the $B$-dependent parts, because artifacts can be rendered in hard-cutoff schemes when $B$ is comparable to the cutoffs~\cite{Fukushima:2010fe,Mao:2016fha,Chao:2013qpa}. The Pauli-Villars regularization is one of the most popular soft-cutoff schemes and the hard-cutoff schemes include three-momentum, four-momentum, and proper-time regularizations~\cite{Klevansky:1992qe}. However, as the second term of Eq.\eqref{Omeg_BT} is quartic divergent, we can isolate the divergence to the vacuum and $m$-independent parts which then can be regularized in a usual way~\cite{Ebert:1999ht,Cao:2014uva,Menezes:2008qt}. For that sake, it is much more convenient to adopt the Schwinger formalism to represent the $B$-dependent divergent part with "vacuum regularization" as
\bea
V_{\rm B}(m)&\equiv&{N_c\over8\pi^2}\sum_{\rm f=u,d}\int_0^\infty{\di s\over s^3}\left(e^{-m^2s}-1\right)\left[{q_{\rm f}Bs\over\tanh(q_{\rm f}Bs)}-1\right]+{N_c\over8\pi^2}\sum_{\rm f=u,d}\int_{\Lambda_s}^\infty{\di s\over s^3}\left[{q_{\rm f}Bs\over\tanh(q_{\rm f}Bs)}-1\right]\nonumber\\
&&-4N_c\int^\Lambda{\di^3p\over(2\pi)^3}(p^2+m^2)^{1\over2}.\label{VB}
\eea
Here, the term with $m\rightarrow0$ is introduced to cancel the left divergence in the pure $B$ part and $\Lambda$ is the three-momentum cutoff for the vacuum part. Usually within the "magnetic field independent regularization", the second term was dropped for the explorations of chiral symmetry since it is $m$-independent~\cite{Ebert:1999ht}, but it has to be kept in order to give qualitatively consistent magnetization as other schemes~\cite{Avancini:2020xqe}. As had been compared to the proper-time regularization~\cite{Gusynin:1994xp,Gusynin:1995nb}, the advantage of "vacuum regularization" was that it guaranteed the Goldstone theorem in chiral limit~\cite{Cao:2014uva}. And it was demonstrated by adopting such regularization that the oscillations of diquark condensate with magnetic field, found in the hard and smooth function regularizations, were nothing but artifacts~\cite{Allen:2015paa}. Note that the first term in $V_{\rm B}$ can be integrated out to give~\cite{Ebert:1999ht}
\bea
-N_c\sum_{\rm f=u,d}{(q_{\rm f}B)^2\over2\pi^2}\left[\zeta^{(1,0)}(-1,x_{\rm f})-{1\over2}(x_{\rm f}^2-x_{\rm f})\ln x_{\rm f}+{x_{\rm f}^2\over4}\right]
\eea
with $x_{\rm f}=m^2/|2q_{\rm f}B|$. Finally, the regularized gap equation follows from $\partial_mV_{\rm BT}^r(m)=0$ as
\bea 
0&=&{m-m_0\over 2G}-4N_c\int^\Lambda{\di^3p\over(2\pi)^3}{m\over (p^2+m^2)^{1\over2}}-{N_c\over4\pi^2}m\sum_{\rm f=u,d}\int_0^\infty{\di s\over s^2}e^{-m^2s}\left[{q_{\rm f}Bs\over\tanh(q_{\rm f}Bs)}-1\right]\nonumber\\
&&+4N_c\sum_{\rm f=u,d}{|q_{\rm f}B|\over2\pi}\sum_{n=0}^\infty\alpha_{\rm n}\int_{-\infty}^\infty{\di p_3\over2\pi} {m\over E_{\rm fn}(p_3)}{1\over1+e^{E_{\rm fn}(p_3)/T}},\label{GapE}
\eea
\end{widetext}
from which we can immediately identify that the second vacuum and third $B$-dependent terms both favor $\chi$SB while the last $T$-dependent one prefers $\chi$SR according to their signs. Here, we note that the greatest advantage of "vacuum regularization" is that it is consistent with the Goldstone theorem in chiral limit~\cite{Cao:2014uva}.

At zero temperature, the $B$-dependent term on the right-hand side of Eq.\eqref{GapE} converges to zero as 
\bea
{N_c\over4\pi^2}m\sum_{\rm f=u,d}|q_{\rm f}B|\ln x_{\rm f}\label{Bm0}
\eea
 in the limit $m\rightarrow0$, so there is always a nontrivial solution $m\neq0$ as long as $G>0$ and even in chiral limit $m_0=0$. Around $m=0$, the curvature
\bea
\partial^2_mV_{\rm BT}^r(m)\approx{N_c\over4\pi^2}\sum_{\rm f=u,d}|q_{\rm f}B|\ln x_{\rm f}
\eea
diverges as $-\infty$, so the trivial solution $m=0$ is a local maximum in chiral limit. In contrary, the nontrivial solution must correspond to the global minimum of $V_{\rm BT}^r(m)$ and the chiral symmetry is meant to be broken by arbitrary $B$~\cite{Miransky:2015ava}. At this point, we recall that there is a finite critical coupling $G_c$ for the spontaneous chiral symmetry breaking in the absence of $B$~\cite{Cao:2014uva,Miransky:2015ava}. Though the logarithmic divergence ensures the MCE at $T=0$ even in chiral limit, we will show later that the chiral symmetry can still be fully restored at finite temperature.

\begin{figure}[!htb]
	\centering
	\includegraphics[width=0.45\textwidth]{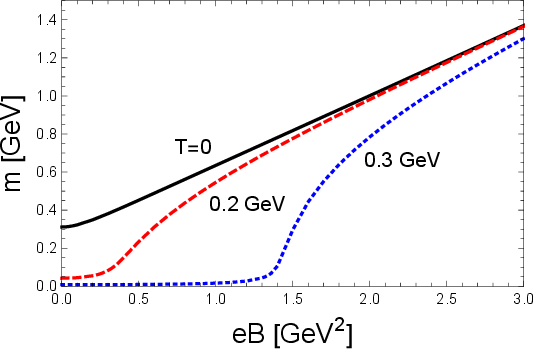}
	\caption{Dynamical quark mass $m$ as a function of magnetic field $eB$ for different temperature: $T=0,0.2$ and $0.3~{\rm GeV}$.}\label{MTB}
\end{figure}
To perform numerical calculations, the three parameters of the NJL model were fixed to $G=4.93~{\rm GeV}^{-2}$, $\Lambda=0.653~{\rm GeV}$ and $m_0=5~{\rm MeV}$ by fitting to the pion mass $m_\pi=134~{\rm MeV}$, pion decay constant $f_\pi=93~{\rm MeV}$ and quark condensate $\langle\bar\psi\psi\rangle=-2\times (0.25~{\rm GeV})^3$~\cite{Zhuang:1994dw}. Then, we demonstrate the numerical results in Fig.\ref{MTB}, where the quark mass is found to increase with $B$ for each $T$. Such feature is just the well-known "magnetic catalysis effect".

\subsubsection{The puzzle of inverse magnetic catalysis effect}\label{ssubsec:IMCE}

As mentioned in Sec.\ref{subsec:Meff}, the discovery of IMCE~\cite{Bali:2011qj,Bali:2012zg} was a watershed for the studies of strong magnetic field effect in QCD. Actually, the IMCE can be understood in another way: for a given temperature comparable to $T_c(B=0)$, the chiral condensate would decrease with larger magnetic field. This feature of course contradicts with the MCE illustrated in Fig.\ref{MTB} thus arouses lots of interests in magnetic effect and deeper thinkings on the properties of QCD. We have pointed out in the introduction that there are roughly three categories of explanations for IMCE. Since no one was yet commonly accepted, we would not lay out too much here but just choose the scenario of running coupling for demonstration, interested readers are referred to reviews Refs.~\cite{Miransky:2015ava,Andersen:2014xxa} for more discussions. Though it is unavoidable to introduce some free parameters in such scheme, the physical basis of asymptotic freedom was found to be quite solid for QCD system~\cite{Miransky:2002rp} and the whole formalism is still self-consistent in MFA. 

There are several forms of running coupling in the market: $B$ dependent Pad\`e ones~\cite{Ferreira:2014kpa,Endrodi:2019whh,Ahmad:2016iez} or $B$ and $T$ dependent ones~\cite{Farias:2014eca,Farias:2016gmy}. To avoid too many free parameters which might affect the physical value of the work, we mainly focus on the $B$ dependent scheme, such as that in Ref.~\cite{Endrodi:2019whh}. In Ref.~\cite{Endrodi:2019whh}, the authors worked with proper-time regularization and extracted the running coupling by fitting to the chiral condensate obtained in LQCD simulations~\cite{Bali:2012zg}. Then, the running coupling was found to reproduce the LQCD results for pseudocritical temperature quite well. However, it is a pity that the fitting process is incorrect as the chiral condensate defined in LQCD is actually half of that defined in NJL model, compare the Gell-Mann--Oakes--Renner relations in Ref.~\cite{Bali:2012zg} and Ref.~\cite{Klevansky:1992qe}. Furthermore, we have checked that the conclusion is really regularization dependent: with our choice of "vacuum regularization", the coupling constant even increases with $B$ thus couldn't reproduce the IMCE at all. 

\begin{figure}[!htb]
	\centering
	\includegraphics[width=0.45\textwidth]{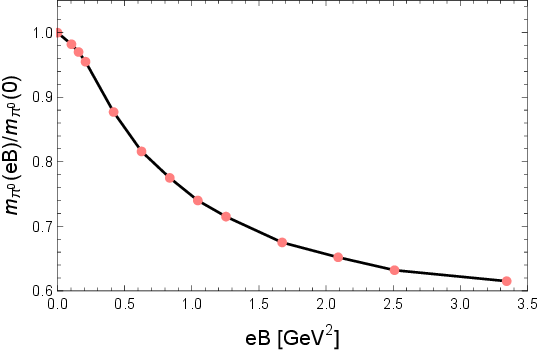}
	\includegraphics[width=0.45\textwidth]{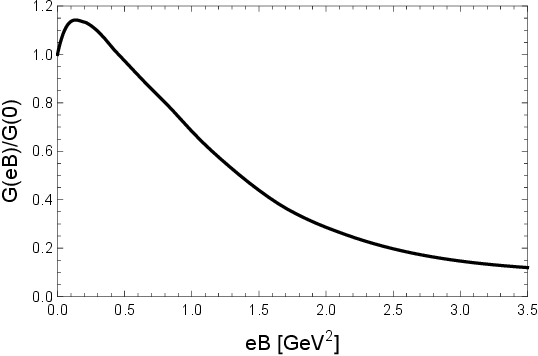}
	\caption{The fitting to the ratio $m_{\pi^0}(eB)/m_{\pi^0}(0)$ obtained in LQCD simulations~\cite{Ding:2020jui} (upper panel) and the associated running coupling $G(eB)$ for NJL model (lower panel).}\label{GB}
\end{figure}

Inspired by the previous scheme, we propose an alternative choice: extracting the running coupling from the $B$-dependent neutral pion mass $m_{\pi^0}$ obtained in LQCD simulations. As $\pi^0$ is still a true Goldstone boson in the presence of magnetic field in chiral limit, the mass is qualitatively more suitable to constrain the coupling constant of the chiral symmetric interactions in NJL model. Recently, only the lattice results with vacuum mass $m_{\pi^0}=220\,{\rm MeV}$ is available in the market~\cite{Ding:2020jui}. The fitting of $m_{\pi^0}$ and thus obtained running coupling are shown together in Fig.\ref{GB}, where the feature of asymptotic freedom can be well identified except for the small $B$ region. 

\begin{figure}
\centering
\begin{overpic}
[width=0.45\textwidth]{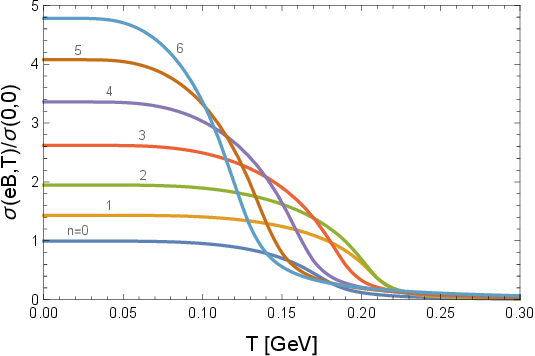}
\put(47,31){\includegraphics[width=0.22\textwidth]{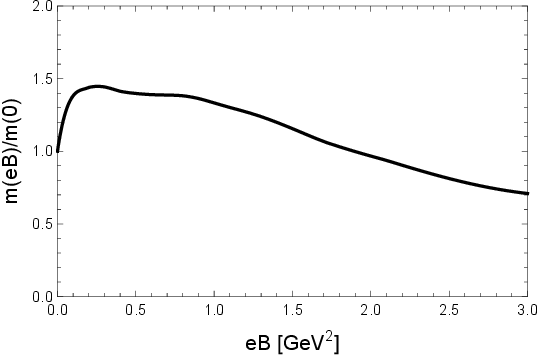}}
\end{overpic}
\caption{By adopting the running coupling in Fig.\ref{GB}, the chiral condensate $\sigma(eB,T)$ is presented as a function of temperature $T$ for several magnetic fields $eB={n/2}~{{\rm GeV}^2}~(n=0,1,\dots,6)$. Here, we insert the results for the dynamical mass $m$ at $T=0$.}\label{sigmBT}
\end{figure}

Finally, we evaluate the change of chiral condensate with magnetic field and temperature by adopting the running coupling, see Fig.\ref{sigmBT}. 
One can immediately tell the IMCE of the pseudocritical temperature for larger magnetic field, while the feature of MCE is still well kept for chiral condensate $\sigma$ at zero temperature. As had been pointed out by many authors~\cite{Kojo:2012js,Miransky:2015ava}, the dynamical mass $m$ is roughly different from $\sigma$ by the multiplying factor $-2G(eB)$; so the MCE doesn't necessarily imply the increasing of $m$, especially when $G(eB)$ decreases with $eB$. In fact, $m$ decreases with larger $eB$ in our scheme, see the insertion in Fig.\ref{sigmBT}. By applying such scheme to the Pauli-Villars regularization~\cite{Fayazbakhsh:2010bh,Cao:2015xja} which also guarantees Goldstone theorem in chiral limit, we are glad to find that these qualitative features remain true, refer to  Ref.~\cite{Mao:2018dqe} for the explicit expressions. Hence, we can safely conclude that the saturation of $m_{\pi^0}$~\cite{Ding:2020jui} and the phenomenon of IMCE~\cite{Bali:2011qj,Bali:2012zg} can be consistently explained by asymptotic freedom at large $B$. Nevertheless, as the effect of asymptotic freedom is not so significant at small $B$, artifacts might show up there due to the overall simpleness of the NJL model compared to QCD, see the increasing part in Fig.\ref{GB}.

Before concluding this section, we'd like to discuss, in the point of view of mathematics, a bit more about why solely the decreasing coupling with $B$ is enough to explain the IMCE at large magnetic field. For large $B$, the LLL dominates the contribution, thus the $T$ dependent term in Eq.\eqref{GapE} can be approximated by 
\bea
\!\!\!\!\!\!\!\!\!&&\!\!\!N_c\sum_{\rm f=u,d}{|q_{\rm f}B|\over\pi^2}\int_{0}^\infty{\di p_3} {m\over \sqrt{p_3^2+m^2}}{1\over1+e^{\sqrt{p_3^2+m^2}/T}}\nonumber\\
\!\!\!\!\!\!\!\!\!&=&\!\!\!N_c\sum_{\rm f=u,d}{m|q_{\rm f}B|\over\pi^2}\int_{0}^\infty \!\!{{\di p_3}\over \sqrt{p_3^2\!+\!\left({m\over T}\right)^2}}{1\over1\!+\!e^{\sqrt{p_3^2\!+\!\left({m\over T}\right)^2}}}.\label{Tpart}
\eea
Around the $T_c$ where $m$ is small, the integral can be roughly estimated as $\sim\ln(T/m)/2$, hence Eq.\eqref{Tpart} becomes simply
\bea
{N_c\over4\pi^2}m\sum_{\rm f=u,d}{|q_{\rm f}B|}\ln\left({T^2\over m^2}\right).\label{Tm0}
\eea
Comparing the logarithmic terms in Eq.\eqref{Bm0} and Eq.\eqref{Tm0}, it is interesting to notice that the role of $T$ in Eq.\eqref{Tm0} is quite similar to the role of $|2q_{\rm f}B|$ in Eq.\eqref{Bm0}, though their contributions are opposite in Eq.\eqref{GapE}. Since the $T$ dependent term can be evaluated as 
$${N_c\over3}mT^2$$
for small $m$ in the absence of magnetic field, one may immediately recognize that it is the magnetic field itself amplifies the effect of relatively small $T$ through the prefactor $|q_{\rm f}B|$.
Collecting the logarithmic terms together, the total contribution would become linear in $m$:
\bea
{N_c\over4\pi^2}m\sum_{\rm f=u,d}{|q_{\rm f}B|}\ln\left({T^2\over |2q_{\rm f}B|}\right),\label{LLLT}
\eea
thus the chiral symmetry can be fully restored at large enough $T$ in chiral limit. 

For $m_0=0$, we are even able to determine the critical temperature for the second-order chiral transition by setting the coefficient of $m$ on the right-hand side of Eq.\eqref{GapE} to zero. In this case, $m$ can be consistently set to $0$ in the coefficient, and we find the critical condition:
\bea
{1\over 2G(eB)}\!-\!{N_c\over\pi^2}\Lambda^2\!+\!{N_c\over4\pi^2}\!\sum_{\rm f=u,d}{|q_{\rm f}B|}\ln\left({T_c^2\over |2q_{\rm f}B|}\right)=0\label{criticalC}
\eea 
according to the Ginzburg-Landau (GL) theory. Actually, the cutoff term is not so important at large $B$, where asymptotic freedom reduces the running coupling $G(eB)$ quite much. Consequently, the critical temperature can be solved analytically from Eq.\eqref{criticalC} as
\bea
T_c\approx {\tilde{q}_{\rm u}}^{{\tilde{q}_{\rm u}}}{\tilde{q}_{\rm d}}^{{\tilde{q}_{\rm d}}}(2eB)^{1\over2}e^{-{2\pi^2\over N_ceB\,G(eB)}},\label{criticalT}
\eea
where $\tilde{q}_{\rm u/d}\equiv|{q}_{\rm u/d}|/e$ are the absolute values of the reduced electric charges. Note that the exponent in Eq.\eqref{criticalT} is in the same form as that given for $m$ in Ref.~\cite{Miransky:2015ava}, see Eq.(164) therein. Now, we can easily tell from Eq.\eqref{criticalT} that: if the decreasing rate of $G(eB)$ is larger than that of $1/eB$ with $eB$, the IMCE is unavoidable at large $B$.

\subsection{de Haas–van Alphen oscillation at finite $\mu_{\rm B}$}\label{subsec:dHvA}

By turning on baryon chemical potential $\mu_{\rm B}$, some intriguing phenomena emerge due to its nontrivial interplay with magnetic field. The effect of $\mu_{\rm B}$ can be introduced through the modification of the temporal derivative term~\cite{Kapusta2006} with the Lagrangian density given by~\cite{Cao:2016fby}
\begin{eqnarray}
{\cal L}_{\rm NJL}\!\!=\!\bar\psi(i\slashed{D}\!-\!\mu\gamma^0\!\!-\!m_0)\psi\!+\!G\!\left[(\bar\psi\psi)^2\!\!+\!\!(\bar\psi i\gamma_5{\tau}\psi)^2\right]\!,\label{LNJL2_mu}
\end{eqnarray}
where we define $\mu=\mu_{\rm B}/3$ for brevity. In this section, we first consider uniform chiral condensate for simplicity. Then, by following a similar procedure as in Sec.\ref{ssubsec:MCE}, the Matsubara summation gives the explicit form of the thermodynamic potential as~\cite{Kapusta2006}
\begin{widetext}
\begin{eqnarray}
V_{\rm BT}(\mu, m)={(m-m_0)^2\over 4G}-2N_c\sum_{\rm f=u,d}{|q_{\rm f}B|\over2\pi}\sum_{n=0}^\infty\alpha_{\rm n}\int_{-\infty}^\infty{\di p_3\over2\pi} \left[E_{\rm fn}(p_3)+\sum_{t=\pm}T\ln\left(1+e^{-(E_{\rm fn}(p_3)+t\,\mu)/T}\right)\right],\label{Omeg_Tmu}
\end{eqnarray}
where $\mu$ is find to only function through the ultraviolet-finite thermal term. So, it can be regularized in the same way as $V_{\rm BT}^r(m)$ and we have
\begin{eqnarray}
V_{\rm BT}^r(\mu, m)&=&{(m-m_0)^2\over 4G}+V_{\rm B}(m)-2N_cT\sum_{t=\pm}\sum_{\rm f=u,d}{|q_{\rm f}B|\over2\pi}\sum_{n=0}^\infty\alpha_{\rm n}\int_{-\infty}^\infty{\di p_3\over2\pi} \ln\left(1+e^{-(E_{\rm fn}(p_3)+t\,\mu)/T}\right).
\end{eqnarray}

At zero temperature, it reduces to
\begin{eqnarray}
V_{\rm B\mu}^r(m)\equiv{(m-m_0)^2\over 4G}+V_{\rm B}(m)-2N_c\sum_{\rm f=u,d}{|q_{\rm f}B|\over2\pi}\sum_{n=0}^\infty\alpha_{\rm n}\int_{-\infty}^\infty{\di p_3\over2\pi} (\mu-E_{\rm fn}(p_3))\theta(\mu-E_{\rm fn}(p_3))\label{Omeg_Bmu}
\end{eqnarray}
with $\theta(x)$ the unit step function. Then, the integral over $p_3$ can be carried out in the last term of Eq.\eqref{Omeg_Bmu} to give
\begin{eqnarray}
V_{\rm B\mu}^r(m)\equiv{(m-m_0)^2\over 4G}+V_{\rm B}(m)-{N_c\over2\pi^2}\sum_{\rm f=u,d}{|q_{\rm f}B|}\sum_{n=0}^\infty\alpha_{\rm n}\left(\mu\,p_{\rm fn}-M_{\rm fn}^2\ln{\mu+p_{\rm fn}\over M_{\rm fn}}\right)\theta(\mu-M_{\rm fn}),\label{Omeg_Bmu1}
\end{eqnarray}
where we have defined the $B$-dependent effective masses $M_{\rm fn}=(2n|q_{\rm f}B|+m^2)^{1/2}$ and $p_{\rm fn}\equiv(\mu^2-M_{\rm fn}^2)^{1/2}$ is the Fermi momentum for the $n$-th Landau level. Note that the contributions of Landau levels are cutoff by $\mu$ due to the step functions in the last term of Eq.\eqref{Omeg_Bmu1} -- an extended application of Silver-Blaze property~\cite{Cohen:2003kd}. For comparison, their contributions are only found to be exponentially small in the last term of Eq.\eqref{Omeg_BT} at finite temperature. Recalling the non-analytic nature of $\theta(x)$, the chiral transition might show a feature of first order, see Ref.\cite{Ebert:1999ht,Zhuang:1994dw,Klevansky:1992qe}.

Next, the gap equation can be obtained by taking $\partial_mV_{\rm B\mu}^r=0$ as:
\bea 
0&=&{m-m_0\over 2G}-4N_c\int^\Lambda{\di^3p\over(2\pi)^3}{m\over (p^2+m^2)^{1\over2}}-{N_c\over4\pi^2}m\sum_{\rm f=u,d}\int_0^\infty{\di s\over s^2}e^{-m^2s}\left[{q_{\rm f}Bs\over\tanh(q_{\rm f}Bs)}-1\right]\nonumber\\
&&+{N_c\over\pi^2}m\sum_{\rm f=u,d}{|q_{\rm f}B|}\sum_{n=0}^\infty\alpha_{\rm n}\ln{\mu+p_{\rm fn}\over M_{\rm fn}}\theta(\mu-M_{\rm fn}).
\eea
\end{widetext}
In the case $(eB)^{1/2}\gg\mu$ and $m\sim0$, we can apply LLL approximation to the last term and get
\bea
{N_c\over4\pi^2}m\sum_{\rm f=u,d}{|q_{\rm f}B|}\ln{(2\mu)^2\over m^2},
\eea
which is the same as Eq.\eqref{LLLT} by taking into account the correspondence $T\leftrightarrow2\mu$. So, the magnetic field can also amplify the effect of chemical potential and the chiral symmetry can be fully restored by $\mu$ in chiral limit. However, as mentioned before, due to the first-order nature of the chiral transition with $\mu$, the GL theory couldn't be applied to determine the critical chemical potential $\mu_{\rm c}$ at all, even in chiral limit. Nevertheless, in the sense that the step function usually delays $\chi$SR compared to that given by GL theory, we can take
\bea
\tilde\mu\approx {1\over2}{\tilde{q}_{\rm u}}^{{\tilde{q}_{\rm u}}}{\tilde{q}_{\rm d}}^{{\tilde{q}_{\rm d}}}(2eB)^{1\over2}e^{-{2\pi^2\over N_ceB\,G(eB)}}
\eea
as the lower boundary of $\mu_{\rm c}$.

At $\mu=M_{\rm fn}^+$, we can check that the curvature becomes
\bea
\partial_m ^2V_{\rm B\mu}^r\approx-{N_c\over\pi^2}{|q_{\rm f}B|}\alpha_{\rm n}{m^2\,\mu\over M_{\rm fn}^2p_{\rm fn}}\theta(\mu\!-\!M_{\rm fn})\sim -\infty,
\eea
which indicates that the point $\mu=M_{\rm fn}^+$ always corresponds to the maximum of the thermodynamic potential for $m\neq0$. Note the effective mass degeneracy $M_{\rm un}=M_{\rm d\,2n}$, so the curvature can be composed of two terms, but this would not change the qualitative conclusion.

\begin{figure}[!htb]
	\centering
	\includegraphics[width=0.45\textwidth]{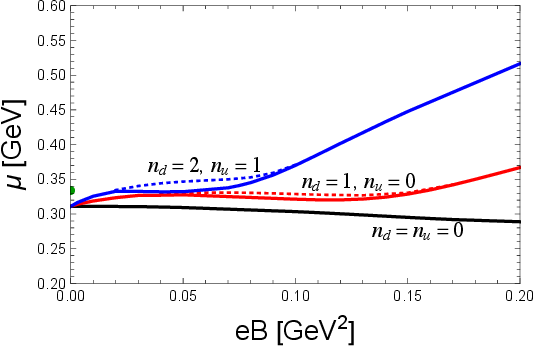}
	\caption{The $\mu-eB$ phase diagram for chiral symmetry breaking and restoration. We only show three lowest first-order transition boundaries crossing the $u$ and $d$ quark Landau levels: $n_{\rm d}=n_{\rm u}=0$ (black); $n_{\rm d}=1,n_{\rm u}=0$ (red) and $n_{\rm d}=2, n_{\rm u}=1$ (blue). The dashed lines are determined by $\mu_{\rm f}^n$ and the green bullet is the critical point at $eB=0$.}\label{muBB}
\end{figure}
\begin{figure}[!htb]
	\centering
	\includegraphics[width=0.45\textwidth]{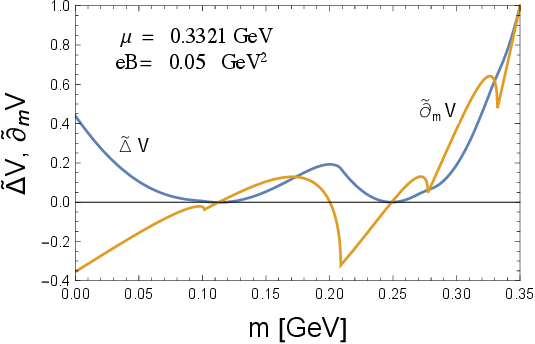}
	\caption{The reduced thermodynamic potential $\tilde{\Delta}V$ (blue) and its derivative $\tilde{\partial}_{\rm m}V$ (yellow) as functions of the dynamical mass $m$ at the third $\mu_c$ for $eB=0.05\, {\rm GeV}^2$. The dHvA oscillation points are more clearly shown by the dips in $\tilde{\partial}_{\rm m}V$, which correspond to $n_{\rm d}=0,1,2,3$ from right to left.}\label{VmuB}
\end{figure}

For a $B$-independent coupling, $\tilde\mu$ increases with $eB$, thus the dHvA effect is only possible when the chiral transition is of first order. In two-flavor NJL model, the $u$ and $d$ quark masses are always equal to each other in MFA due to the lack of isovector interaction channels. This helps to simplify the calculations and discussions as the dispersion of $u$ quark is the same as that of $d$ quark with even Landau levels. We illuminate three lowest phase boundaries together with the dHvA oscillation boundaries in Fig.\ref{muBB}. One should notice that not all the critical transition points are well determined by the dHvA oscillation points $\mu_{\rm f}^n=(2n|q_{\rm f}B|+m^2)^{1\over2} ~(n\in{\mathbb N})$, see the difference between dashed and solid lines for some given magnetic field. To demonstrate that more clearly, we depict in Fig.\ref{VmuB} the reduced thermodynamic potential and its derivative defined as
\bea
\tilde{\Delta}V&\equiv& \Delta V_{\rm B\mu}^r(m)/[\Delta V_{\rm B\mu}^r(m)|_{\rm m=0.35}],\nonumber\\
\tilde{\partial}_{\rm m}V&\equiv& {\partial}_{\rm m}V_{\rm B\mu}^r(m)/[{\partial}_{\rm m}V_{\rm B\mu}^r(m)|_{\rm m=0.35}],\nonumber
\eea
where the degenerate minima of $\tilde{\Delta}V$ are set to zero for convenience. With the minima widely separated, this critical chemical potential is quite different from the third dHvA oscillation point of $d$ quark $\mu_{\rm d}^2$. Nevertheless, it is true that each critical point only crosses one oscillation point.

In Ref.~\cite{Ebert:1999ht}, the  one-flavor NJL model respects exact chiral symmetry; but the symmetry is explicitly broken by the tiny $m_0$ in present study, which then physically gives rise to a finite pion mass. Such difference induces two consequences for Fig.\ref{muBB}: First, it is not valid anymore to categorize the number of massive boundaries according to the relative magnitudes of $G$ and $G_{\rm c}$~\cite{Ebert:1999ht}, since $G_{\rm c}$ is almost zero here but there are still many nontrivial phase boundaries. Second, the second-order transitions separating the $\chi$SB and $\chi$SR phases and inside the $\chi$SR phase~\cite{Ebert:1999ht} are taken over by smooth crossovers and very weak first-order transitions, respectively. For explicitness, only small magnetic field region is shown in Fig.\ref{muBB}, where one can easily identify the dHvA effect through the decreasing feature of all the three branches of $\mu_c$ for some $eB$. This is similar to that found for the case $G_{\rm c}<G<1.225\,G_{\rm c}$ in Ref.~\cite{Ebert:1999ht} if the second-order transitions are taken over by the smooth crossovers. Note that the dHvA oscillations were maintained by the spinodal and binodal boundaries in the study of the realistic three-flavor NJL model~\cite{Ferreira:2017wtx}.

\subsection{Rotational magnetic inhibition}\label{subsec:RMI}

Similar to that of baryon chemical potential, the effect of rotation or vorticity can be introduced by modifying the temporal derivative~\cite{Chen:2015hfc,Jiang:2016wvv} and the Lagrangian density is given by
\begin{eqnarray}
{\cal L}_{\rm NJL}&=&\bar\psi[i\slashed{D}-{\Omega(\hat{L}_3+\hat{S}_3)}\gamma^0-m_0]\psi\nonumber\\
&&+G\left[(\bar\psi\psi)^2+(\bar\psi i\gamma_5{\tau}\psi)^2\right].\label{LNJL2_R}
\end{eqnarray}
Here, we choose the angular velocity in the same direction as the magnetic field ${\bf B}$, and the orbital angular momentum and spin operators are defined as $\hat{L}_3\equiv-i(x\partial_y-y\partial_x)$ and $\hat{S}_3\equiv \sigma^{12}/2$, respectively. The symmetric gauge $A_i=(0,By/2,-Bx/2,0)$ will be chosen for the vector potential in the tangent space in order to facilitate solving the eigenstates. Note that in the rotating frame, the vector potential is given by $A_\mu=A_ie_\mu^i=(-B\Omega r^2/2,By/2,-Bx/2,0)$, which leads to an electric field. However,  $A_0=-B\Omega r^2/2$ does not appear in Eq.(\ref{LNJL2_R}) because the gamma matrix $\gamma^0=\gamma^i e_i^t$ cancels it out~\cite{Chen:2015hfc}. Then, the eigenstates can be characterized by the eigenvalues of $\hat{L}_3$ and $\hat{S}_3$, and the rotational term becomes $\Omega(l_3+S_3)\gamma^0$. Comparing it to Eq.\eqref{LNJL2_mu}, we immediately recognize that it is quite similar to the $\mu_{\rm B}$ term except for the dependence on the total angular momentum $j_3\equiv l_3+S_3$, thus we expect the rotation also to restore chiral symmetry. 

 Actually, the discussion on rotational effect is more involved: one must work in a finite system, otherwise the particle velocity would exceed that of light which breaks causality~\cite{Chen:2015hfc,Jiang:2016wvv}. For that sake, we consider a cylinder system with radius $R$ and the rotation effect is transmitted from the boundary to the bulk if possible. It is clear that the boundary would induce gaps to the particle energies for different $l_3$ and $S_3$, which then avoid the disaster of meson condensations of all kinds in rotational system~\cite{Chen:2017xrj,Cao:2020pmm}. Here, we neglect the boundary effect on quarks for simplicity, but one should keep in mind that there can be no particles in the system when the container is rotating. At zero temperature, it seems that the system with small $\Omega$ is the same as the vacuum with $\Omega=0$ according to the silver-Blaze property~\cite{Cohen:2003kd}; but they're surely different at finite temperature when medium with real particles is involved. 
 
Now, considering one flavor and color fermion field with $qB>0$ first for simplicity, we devote to solving the modified Dirac equation:
\bea
[i\slashed{D}-{\Omega(\hat{L}_3+\hat{S}_3)}\gamma^0-m]\psi=0
\eea
by adopting the Ritus formalism in the following. It is more convenient to work in cylindrical coordinate system with
$x=r\cos\theta, y=r\sin\theta$, thus
\begin{eqnarray}
\partial_x=\cos\theta\,\partial_r-{\sin\theta\over r}\,\partial_\theta,\ \
\partial_y=\sin\theta\,\partial_r+{\cos\theta\over r}\,\partial_\theta.
\end{eqnarray}
Then, using the identity 
$$i\gamma^1D_x+i\gamma^2D_y=\left[{\cal P}_\downarrow(D_x+iD_y)-{\cal P}_\uparrow(D_x-iD_y)\right]\gamma^2$$
with the spin projectors ${\cal P}_{\uparrow,\downarrow}\equiv{1\over2}(1\pm\sigma^{12})$, the Dirac equation can be rewritten explicitly as~\cite{Cao:2019ctl}  
\begin{eqnarray}\label{L1f}
	&&\Big\{\gamma^0\left[i\partial_t+\Omega \left(\hat{L}_3+\hat{S}_3\right)\right]+\left(e^{i\theta}D_r^\downarrow{\cal P}_\downarrow+e^{-i\theta}D_r^\uparrow{\cal P}_\uparrow\right)\gamma^2\nonumber\\
	&&\ \ +\ i\gamma^3\partial_3-m\Big\}\psi=0,\ D_r^{\downarrow/\uparrow}\equiv\partial_r\pm{qBr\over2}\pm{i\over r}\partial_\theta.
	\end{eqnarray}
As mentioned in the appendix of Ref.~\cite{Chen:2015hfc}, $D_r^{\downarrow/\uparrow}$ plays a role of lowering/raising ladder operator to the Landau levels.

Notice that $\gamma^2{\cal P}_{\downarrow/\uparrow}={\cal P}_{\uparrow/\downarrow}\gamma^2$, then the solution to the Dirac equation is found to be
\begin{eqnarray}
&&\psi=e^{-i\, (E t-p_3)}H_{\rm n,l}(\theta,r)~u_{\rm n,l}(p_3),\nonumber\\
&&H_{\rm n,l}(\theta,r)={\cal P}_\uparrow\chi_{\rm n,l}^+(\theta,r)+{\cal P}_\downarrow\chi_{\rm n-1,l+1}(\theta,r)\label{EigF}
\end{eqnarray}
by trying with the radial functions given in Ref.~\cite{Chen:2015hfc}. Here, $E$ and $p_3$ are the energy and longitudinal momentum, and the normalized radial function $\chi_{\rm n,l}$ is given by
\begin{eqnarray}
\chi_{\rm n,l}^+(\theta,r)=\left[{qB\over2\pi}{ n!\over(n+l)!}\right]^{1\over2}{e^{i\, l\theta}}~\tilde{r}^le^{-\tilde{r}^2/2}L_{\rm n}^l\left(\tilde{r}^2\right)\label{chi+}
\end{eqnarray}
with $\tilde{r}^2=|qB|r^2/2$. The Laguerre polynomial $L_{\rm n}^l(x)$ is nonvanishing only for $n\ge0$, and we set $l\in\left[-n,N-n\right]$ where the degeneracy factor $N$ for each Landau level reads
\begin{eqnarray}
N=\left \lfloor \frac{qBS}{2\pi} \right \rfloor,
\end{eqnarray}
with $S$ being the area of the $xy$-plane. Even though $N$ is not necessary when boundary condition is taken into account~\cite{Cao:2020pmm}, it still serves a good approximation in present simplified discussions~\cite{Chen:2017xrj}. As the action of the ladder operators on the pre-matrix of the eigenfunction gives
\begin{eqnarray}
\left(e^{i\theta}\!D_r^\downarrow{\cal P}_\downarrow\!\!+\!e^{-i\theta}\!D_r^\uparrow{\cal P}_\uparrow\right)\!\gamma^2\!H_{\rm n,l}(\theta,r)\!=\!\!-\!H_{\rm n,l}(\theta,r)\gamma^2\!\sqrt{2n qB},\nonumber
\end{eqnarray}
the eigen-equation for $u_{\rm n,l}(p_3)$ follows directly as:
\begin{eqnarray}
\left(\gamma^0\varepsilon^+-\gamma^3p_3-\gamma^2\sqrt{2n qB}-m\right)u_{\rm n,l}(p_3)=0,\label{EigE}
\end{eqnarray}
with $\varepsilon^+\equiv E+\Omega \left(l+{1\over2}\right)$. Then, one branch of solutions, corresponding to the positive-energy ones in the absence of rotation, can be expressed in a compact form as~\cite{Peskin1995}
\begin{eqnarray}
u_{\rm n,l}^s(p_3)=\left(\begin{array}{c}
\sqrt{p\cdot{\sigma}}\xi^s\\\sqrt{p\cdot\bar{\sigma}}\xi^s\end{array}\right),   
\end{eqnarray}
where $s=\pm$, $p_\mu=(E_{\rm n},0,\sqrt{2n qB},p_3)$ with $E_{\rm n}\equiv(2n qB+p_3^2+m^2)^{1/2}$, $\sigma^\mu=(1,\mbox{\boldmath{$\sigma$}})$, and $\bar{\sigma}^\mu=(1,-\mbox{\boldmath{$\sigma$}})$. The two-component spinors $\xi^\pm$ are given by 
$\xi^+=(1,0)^{\rm T}$ and $\xi^-=(0,1)^{\rm T}$.

The eigenenergy can be solved from Eq.\eqref{EigE} as $$\pm E_{n}(p_3)-\Omega \left(l+{1\over2}\right),$$ then the thermodynamic potential for two-flavor NJL model follows directly as~\cite{Cao:2019ctl} 
\begin{widetext}
\begin{eqnarray}
V_{\rm BT}(\Omega, m)={(m-m_0)^2\over 4G}-{2N_c\over S}\sum_{\rm f=u,d}\sum_{l=0}^{N_{\rm f}}\sum_{n=0}^\infty\alpha_{\rm n}\int_{-\infty}^\infty{\di p_3\over2\pi} \left[E_{\rm fn}(p_3)+\sum_{t=\pm}T\ln\left(1\!+\!e^{-(E_{\rm fn}(p_3)+t\,\Omega_{\rm nl} )/T}\right)\right]\label{Omeg_TOmg}
\end{eqnarray}
with $\Omega_{\rm nl}\equiv|\Omega\left(l-n+{1\over2}\right)|$. Here, we assume uniform chiral condensate as the previous section, which is proper when only rotation is present~\cite{Wang:2018zrn}. By performing "vacuum regularization", it becomes
\begin{eqnarray}
V_{\rm BT}^r(\Omega, m)={(m-m_0)^2\over 4G}+V_{\rm B}(m)-{2N_c\over S}T \sum_{t=\pm}\sum_{\rm f=u,d}\sum_{l=0}^{N_{\rm f}}\sum_{n=0}^\infty\alpha_{\rm n}\int_{-\infty}^\infty{\di p_3\over2\pi}\ln\left(1\!+\!e^{-(E_{\rm fn}(p_3)+t\,\Omega_{\rm nl})/T}\right).\label{Omeg_BOmgT}
\end{eqnarray}
At zero temperature, the integral in the last term can be carried out and the result follows Eq.\eqref{Omeg_Bmu1} as
\begin{eqnarray}
V_{\rm B\Omega}^r(m)\equiv{(m-m_0)^2\over 4G}+V_{\rm B}(m)-{N_c\over\pi S}\sum_{\rm f=u,d}\sum_{l=0}^{N_{\rm f}}\sum_{n=0}^\infty\alpha_{\rm n}\left(\Omega_{\rm nl}\,p_{\rm fnl}-M_{\rm fn}^2\ln{\Omega_{\rm nl}+p_{\rm fnl}\over M_{\rm fn}}\right)\theta(\Omega_{\rm nl}-M_{\rm fn})\label{Omeg_BOmg1}
\end{eqnarray}
with $p_{\rm fnl}\equiv(\Omega_{\rm nl}^2-M_{\rm fn}^2)^{1/2}$ the Fermi momentum for different $n,l$ and $S$. Then, the gap equation becomes
\bea 
0&=&{m-m_0\over 2G}-4N_c\int^\Lambda{\di^3p\over(2\pi)^3}{m\over (p^2+m^2)^{1\over2}}-{N_c\over4\pi^2}m\sum_{\rm f=u,d}\int_0^\infty{\di s\over s^2}e^{-m^2s}\left[{q_{\rm f}Bs\over\tanh(q_{\rm f}Bs)}-1\right]\nonumber\\
&&+{2N_c\over\pi S}m\sum_{\rm f=u,d}\sum_{l=0}^{N_{\rm f}}\sum_{n=0}^\infty\alpha_{\rm n}\ln{\Omega_{\rm nl}+p_{\rm fnl}\over M_{\rm fn}}\theta(\Omega_{\rm nl}-M_{\rm fn}).\label{Gm_Omg}
\eea
So, according to Eq.\eqref{Omeg_BOmg1}, it seems that the chiral symmetry can be fully restored by large enough $\Omega$ in chiral limit; but when the boundary condition is self-consistently taken into account, it is impossible since there are always gaps in the excitation energies.

For later use, we can construct the fermion Green's function from the eigenfunctions Eq.\eqref{EigF} by following the schemes presented in Ref.~\cite{Cao:2014uva,Cao:2019ctl}. We worked out the retarded Green's functions first and then found the Feynman Green's functions for positive and negative charged fermions as~\cite{Cao:2019ctl}
\begin{eqnarray}
S_{\rm F}(x,x')&=&\sum_{n=0}^\infty\sum_l\int\int{dp_0dp_3\over(2\pi)^2}{i~e^{-ip_0(t-t^\prime)+ip_3(z-z^\prime)}\over\left({p}_0^{l+}\right)^2-E_{\rm n}^2+i\epsilon}
\Bigg\{\left[{\cal P}_\uparrow\chi_{\rm n,l}^+(\theta,r)\chi_{\rm n,l}^{+*}(\theta',r')+{\cal P}_\downarrow\chi_{\rm n-1,l+1}^{+}(\theta,r)\chi_{\rm n-1,l+1}^{+*}(\theta',r')\right]\nonumber\\
&&\left(\gamma^0{p}_0^{l+}-\gamma^3p_3+m\right)-\left[{\cal P}_\uparrow\chi_{\rm n,l}^+(\theta,r)\chi_{\rm n-1,l+1}^{+*}(\theta',r')+{\cal P}_\downarrow\chi_{\rm n-1,l+1}^{+}(\theta,r)\chi_{\rm n,l}^{+*}(\theta',r')\right]\sqrt{2n qB}\gamma^2\Bigg\},\label{propagator+}\\
S_{\rm F}(x,x')&=&\sum_{n=0}^\infty\sum_l\int_{-\infty}^{\infty}{dp_0dp_3\over(2\pi)^2}{i~e^{-ip_0(t-t^\prime)+ip_3(z-z^\prime)}\over\left({p}_0^{l-}\right)^2-E_{\rm n}^2+i\epsilon}
	\left\{\left[{\cal P}_\uparrow\chi_{\rm n-1,l-1}^{-}(\theta,r)\chi_{\rm n-1,l-1}^{-*}(\theta',r')+{\cal P}_\downarrow\chi_{\rm n,l}^{-}(\theta,r)\chi_{\rm n,l}^{-*}(\theta',r')\right]\right.\nonumber\\
	&&\left.\left(\gamma^0{p}_0^{l-}-\gamma^3p_3+m\right)+\left[{\cal P}_\uparrow\chi_{\rm n-1,l-1}^{-}(\theta,r)\chi_{\rm n,l}^{-*}(\theta',r')+{\cal P}_\downarrow\chi_{\rm n,l}^{-}(\theta,r)\chi_{\rm n-1,l-1}^{-*}(\theta',r')\right]\sqrt{2n |qB|}\gamma^2\right\}.\label{propagator-}
\end{eqnarray}
\end{widetext}
Here, ${p}_0^{ls}=p_0+\Omega \left(l+s{1\over2}\right)$ and Eq.\eqref{propagator-} is the fermion propagator for $qB<0$ with the radial function defined as
\begin{eqnarray}
\chi_{\rm n,l}^{-}(\theta,r)\!=\!\left({|qB|\over2\pi}{ n!\over(n\!-\!l)!}\right)^{1\over2}{e^{i\, l\theta}}~\tilde{r}^{-l}e^{-\tilde{r}^2/2}L_{\rm n}^{-l}\left(\tilde{r}^2\right)\!.\label{chi-}
\end{eqnarray}
 The LLL contribution comes from the term $\chi_{\rm n,l}^\pm(\theta,r)\chi_{\rm n,l}^{\pm*}(\theta',r')$ as should be, and it had been demonstrated in Ref.~\cite{Cao:2019ctl} that the vanishing $\Omega$ limit of Eq.\eqref{propagator+} is consistent with that given in Ref.~\cite{Miransky:2015ava}. 
 
\begin{figure}[!htb]
	\centering
	\includegraphics[width=0.45\textwidth]{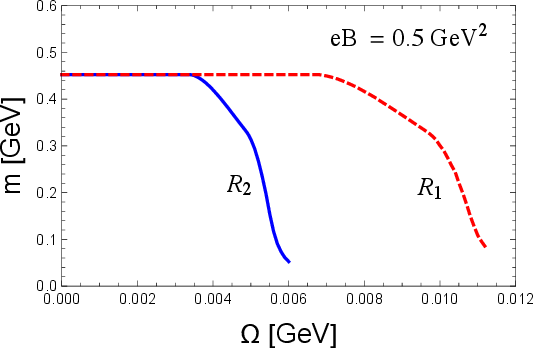}
	\caption{The dynamical quark mass $m$ as a function of the angular velocity $\Omega$ for two system sizes $R_1=20/\sqrt{2eB}$ (red dashed lines) and $R_2=20/\sqrt{eB}$ (blue solid lines) at $eB=0.5~{\rm GeV}^2$. The plots are from Ref.~\cite{Cao:2019ctl}.}\label{m_omg}
\end{figure}

\begin{figure}[!htb]
	\centering
	\includegraphics[width=0.45\textwidth]{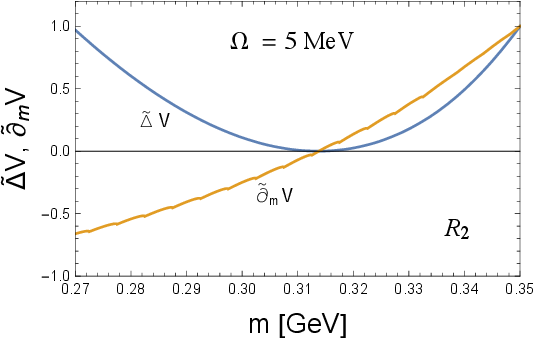}
	\caption{The reduced thermodynamic potential $\tilde{\Delta}V$ (blue) and its derivative $\tilde{\partial}_{\rm m}V$ (yellow) as functions of the dynamical mass $m$ at $\Omega=5\, {\rm MeV}$.  The tiny dips in $\tilde{\partial}_{\rm m}V$ correspond to dHvA points with different $l$.}\label{V_omg}
\end{figure}

To demonstrate the effect of angular velocity $\Omega$, we choose a magnetic field $eB=0.5~{\rm GeV}^2$ for example and consider two system sizes $R_1=20/\sqrt{2eB}$ and $R_2=20/\sqrt{eB}$ such that $N_{\rm f}$ are independent of $eB$. Then, confining ourselves to the causality region $\Omega<1/R$, the inhibition effect of $\Omega$ is well illustrated in Fig.\ref{m_omg}, that is, the dynamical mass $m$ decreases with $\Omega$. 
Similar to that illustrated in Ref.~\cite{Chen:2015hfc}, it seems that $m$ changes continuously with $\Omega$. However, from the reduced derivative of the thermodynamic potential shown in Fig.\ref{V_omg}, it is easy to find lots of tiny dips for different $l$, which then indicate that the change of $m$ is invisibly discontinuous at some $\Omega$. Besides, one can tell from Fig.\ref{m_omg} that $m$ is not analytic with $\Omega$ at the bumps. The three dimensional diagram of $m$ with respect to $eB$ and $\Omega$ has been depicted in Ref.~\cite{Chen:2015hfc} but no dHvA oscillation had been found in the $\Omega-eB$ phase boundary. So we are not going to plot the trivial phase boundary anymore here and Fig.\ref{V_omg} will serve as the basis for further calculations in Sec.\ref{subsec:NJLcpi}.

\subsection{Inhomogeneous phases at finite $\mu_{\rm B}$}\label{subsec:inhom}

As mentioned in the introduction, the case with finite baryon chemical potential is quite nontrivial as inhomogeneous phases and color superconductivity are all possible for adequate $\mu_{\rm B}$. Here, we present the studies on two kinds of inhomogeneous chiral condensate, chiral density wave (CDW) and solitonic modulation (SM)~\cite{Nickel:2009wj}, in the presence of $\mu_{\rm B}$ and $B$. To facilitate numerical calculations, we work in chiral limit with $m_0=0$ in this section. Then, in mean field approximation, the order parameters are assumed to be 
\bea
\langle\sigma(x)\rangle=m_{\rm r}(x),\ \langle\pi^3(x)\rangle=m_{\rm i}(x),\
\langle\pi^{1,2}(x)\rangle=0
\eea
in Eq.\eqref{action}, which give the thermodynamic potential as
\begin{eqnarray}
V\!\!\!&=&\!\!\!{T\over V_{\rm 3}}\left[\frac{1}{4G}\int d^4x|M(x)|^2\!-{\rm Tr\ln}\left(i\slashed{D}\!-\!{1\!+\!\gamma^5{\tau^3}\over2}M(x)\right.\right.\nonumber\\
&&\left.\left.\qquad-{1\!-\!\gamma^5{\tau^3}\over2}M^*(x)\!-\!\mu\gamma^0\right)\right].
\end{eqnarray}
Here, we define the mass gap $M(x)\equiv m_{\rm r}(x)+i\,m_{\rm i}(x)$ and work in Euclidean space with the integral of the imaginary time $x_4=ix_0$ in the region $[0,1/T]$. The explicit forms of the inhomogeneous chiral condensate~\cite{Dautry:1979bk,Schnetz:2005ih} are
 \begin{eqnarray}
 M_{\rm CDW}(m,q)\!\!\!&=&\!\!\!m\,e^{2ikz},\label{MCDW}\\
M_{\rm SM}(m,\nu,{\rm z})
\!\!\!&=&\!\!\!m\Big(\nu\,{\rm sn}(\mathbf{K}_\nu|\nu)\,{\rm sn}(m {\rm z}|\nu)\,{\rm sn}(m {\rm z}\!+\!\mathbf{K}_\nu|\nu)\nonumber\\
&&+\frac{{\rm cn}(\mathbf{K}_\nu|\nu)\,{\rm dn}(\mathbf{K}_\nu|\nu)}{{\rm sn}(\mathbf{K}_\nu|\nu)}\Big),\label{MSM}
\end{eqnarray}
where $\mathbf{K}_\nu$ is the quarter period, and ${\rm sn}(u|\nu), {\rm cn}(u|\nu)$ and ${\rm dn}(u|\nu)$ are elliptic Jacobi functions with elliptic modulus $\sqrt{\nu}$. As it had been demonstrated that the inhomogeneity along the magnetic field is more favored over the transverse ones~\cite{Frolov:2010wn}, we only focus on the longitudinal inhomogeneity in this section, see the $z$-dependence in Eqs.\eqref{MCDW} and \eqref{MSM}.

Then, the most important mission is to evaluate the eigenvalues of the following Hamiltonian:
\begin{eqnarray}\label{Hf}
H_{\rm f}&=&-\gamma^0\big[i\slashed{\boldsymbol D}_{\rm f}-{1+\gamma^5\over2}M({\rm z})-{1-\gamma^5\over2}M^*({\rm z})\big]\nonumber\\
&=&-\gamma^0\big[i\gamma^1\partial_{\rm x}+i\gamma^2(\partial_{\rm y}-iq_{\rm f}B{\rm x})+i\gamma^3\partial_{\rm z}-\nonumber\\
&&\!\!\!\!{1\!+\!{\cal S}(q_{\rm f}B)\gamma^5\over2}M({\rm z})-{1\!-\!{\cal S}(q_{\rm f}B)\gamma^5\over2}M^*({\rm z})\big].
\end{eqnarray}
Suppress ${\cal S}(q_{\rm f}B)$ for a while. If we expand the eigenstates with Ritus method by separating variables, the first two terms in the square bracket give rise to a Landau level term $i\sqrt{2n |q_{\rm f}B|}\gamma^2$, see the previous section. The left terms actually correspond to a one-dimensional NJL model or generalized Gross-Neveu (GN) model with the following Hamiltonian:
\begin{eqnarray}
H_{\rm GN}&=&-\gamma^0\big[i\gamma^3\partial_{\rm z}-{1+\gamma^5\over2}M({\rm z})-{1-\gamma^5\over2}M^*({\rm z})\big]\nonumber\\
&=&\qquad\left(\begin{array}{cccc}  i\partial_{\rm z}&0&M({\rm z})&0\\ 0&-i\partial_{\rm z}&0&M({\rm z})\\M^*({\rm z})&0&-i\partial_{\rm z}&0\\0&M^*({\rm z})&0&i\partial_{\rm z}\end{array}\right).
\end{eqnarray}
The Hamiltonian can be brought to a block diagonal form by performing a similitude transformation, that is,
\begin{eqnarray}
U^{-1}H_{\rm GN}U=\left(\begin{array}{cc}  H_{{\rm z}}\big(M({\rm z})\big)&0\\ 0&H_{\rm z}\big(M^*({\rm z})\big)\end{array}\right),
\end{eqnarray}
where the involved matrices are respectively:
\begin{eqnarray}
H_{\rm z}\big(M({\rm z})\big)\!=\!\left(\begin{array}{cc}  i\partial_{\rm z}&M^*({\rm z})\\ M({\rm z})&-i\partial_{\rm z}\end{array}\right),U\!=\!\left(\begin{array}{cccc}  1&0&0&0\\ 0&0&0&1\\0&1&0&0\\0&0&1&0\end{array}\right).\label{Hamiltonian}
\end{eqnarray}

For the inhomogeneous states, it was found that the explicit forms of the thermodynamic potential can be conveniently evaluated with the help of the densities of states~\cite{Schnetz:2004vr,Schnetz:2005ih,Basar:2008im,Basar:2008ki}, see the extensions to $3+1$ dimensions case in Ref.~\cite{Nickel:2009wj}. Here, one should remember that the spinors are actually half valid at the LLL and take the following forms 
\bea
u(x)&=&\big(u_1(x),0,u_3(x),0\big)^T,\nonumber\\
d(x)&=&\big(0,d_2(x),0,d_4(x)\big)^T
\eea
for $u$ and $d$ quarks, respectively~\cite{Warringa:2012bq}. Then, in the case that $M({\rm z})$ is not real, the spectra $\{\varepsilon\}$ of $H_{\rm GN}$ are not symmetric with $\{-\varepsilon\}$ at all thus should be treated more carefully. Take the CDW phase for example,  the form of chiral condensate is quite like that of electronic pairing in the Fulde-Ferrell phase of a non-relativistic electron system~\cite{Fulde:1964zz}. By adopting the same trick, the eigenenergies follow directly as
\bea
E_{\rm fnts}(p_3)=\left\{\!\!\!\!\!\!\begin{array}{ll}
&t\,E_0(p_3)+k,\qquad\qquad\qquad\ \ \ \ \ n=0,\\
&t[(s\,E_0(p_3)+k)^2+2n|q_{\rm f}B|]^{1\over2},\ n>0
\end{array}\right.
\eea
with $t,s=\pm$. So, the spectra at LLL are $\{\pm\sqrt{p_{\rm 3}^2+m^2}+k\}$ where the vector number $k$ plays a role quite like the chemical potential. For the SM phase, $M({\rm z})$ is real; it can be checked that the LLL spectra are sign symmetric, thus the regularization can be self-consistently performed~\cite{Cao:2016fby}.

Now, recalling the densities of state in Gross-Neveu model~\cite{Schnetz:2004vr,Basar:2008im,Basar:2008ki}:
\begin{eqnarray}
\!\!\rho_{\rm CDW}(\varepsilon;m,k)\!\!\!&=&\!\!\!{1\over\pi}{|\varepsilon-k|\,\theta[(\varepsilon-k)^2-m^2]\over\sqrt{(\varepsilon-k)^2-m^2}},\\
\!\!\rho_{\rm SM}(\varepsilon;m,\nu)\!\!\!&=&\!\!\!{1\over\pi}\frac{[\varepsilon^2\!-\!m^2\mathbf{E}(\nu)/\mathbf{K}(\nu)]{\Theta}(\varepsilon,m,\nu)}
{\sqrt{(\varepsilon^2-m^2)\big(\varepsilon^2-(1\!-\!\nu)m^2\big)}},\\
 {\Theta}(\varepsilon,m,\nu)\!\!\!&=&\!\!\!\theta(\varepsilon^2\!-\!m^2)\!-\!\theta\big(-\varepsilon^2\!+\!(1\!-\!\nu)m^2\big)\label{DOS}
\end{eqnarray}
with $\mathbf{E}(\nu)$ the incomplete elliptic integral and the signs ${\cal S}(q_{\rm f}B)$ in Eq.\eqref{Hf}, the thermodynamic potentials can be given as~\cite{Cao:2016fby}
\begin{widetext}
\begin{eqnarray}
V_{\rm BT}^{\rm CDW}(\mu,m,k)\!\!&=&\!\!{m^2\over4G}\!-\!N_c\sum_{\rm f=u,d}{|q_{\rm f}B|\over4\pi}\int_{-\infty}^\infty \di\varepsilon\sum_{t=\pm}\rho_{\rm CDW}(\varepsilon;m,{\cal S}({q}_{\rm f}B)\,k)T\ln\left(1+e^{{t\over T}(\varepsilon+\mu)}\right)\nonumber\\
&&-N_c\sum_{\rm f=u,d}{|q_{\rm f}B|\over2\pi}\sum_{n=1}\int_{-\infty}^\infty d\varepsilon\rho_{\rm CDW}(\varepsilon;m,{\cal S}({q}_{\rm f}B)\,k)\,f(T,\mu,n,q_{\rm f}B,\varepsilon),\label{VCDW}\\
V_{\rm BT}^{\rm SM}(\mu,m,\nu)\!\!&=&\!\!{m^2\over4G}\left({1\over{\rm sn}^2(\mathbf{K}(\nu)|\nu)}-{2\mathbf{E}(\nu)\over\mathbf{K}(\nu)}+1-\nu\right)\!-\!N_c\!\!\sum_{\rm f=u,d}\!\!{|q_{\rm f}B|\over2\pi}\!\!\sum_{n=0}\!\alpha_{\rm n}\!\int_{-\infty}^\infty\!\!\di\varepsilon\,\rho_{\rm SM}(\varepsilon;m,\nu)f(T,\mu,n,q_{\rm f}B,\varepsilon).\nonumber\\
\label{VSM}
\end{eqnarray}
Here, the auxiliary function is defined as
$$f(T,\mu,n,q_{\rm f}B,\varepsilon)=\epsilon(n,q_{\rm f}B,\varepsilon)+\sum_{t=\pm}T\ln(1+e^{-(\epsilon(n,q_{\rm f}B,\varepsilon)+t\,\mu)/T})$$
with the excitation energy $\epsilon(n,q_{\rm f}B,\varepsilon)=\big(2n|q_{\rm f}B|+\varepsilon^2\big)^{1/2}$. We can immediately check that both Eq.\eqref{VCDW} and Eq.\eqref{VSM} consistently reduce to Eq.\eqref{Omeg_Tmu} in the limits $k\rightarrow0$ and $\nu\rightarrow1$, respectively. For the CDW phase, it had been shown that: to derive the eigenenergies, the trick of momentum shift $p_3\rightarrow p_3\pm k$ usually introduces an artificial $k$-dependence of the thermodynamic potentials in the limit $m\rightarrow0$~\cite{Frolov:2010wn}. But the alternative presentation Eq.\eqref{VCDW} here avoids such problem with the help of the density of state thus is a much better form to start with. Moreover, note that the $\chi$SR phase can also be self-consistently reproduced by the limit $m\rightarrow0$ or $\nu\rightarrow0$ of Eq.\eqref{VSM}.
 
 We regret that $V_{\rm BT}^{\rm CDW}(\mu,m,k)-V_{\rm BT}^{\rm CDW}(\mu,m,0)$ and $V_{\rm BT}^{\rm SM}(\mu,m,\nu)-V_{\rm BT}^{\rm SM}(\mu,m,1)$ are not convergent, so the "vacuum regularization" cannot be applied to the inhomogeneous phases. Instead, we refer to the PV regularization and the thermodynamic potentials become 
\bea
V_{\rm BT}^{\rm CDWr}(\mu,m,k)\!\!&=&\!\!{m^2\over4G}\!-\!N_c\sum_{\rm f=u,d}{|q_{\rm f}B|\over4\pi}\int_{0}^\infty \di\varepsilon\sum_{t,l=\pm}\rho_{\rm CDW}(\varepsilon;m,l\,{\cal S}({q}_{\rm f}B)\,k)\sum_{i=0}^3 (-1)^iC_3^i\,T\ln{\left(1+e^{{t\over T}(l\,\varepsilon^i+\mu)}\right)}\nonumber\\
&&-N_c\sum_{\rm f=u,d}{|q_{\rm f}B|\over2\pi}\sum_{n=1}\int_{-\infty}^\infty \di\varepsilon\,\rho_{\rm CDW}(\varepsilon;m,{\cal S}({q}_{\rm f}B)\,k)\sum_{i=0}^3 (-1)^iC_3^i\,f(T,\mu,n,q_{\rm f}B,\varepsilon^i),\label{VCDWr}
\eea
\bea
V_{\rm BT}^{\rm SMr}(\mu,m,\nu)\!\!&=&\!\!{m^2\over4G}\left({1\over{\rm sn}^2(\mathbf{K}(\nu)|\nu)}-{2\mathbf{E}(\nu)\over\mathbf{K}(\nu)}+1-\nu\right)\!-\!N_c\!\!\sum_{\rm f=u,d}\!\!{|q_{\rm f}B|\over2\pi}\!\!\sum_{n=0}\!\alpha_{\rm n}\!\int_{-\infty}^\infty\!\!\di\varepsilon\,\rho_{\rm SM}(\varepsilon;m,\nu)\nonumber\\
&&\sum_{i=0}^3 (-1)^iC_3^i\,f(T,\mu,n,q_{\rm f}B,\varepsilon^i)\label{VSMr}
\eea
with $\epsilon_{\rm fn}^i\equiv\epsilon(n,q_{\rm f}B,\varepsilon^i)$ and $\varepsilon^i=\sqrt{\varepsilon^2+i\,\Lambda^2}$. Since they are regularized in the same way, we can compare their values to find out the favored ground state. Here, the advantages of the PV regularization scheme are ready to be seen: In the limit $m\rightarrow0$ or $\nu\rightarrow0$, Eqs.\eqref{VCDWr} and \eqref{VSMr} maintain the capability to reproduce the $\chi$SR result; in the limit $k\rightarrow0$ or $\nu\rightarrow1$, the  correct form for the $\chi$SB phase can still be reproduced. 

These forms are not so useful for deriving the gap equations that can be evaluated numerically, as the derivatives with respect to the order parameters, that is, $\partial_{m,k}\rho_{\rm CDW}(\varepsilon;m,k)$ and $\partial_{m,\nu}\rho_{\rm SM}(\varepsilon;m,\nu)$, would give the form of $\infty-\infty$ around the integral boundary of $\varepsilon$. Now, we try to transfer the thermodynamic potentials to more friendly forms by taking integral variable transformations or partial integrations. For the CDW phase, the partial integration alters $V_{\rm BT}^{\rm CDWr}$ to
\bea
V_{\rm BT}^{\rm CDWr}(\mu,m,k)\!\!&=&\!\!{m^2\over4G}\!+\!N_c\sum_{\rm f=u,d}{|q_{\rm f}B|\over4\pi}\sum_{t,l=\pm}{\cal P}_{\rm CDW}(0;m,l\,{\cal S}({q}_{\rm f}B)\,k)\sum_{i=0}^3 (-1)^iC_3^i\,T\ln{\left(1+e^{{t\over T}(l\,\sqrt{i}\Lambda+\mu)}\right)}\nonumber\\
&&\!+\!N_c\sum_{\rm f=u,d}{|q_{\rm f}B|\over4\pi}\int_{0}^\infty \di\varepsilon\sum_{t,l=\pm}{\cal P}_{\rm CDW}(\varepsilon;m,l\,{\cal S}({q}_{\rm f}B)\,k)\sum_{i=0}^3 (-1)^iC_3^i{t\,l\,\varepsilon/\varepsilon^i\over1+e^{-{t\over T}(l\,\varepsilon^i+\mu)}}\nonumber\\
&&+N_c\sum_{\rm f=u,d}{|q_{\rm f}B|\over4\pi}\sum_{n=1}\int_{-\infty}^\infty \di\varepsilon\sum_{t=\pm}{\cal P}_{\rm CDW}(\varepsilon;m,{\cal S}({q}_{\rm f}B)\,k)\sum_{i=0}^3 (-1)^iC_3^i{\varepsilon\over\epsilon_{\rm fn}^i}\tanh\left({\epsilon_{\rm fn}^i+t\,\mu\over2T}\right)\label{VCDWr1}
\eea
with the auxiliary function ${\cal P}_{\rm CDW}(\varepsilon;m,k)\equiv{\cal S}(\varepsilon-k)\sqrt{(\varepsilon-k)^2-m^2}\,\theta[(\varepsilon-k)^2-m^2]$; for the SM phase, the variable transformation of the first branch and partial integration of the second branch alter $V_{\rm BT}^{\rm SMr}$ to
\begin{eqnarray}
V_{\rm BT}^{\rm SMr}(\mu,m,\nu)\!\!&=&\!\!{m^2\over4G}\left({1\over{\rm sn}^2(\mathbf{K}(\nu)|\nu)}-{2\mathbf{E}(\nu)\over\mathbf{K}(\nu)}+1-\nu\right)\!-\!N_c\!\!\sum_{\rm f=u,d}\!\!{|q_{\rm f}B|\over\pi}\!\!\sum_{n=0}\!\alpha_{\rm n}\!\int_{0}^\infty\!\!\di x {1\over\pi}\frac{[x^2+m^2(1-\mathbf{E}(\nu)/\mathbf{K}(\nu))]}
{\sqrt{(x^2+m^2)\big(x^2+\nu\, m^2\big)}}\nonumber\\
&&\sum_{i=0}^3 (-1)^iC_3^i\,f(T,\mu,n,q_{\rm f}B,\sqrt{x^2+m^2+i\,\Lambda^2})-N_c{m\over2}{\mathbf{E}(\nu)\over\mathbf{K}(\nu)}\!\!\sum_{\rm f=u,d}\!\!{|q_{\rm f}B|\over\pi}\!\!\sum_{n=0}\!\alpha_{\rm n} \sum_{i=0}^3 (-1)^iC_3^i\times\nonumber\\
&&f(T,\mu,n,q_{\rm f}B,\sqrt{i}\,\Lambda)+N_c\!\!\sum_{\rm f=u,d}\!\!{|q_{\rm f}B|\over\pi}\!\!\sum_{n=0}\!\alpha_{\rm n}\!\int_{0}^{\sqrt{1-\nu}\,m} \!\!\di\varepsilon\,{\cal P}_{\rm SM}^2(\varepsilon;m,\nu)\sum_{i=0}^3 (-1)^iC_3^i\,f(T,\mu,n,q_{\rm f}B,\varepsilon^i)\nonumber\\
&&-N_c\sum_{\rm f=u,d}{|q_{\rm f}B|\over2\pi}\sum_{n=0}\alpha_{\rm n}\int_{0}^{\sqrt{1-\nu}\,m} \di\varepsilon\sum_{t=\pm}{\cal P}_{\rm SM}^1(\varepsilon;m,\nu)\sum_{i=0}^3 (-1)^iC_3^i{\varepsilon\over\epsilon_{\rm fn}^i}\tanh\left({\epsilon_{\rm fn}^i+t\,\mu\over2T}\right)\label{VSMr1}
\end{eqnarray}
with the auxiliary functions
\bea
{\cal P}_{\rm SM}^1(\varepsilon;m,\nu)&=&{1\over\pi}\left[\arcsin\left({\varepsilon\over\sqrt{1-\nu}\,m}\right)-{\pi\over2}\right]\frac{\varepsilon^2-m^2\mathbf{E}(\nu)/\mathbf{K}(\nu)}
{\sqrt{m^2-\varepsilon^2}},\\
{\cal P}_{\rm SM}^2(\varepsilon;m,\nu)&=&{1\over\pi}\left[\arcsin\left({\varepsilon\over\sqrt{1-\nu}\,m}\right)-{\pi\over2}\right]\frac{\varepsilon\,m^2\big(-2+\mathbf{E}(\nu)/\mathbf{K}(\nu))+\varepsilon^3]}
{\big(m^2-\varepsilon^2\big)^{3/2}}.
\eea
\end{widetext}
Eventually, the ultraviolet convergent gap equations can be given by the derivatives of the altered thermodynamic potentials, that is, $\partial_{m,k}V_{\rm BT}^{\rm CDWr}(\mu,m,k)=\partial_{m,\nu}V_{\rm BT}^{\rm SMr}(\mu,m,\nu)=0$.
The derivatives are easy to be carried out, but the expressions are so lengthy that we would not present them explicitly here. It can be checked that $m=0$ is the general trivial solution to the gap equations for both phases, but the cases with $k=0$ and $\nu=0,1$ are not trivial ones at all.

\begin{figure}[!htb]
	\centering
	\includegraphics[width=0.45\textwidth]{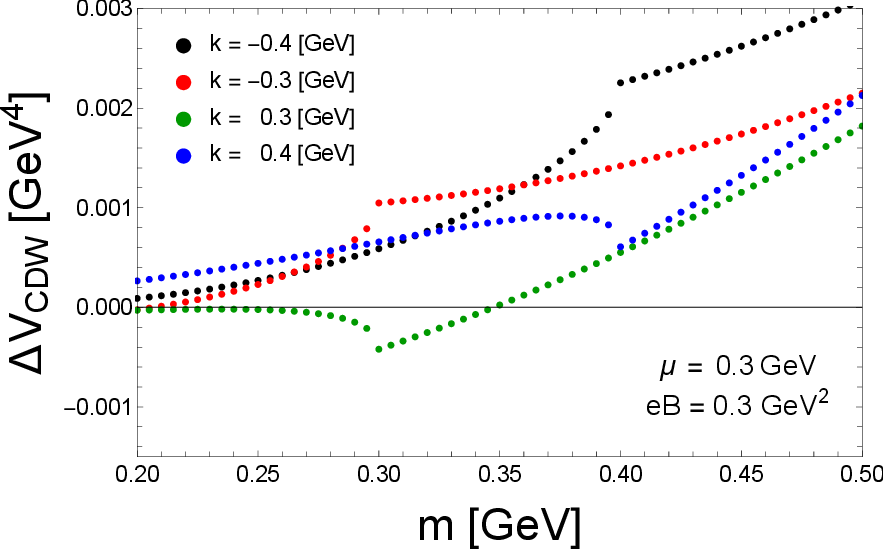}
	\caption{The thermodynamic potential $\Delta V\equiv V_{\rm BT}^{\rm CDWr}-V_{\rm \chi SR}^r$ as functions of the chiral magnitude $m$ for given wave vectors $k=\pm0.3, \pm0.4~{\rm GeV}$.}\label{Vcdw3}
\end{figure}

To carry out numerical calculations, the model parameters can be fixed as $\Lambda=0.786~\text{GeV}$ and $G\Lambda^2=6.24$, corresponding to dynamical quark mass $m=0.33~{\rm GeV}$ and pion decay constant $f_\pi=93~{\rm MeV}$~\cite{Cao:2016fby}. These parameters will be adopted whenever we refer to the PV regularization in this review. For the CDW phase, the calculations show that the points with $m=k$ are always the local minima of the thermodynamic potential at finite magnetic field and chemical potential, see Fig.\ref{Vcdw3} for example. 
Here, it turns out that the points with $m=-k$ are not extrema at all and the points with $m=k$ determine the global minimum for the chosen parameters. Nevertheless, the homogeneous phase with $k=0$ is still favored for smaller $\mu$, which contradicts with the conclusion from Ref.\cite{Frolov:2010wn} that CDW phase is immediately favored for nonzero $eB$ and $\mu$. One should note that the limit $m\rightarrow 0$ of $V_{\rm CDW}$ does not give the correct $V_{\rm \chi SR}$ in Ref.\cite{Frolov:2010wn} and this drawback is cured with the help of density of state here. Then for the study of CDW phase, it is enough to compare the thermodynamic potentials with $k=0$ and $k=m$. For illustration, the $m$-dependences of $V_{\rm CDW}(k=m)$ and homogeneous $V_{\rm \chi SB}(m)$ are shown for the critical chemical potential in Fig.\ref{Vcdw-Vn}, where one can immediately recognize the transition to be of first order.
\begin{figure}[!htb]
	\centering
	\includegraphics[width=0.45\textwidth]{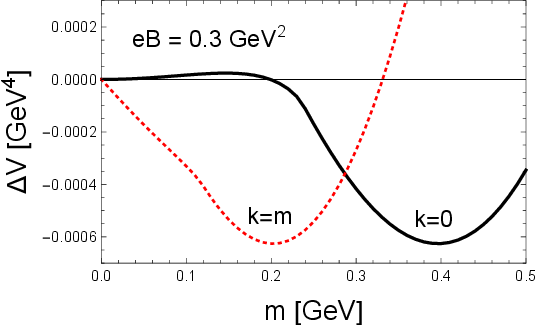}
	\caption{The thermodynamic potentials of $\chi$SB and CDW (with $k=m$) phases as functions of the chiral magnitude $m$ at the critical chemical potential $\mu_c=0.2376~{\rm GeV}$. The definition of  $\Delta V$ is the same as that in Fig.\ref{Vcdw3}.}\label{Vcdw-Vn}
\end{figure}

\begin{figure}[!htb]
	\centering
	\includegraphics[width=0.45\textwidth]{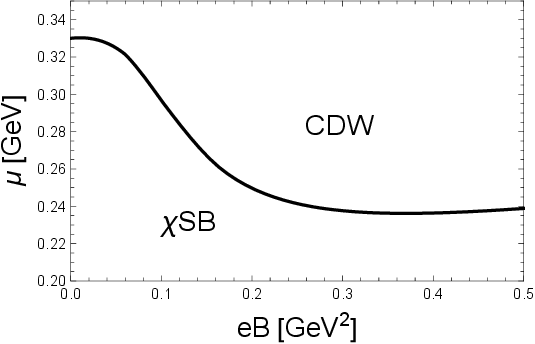}
	\caption{The $\mu-eB$ phase boundary for the first-order $\chi$SB-CDW phase transition.}\label{critical_CDW}
\end{figure}

Now, we take the SM phase into account. In Ref.~\cite{Cao:2016fby}, we found both lower and upper boundaries for such phase, which oscillate with the magnetic field. Right now, with the help of more advanced computers, the upper boundary is actually not found to exist, that is, $\nu$ never becomes vanishingly small for finite $eB$. Thus, it seems that the so-called "Ginzburg-Landau" approximation is not correct in Ref.~\cite{Cao:2016fby}. For the chosen magnetic field, we compare the minima of the thermodynamic potentials for $\chi$SB, CDW and SM phases in Fig.\ref{Vcdw-Vn}.  
\begin{figure}[!htb]
	\centering
	\includegraphics[width=0.45\textwidth]{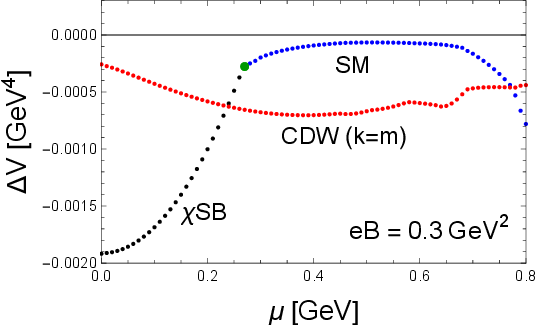}
	\caption{The minima of thermodynamic potentials for $\chi$SB, SM and CDW (with $k=m$) phases, as functions of $\mu$ at given magnetic field $eB=0.3~{\rm GeV}^2$. The definition of  $\Delta V$ follows that in Fig.\ref{Vcdw3}.}\label{Vcdw-Vn}
\end{figure}
Here, we can clearly identify the second-order transition from $\chi$SB phase to SM phase~\cite{Cao:2016fby},  but the latter is still not favored since the first-order transition from $\chi$SB phase to CDW phase occurs earlier. However, it seems that SM phase would overwhelm CDW phase when $\mu$ is large enough. But the critical value is already as large as the cutoff $\Lambda$, this might be an artifact. Hence, we need a better regularization scheme to clarify the situation. Forgetting that at this moment, we're satisfied to figure out the $\chi$SB-CDW transition boundary in Fig.\ref{critical_CDW}. Similar to the $\chi$SB-SM transition boundary in Ref.~\cite{Cao:2016fby}, we can also identify the feature of dHvA oscillation here.

\section{Neutral pseudoscalar superfluidity}\label{sec:pseudo}
Neutral pion $\pi^0$ is the lightest pseudoscalar meson in QCD and the uniform condensation of $\pi^0$ had been early explored in the context of disoriented chiral condensate (DCC)~\cite{Mohanty:2005mv}. The study found within the linear sigma model that large parallel EM (PEM) field, generated in heavy-ion collisions, could possibly catalyze the formation of the DCC~\cite{Minakata:1995gq,Asakawa:1998st}. Actually, DCC is only a far-from-equilibrium transient phenomenon, but the same result was rediscovered by us in a nearly equilibrium (quasi-)stationary state by adopting both chiral perturbation theory (ChPT) and NJL model~\cite{Cao:2015cka}. According to our study, the formula for $\pi^0$ condensate is general and model independent for two-flavor QCD. The system with PEM is also interesting for other considerations, such as the Schwinger mechanism~\cite{Copinger:2016llk,Ruggieri:2016lrn}, critical temperature~\cite{Ruggieri:2016xww} and real-time chiral magnetic effect~\cite{Copinger:2018ftr}, see the review Ref.~\cite{Copinger:2020nyx}. In particular, the in-in and in-out formalisms were compared to understand the axial Ward identity and the former was find to be more physical~\cite{Copinger:2018ftr}. We note that the expectation values of operators are evaluated with the states at the initial time in in-in scheme but with states from initial to final time in in-out scheme -- they are different in electric field~\cite{Fradkin:1991}.

Before moving on with explicit studies, we'd like to show that the PEM field induced neutral pion condensation does not violate the Vafa-Witten theorem at all. Actually, the proof of the Vafa-Witten theorem relies on the positivity of the Dirac determinant~\cite{Vafa:1983tf,Vafa:1984xg}, that is, $\det {\cal D}>0$ with ${\cal D}$ the Euclidean Dirac operator. However, the positivity is lost in the presence of a real electric field. For two-flavor quarks in PEM field, $\cal D$ is given by
\begin{eqnarray}
{\cal D}=\gamma_\mu(\partial_\mu-ig{\cal A}_\mu+iQ A_\mu)+m,
\end{eqnarray}
where ${\cal A}_\mu$is the gluon field, $A_\mu$ is the background EM field, and the charge matrix is $Q/e={\rm diag}(2/3,-1/3)=1/6+\tau_3/2$. If we choose the gauge such that ${\bf E}=-{\mathbf \nabla} A_0$ and $A_i$ is time independent, the crucial observation is that $Q$ is {\it not} traceless, thus the role of $Q A_0$ is partially similar to $\mu_{\rm B}$ which destroys the positivity of $\det {\cal D}$. Shifting to another gauge with $A_i=-E_i t$ would not change the conclusion as it still spoils the positivity after Wick rotation. Note that if $q_{\rm u}=-q_{\rm d}$, $\det {\cal D}$ is semi-positive~\cite{Yamamoto:2012bd} and the $\pi^0$ condensation is indeed gone, refer to Eqs.\eqref{Sphi} and \eqref{pi0II}. Physically, the Dirac determinant has to be complex in the presence of ${\bf E}$ in order to be consistent with the Schwinger mechanism of particle-antiparticle pair production~\cite{Schwinger:1951nm}. For the three-flavor case, the above discussions still apply and $\eta$ and $\eta'$ condensations are allowed in PEM.

\subsection{Two-flavor case in parallel EM field}\label{subsec:2PEM}

\subsubsection{Intuition from chiral perturbation theory }
As we know, ChPT is a very precise approximation of QCD at low energy scale. So firstly, we'd like to utilize ChPT to demonstrate how $\pi^0$ superfluidity can be developed in PEM and what is its nature. The Lagrangian density of two-flavor ChPT is given by
\begin{eqnarray}
{\cal L}\!\!\!&=&\!\!\!{\cal L}_0+{\cal L}_{\rm WZW};\label{chirall}\\
{\cal L}_0\!\!\!&=&\!\!\!\frac{f_\pi^2}{4}{\rm Tr}\left[ D_\mu U^\dagger D^\mu U+m_\pi^2 (U+U^\dagger)\right],\nonumber\\
&&D_\mu U=\partial_\mu U+A_\mu[Q,U];\label{chirall0}\\
{\cal L}_{\rm WZW}\!\!\!&=&\!\!\!\frac{N_c}{48\pi^2}A_\mu \epsilon^{\mu\nu\alpha\beta}\Big\{{\rm Tr}\left( QL_\nu L_\alpha L_\beta\!+\!Q R_\nu R_\alpha R_\beta\right)-\nonumber\\
&&\!\!\!\!\!\!\!\!\!\!\!\!\!\!\!\!\!\!i{F_{\alpha\beta}\over2}{\rm Tr}\left[2Q^2(L_\nu\!+\!R_\nu)\!+\!QUQU^\dagger L_\nu\!+QU^\dagger QUR_\nu\right]\Big\},\nonumber\\
&&L_\mu= U\partial_\mu U^\dagger,\ R_\mu= \partial_\mu U^\dagger U;\label{LWZW}
\end{eqnarray}
where ${\cal L}_0$ is the kinetic part and ${\cal L}_{\rm WZW}$ is the Wess-Zumino-Witten (WZW) term induced by chiral anomaly~\cite{Wess:1971yu,Witten:1983tw,Scherer:2012xha,Son:2007ny,Fukushima:2012fg}.
With Weinberg parametrization, the $2\times2$ unitary matrix for the chiral fields can be presented as
\begin{eqnarray}
U=\frac{1}{f_\pi}(s+i{{\bm\tau}\cdot{\bf t}}),
\end{eqnarray}
where the fields satisfy $s^2+{\bf t}^2=f_\pi^2$. Note that $s$ and ${\bf t}$ are proportional to the $\sigma$ and $\boldsymbol{\pi}$ fields in NJL model, respectively.

In the vacuum, it is well-known that the CSB takes place along the $\sigma$-direction so that $s=f_\pi$ and ${\bf t}={\bf 0}$. In constant EM field, we assume all the chiral condensates to be uniform in Eq.(\ref{chirall}), and the coupling between PEM field, with the second Lorentz invariant $I_2\equiv\bar{I}_2^2\equiv{\bf E}\cdot{\bf B}\neq 0$, and the $\pi^0$ field raises the possibility of $\pi^0$ condensation via the triangle anomaly. In this case, the CSB could be rotated from $\sigma$-direction to $\pi^0$-direction. To explore that, we set
\begin{eqnarray}
s= f_\pi \cos\theta,\ t_3=f_\pi\sin\theta,\
t_1= t_2=0,
\end{eqnarray}
where the chiral angle $\theta$ is restricted to $[-\pi/2,\pi/2]$ without loss of generality so that $s\geq0$. In mean field approximation, we drop all the regular derivatives with respect to $U$ to reduce the Lagrangian ${\cal L}$ simply to
\begin{eqnarray}
{\cal L}(\theta)&=&f^2_\pi m_\pi^2 \cos\theta+\frac{N_c I_2}{4\pi^2}{\rm Tr}(Q^2\tau_3) \theta,
\end{eqnarray}
where the second term is modified from ${\cal L}_{\rm WZW}$ by compensating a total derivative. It is obvious that the second term is responsible for the triangle anomaly processes such as the decay: $\pi^0\rightarrow \gamma\gamma$. Then, we can minimize the thermodynamic potential $-{\cal L}(\theta)$ and obtain
\begin{eqnarray}
\sin\theta= \frac{N_c I_2}{4\pi^2 f_\pi^2 m_\pi^2}{\rm Tr}(Q^2\tau_3)=\frac{N_c I_2}{4\pi^2 f_\pi^2 m_\pi^2}(q_{\rm u}^2-q_{\rm d}^2)\label{Sphi}
\end{eqnarray}
for $I_2$ smaller than the critical strength $$I_2^c=4\pi^2 f_\pi^2 m_\pi^2/[N_c(q_{\rm u}^2-q_{\rm d}^2)].$$ Beyond $I_2^c$, no local minima can be found and the chiral angle $\theta$ would keep rotating all the time, like a "chaotic" phase~\cite{Cao:2020pjq}.

\begin{figure}[!htb]
\begin{center}
\includegraphics[width=5cm]{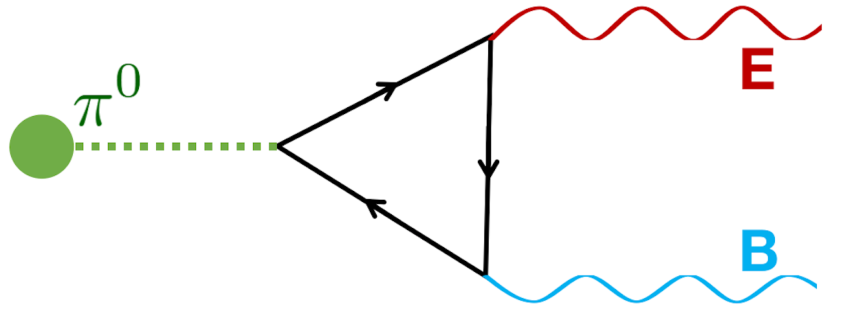}
\caption{The electromagnetic triangle anomaly.}\label{EMTA}
\label{triangle}
\end{center}
\end{figure}

Some comments are in order. (1) As $\pi^0$ condensate is only possible when WZW term is present, its origin can be attributed to triangle anomaly, see the illustration in Fig.\ref{EMTA}. 
(2) The PEM induced $\pi^0$ condensation is merely a chiral rotation but not a phase transition, as no extra symmetry is broken in this process. The point can be well verified by noticing the full analyticity of ${\cal L}(\theta)$ around $\theta\sim 0$.
(3) As a low-energy effective theory of QCD, the ChPT is reliable only when all the involved energy scales, such as  $m_\pi$ and $(e\bar{I}_2)^{1/2}$ here, are much smaller than the typical hadronic one $\Lambda_\chi\sim 1\,{\rm  GeV}$. Recalling the explicit form of $I_2^c$, such validity condition for the ChPT is well satisfied. 
(4) As well-known, for $e{E}\gg m_\pi^2$, the QCD vacuum will be unstable against the Schwinger pair production of $\pi^+\pi^-$  or $q\bar{q}$, and the feedback should be properly taken into account through the real-time formalism~\cite{Copinger:2018ftr,Cao:2019hku}.
(5) In the chiral limit, $(e\bar{I}_2^c)^{1/2}=m_\pi=0$, then an infinitesimal $I_2$ will drive the system to a "chaotic" phase~\cite{Cao:2020pjq}.

\subsubsection{Nambu--Jona-Lasinio Model calculations}
The ChPT is built on hadronic degrees of freedom, we now adopt the NJL model, which is based on more fundamental quark degrees of freedom, to carry out both analytical and numerical calculations. With uniform $\pi^0$ condensate, the quark propagator is formally modified from Eq.\eqref{Green's} to~\cite{Cao:2015cka}
\begin{eqnarray}
G(x,x')=-(i\slashed{D}-m-i\pi^0\gamma_5\tau_3)^{-1}.
\end{eqnarray}
It is diagonal in flavor space which allows us to calculate it analytically by adopting the Schwinger proper-time formalism~\cite{Schwinger:1951nm}, which is actually the in-out scheme~\cite{Copinger:2018ftr}. In PEM field, it is actually more convenient to explicitly obtain the gap equations first according to the formal expressions~\cite{Cao:2015cka}:
\bea
{m-m_0\over 2G}-{1\over V_4}\text {Tr}\;{G}(x,x')&=&0,\nonumber\\
 {\pi^0\over 2G}-{1\over V_4}\text {Tr}\;{G}(x,x')i\gamma^5&=&0,\label{Gmpi}
 \eea
and then integrate over the order parameters $m$ and $\pi^0$ to find the thermodynamic potential.

Now, we try to work out the whole formalism with the help of the explicit quark propagator~\cite{Cao:2015cka}:
\begin{widetext}
\begin{eqnarray}
	G_{\rm f}(x,x')
	&=&{-i\,q_{\rm f}Bq_{\rm f}E\over(4\pi)^2}\int_0^\infty\di s\;e^{-iq_{\rm f}\int_{x'}^xA\cdot dx}\exp\Big\{-im^{*2}s+{i\over4}\left[{q_{\rm f}B\over\tan(q_{\rm f}Bs)}(y_1^2+y_2^2)+{q_{\rm f}E\over\tanh(q_{\rm f}Es)}(y_3^2-y_0^2)\right]\Big\}\nonumber\\
	&&\left\{m\!-\!i\,{\cal S}(q_{\rm f})\gamma^5\pi^0\!-\!{q_{\rm f}B\over2}\Big[\big(\cot(q_{\rm f}Bs)\gamma^1\!+\!\gamma^2\big)y_1\!+\!\big(\cot(q_{\rm f}Bs)\gamma^2-\gamma^1\big)y_2\Big]\!-\!{q_{\rm f}E\over2}\Big[\big(\coth(q_{\rm f}Es)\gamma^3\!-\!i\gamma^4\big)y_3\right.\nonumber\\
	&&\left.+\big(\coth(q_{\rm f}Es)\gamma^4+i\gamma^3\big)i\,y_0\Big]\right\}\Big[\cot(q_{\rm f}Bs)\coth(q_{\rm f}Es)+i\gamma^5
	+\coth(q_{\rm f}Es)\gamma^1\gamma^2+i\cot(q_{\rm f}Bs)\gamma^4\gamma^3\Big]\label{GEM}
	\end{eqnarray}
with variables $y\equiv x-x'$.  Apart for the Schwinger phase, the effective fermion propagator to $G_{\rm f}(x,x')$ is a function of the relative coordinate $y$, so we can take Fourier transformation to give the form in energy-momentum space as~\cite{Wang:2017pje}
\begin{eqnarray}
\hat{G}_{\rm f}({p})
&=&i\int {\di s}\exp\Big\{-i (m^{*2}-i\eta)s-i{\tan(q_{\rm f}Bs)\over q_{\rm f}B}(p_1^2+p_2^2)-i{\tanh(q_{\rm f}Es)\over q_{\rm f}E}(-{p}_0^2+p_3^2)\Big\}\nonumber\\
&&\big[m\!-\!i\,{\cal S}(q_{\rm f})\gamma^5\pi^0+i\,\gamma^4(p_0+{\tanh(q_{\rm f}Es)}p_3)\!-\!\gamma^3(p_3+\,{\tanh(q_{\rm f}Es)}p_0)\!-\!\gamma^2(p_2+{\tan(q_{\rm f}Bs)}p_1)\nonumber\\
&&-\gamma^1(p_1-{\tan(q_{\rm f}Bs)}p_2)\big]\Big[1+{i\gamma^5\tanh(q_{\rm f}Es)\tan(q_{\rm f}Bs)}
+{\gamma^1\gamma^2\tan(q_{\rm f}Bs)}-i{\gamma^4\gamma^3\tanh(q_{\rm f}Es)}\Big].\label{prop_em}
\end{eqnarray}
Taking the limits $E,\pi^0\rightarrow0$, the effective propagator in pure magnetic field is then given by
\begin{eqnarray}
\hat{G}_{\rm f}({p})
&=&i\int {\di s}\exp\Big\{-i (m^{2}-{p}_0^2+p_3^2-i\eta)s-i{\tan(q_{\rm f}Bs)\over q_{\rm f}B}(p_1^2+p_2^2)\Big\}\left[m+i\,\gamma^4p_0\!-\!\gamma^3p_3\!-\!\gamma^2(p_2+{\tan(q_{\rm f}Bs)}p_1)\right.\nonumber\\
&&\left.-\gamma^1(p_1-{\tan(q_{\rm f}Bs)}p_2)\right]\Big[1
+{\gamma^1\gamma^2\tan(q_{\rm f}Bs)}\Big].\label{prop_m}
\end{eqnarray}

To facilitate the discussions, we give the thermodynamic potential first as~\cite{Cao:2015cka}
\bea
V_{\rm EM}(m,\pi^0)&=&{(m-m_0)^2+(\pi^0)^2\over 4G}+{N_c\over8\pi^2}\sum_{f=u,d}\int_0^\infty {ds\over s^3}e^{-\left[m^2+(\pi^0)^2\right]s}{q_{\rm f}Es
	\over\tan(q_{\rm f}Es)}{q_{\rm f}Bs
	\over\tanh(q_{\rm f}Bs)}\nonumber\\
	&&-{N_c\over4\pi^{2}}\tan^{-1}\Big({\pi^0\over m}\Big)
	(q_{\rm u}^2-q_{\rm d}^2)EB,
\eea
or alternatively in a more general form with polar variables as~\cite{Wang:2017pje}
\bea
V_{\rm EM}(m^*,\theta)={m^{*2}\!-\!2m^*m_0\cos\theta\!+\!m_0^2\over 4G}\!+\!{N_c\over8\pi^2}\sum_{f=u,d}\int_0^\infty {ds\over s^3}e^{-m^{*2}s}{q_{\rm f}Es
	\over\tan(q_{\rm f}Es)}{q_{\rm f}Bs
	\over\tanh(q_{\rm f}Bs)}-{N_c\over4\pi^{2}}\theta
	(q_{\rm u}^2-q_{\rm d}^2)EB.
\eea
According to the variable transformations: $m^*=\sqrt{m^2+(\pi^0)^2}$ and $\tan\theta=\pi^0/m$, there should be $2\pi$-periodicity of $V_{\rm EM}(m^*,\theta)$, so the $\theta$-linear term can be understood by constraining $\theta$ to a given period. Only the second term is divergent and can be regularized in a similar way as Eq.\eqref{VB}, that is,
\bea
V_{\rm EM}(m^*)\!\!\!&\equiv&\!\!\!{N_c\over8\pi^2}\!\sum_{\rm f=u,d}\int_0^\infty\!{\di s\over s^3}\left(e^{-m^{*2}s}\!-\!1\right)\left[\!{q_{\rm f}Es
	\over\tan(q_{\rm f}Es)}{q_{\rm f}Bs\over\tanh(q_{\rm f}Bs)}-\!1\!\right]\!+\!{N_c\over8\pi^2}\!\sum_{\rm f=u,d}\int_{\Lambda_{\rm s}}^\infty\!{\di s\over s^3}\left[\!{q_{\rm f}Es
	\over\tan(q_{\rm f}Es)}{q_{\rm f}Bs\over\tanh(q_{\rm f}Bs)}-\!1\!\right]\nonumber\\
&&-4N_c\int^\Lambda{\di^3p\over(2\pi)^3}(p^2+m^{*2})^{1\over2}.\label{VEM}
\eea

Eventually, the regularized thermodynamic potential is 
\bea
V_{\rm EM}^r(m^*,\theta)&=&{m^{*2}\!-\!2m^*m_0\cos\theta\!+\!m_0^2\over 4G}+V_{\rm EM}(m^*)-{N_c\over4\pi^{2}}\theta
	(q_{\rm u}^2-q_{\rm d}^2)EB,
\eea
and the associated gap equations are given through $\partial_{m^*,\theta}V_{\rm EM}^r(m^*,\theta)=0$ as:
\bea
0&=&{m^{*}\!-\!m_0\cos\theta\over 2G}-{N_c\over4\pi^2}m^{*}\!\sum_{\rm f=u,d}\int_0^\infty\!{\di s\over s^2}e^{-m^{*2}s}\left[\!{q_{\rm f}Es
	\over\tan(q_{\rm f}Es)}{q_{\rm f}Bs\over\tanh(q_{\rm f}Bs)}-\!1\!\right]-4N_c\int^\Lambda\!{\di^3p\over(2\pi)^3}{m^*\over(p^2\!+\!m^{*2})^{1\over2}},\\
0&=&{2m^*m_0\sin\theta\over 4G}-{N_c\over4\pi^{2}}(q_{\rm u}^2-q_{\rm d}^2)EB.\label{Gtheta}
\eea
\end{widetext}
By assuming the NJL version of the Gell-Mann-Oakes-Renner (GMOR) relation~\cite{Klevansky:1992qe}
: $m^2_\pi f_\pi^2=m_0m^*(2G)^{-1}$, Eq.\eqref{Gtheta} exactly reduces to Eq.\eqref{Sphi} which justifies its model independence. As a matter of fact, $m^*, G$ and the GMOR relation might be altered in external EM field, thus one should keep in mind that the expression Eq.\eqref{Sphi} for $\pi^0$ condensation is only valid when the EM field is not too strong. In principle, the in-in scheme is the correct one to treat the electric effect~\cite{Copinger:2018ftr}, but the in-out one presented here gives almost the same results when the Schwinger pair production (SPP) is exponentially suppressed for $I_2<I_2^c$, see Eq.\eqref{pi0II}. So within that valid region, Eq.\eqref{Sphi} is even a very precise approximation to Eq.\eqref{pi0II}.

To explore the phase diagram correctly and completely, next we'd like to develop the formalism with in-in scheme as the Schwinger mechanism might be important for $I_2> I_2^c$. The fermion propagator in in-in scheme is mainly different from that in in-out one by the integral region of the proper-time, thus the explicit form can be adjusted from Eq.\eqref{GEM} by shifting the integral region $\int_{0}^{\infty}$ to $\int_{\Gamma_{{\rm f}\epsilon}}\equiv\int_{0}^{\infty}-\int_{-i(S_{\rm fE}-\epsilon)}^{\infty-i(S_{\rm fE}-\epsilon)}$ for the study of phase diagram~\cite{Cao:2019hku}. Here, the SPP parameter is defined as $S_{\rm fE}=\pi/|q_{\rm f}E|$.

\begin{widetext}
Then, the "gap equations" can be evaluated according to Eq.\eqref{Gmpi} as~\cite{Cao:2019hku}
\begin{eqnarray}
0&=&{m-m_0\over 2G}-N_cm\sum_{\rm f=u,d}K_{\rm f}(m^{*},E,B)+N_c\sum_{\rm f=u,d}{\cal S}(q_{\rm f}){\pi^0q_{\rm f}Eq_{\rm f}B\over 4\pi^2m^{*2}}\left(1-e^{-m^{*2}S_{\rm fE}}\right),\label{gapm}\\
0&=&{\pi^0\over 2G}-N_c\pi^0\sum_{\rm f=u,d}K_{\rm f}(m^{*},E,B)-N_c\sum_{\rm f=u,d}{\cal S}(q_{\rm f}){mq_{\rm f}Eq_{\rm f}B\over 4\pi^2m^{*2}}\left(1-e^{-m^{*2}S_{\rm fE}}\right),\label{gappi0}\\
K_{\rm f}(m^{*},E,B)&=&\Re\left\{\int_{0}^{\infty}{\di s\over 4\pi^2}\;e^{-m^{*2}s}\left[{q_{\rm f}B\over\tanh(q_{\rm f}Bs)}{q_{\rm f}E\over\tan(q_{\rm f}Es)}-e^{-m^{*2}S_{\rm fE}}{q_{\rm f}B\over\tanh(q B(s+S_{\rm fE})}{q_{\rm f}E\over\tan(q_{\rm f}Es)}\right]\right\},\label{Kfunc}
\end{eqnarray}
where we can easily identify the feedback of SPP through the term $e^{-m^{*2}S_{\rm fE}}$. 
Similar to the "vacuum regularization" scheme, we introduce counter terms to deal with the divergence of the proper time integral in Eq.\eqref{Kfunc} and find
\begin{eqnarray}
K_{\rm f}^r(m^{*},E,B)&=&\Re\left\{\int_{0}^{\infty}{\di s\over 4\pi^2}\;e^{-m^{*2}s}\left[{q_{\rm f}B\over\tanh(q_{\rm f}Bs)}{q_{\rm f}E\over\tan(q_{\rm f}Es)}-{1\over s^2}-e^{-m^{*2}S_{\rm fE}}\left({q_{\rm f}B\over\tanh(q B(s+S_{\rm fE})}{q_{\rm f}E\over\tan(q_{\rm f}Es)}\right.\right.\right.\nonumber\\
&&\left.\left.\left.-{1\over s(s+\tilde{S}_{\rm fE}(1))}\right)\right]\right\}+F_{\Lambda}^{0^+}(m^{*})-e^{-m^{*2}S_{\rm fE}}F_{\Lambda}^{\tilde{S}_{\rm fE}(1)}(m^{*}).\label{Kfuncr}
\end{eqnarray}
Here, we define $\tilde{S}_{\rm fE}(x)=\tanh(x\,q_{\rm f}BS_{\rm fE})/q_{\rm f}B$ for simplicity and the auxiliary function is given by
\begin{eqnarray}
F_{\Lambda}^{S}(m^{*})\equiv \int^{\Lambda}{\di^3 p\over 2\pi^3}\int_{-\infty}^{\infty}{\di p_4\over 2\pi}{e^{-p_\bot^2S}\over p_4^2+E_{\rm p}^2}=\int_0^{\Lambda}{p\di p\over \pi^{2}}{D\left(pS^{1/2}\right)\over S^{1/2}E_{\rm p}}
\end{eqnarray}
with $p_\bot^2=p_1^2+p_2^2, E_{\rm p}=\sqrt{{\bf p}^2+m^{*2}}$ and $D(x)$ the Dawson's integral function. In the large $eB$ limit, $\tilde{S}_E(1)\rightarrow 0^+$, thus $K_{\rm f}^r\approx(1-e^{-m^{*2}S_{\rm fE}})\tilde{K}_{\rm f}^r$ with $\tilde{K}_{\rm f}^r$  exactly the same as that in the IOF, which is consistent with that found in Ref.~\cite{Copinger:2018ftr}. With the substitution $K_{\rm f}\rightarrow K_{\rm f}^r$ in both "gap equations" Eqs.\eqref{gapm} and \eqref{gappi0}, the neutral pion condensate can be self-consistently  solved as
  \bea
\pi^0=\sum_{\rm f=u,d}{\cal S}(q_{\rm f}){q_{\rm f}Eq_{\rm f}B\over 4\pi^2}\left(1-e^{-m^{*2}S_{\rm fE}}\right){2G\over m_0},\label{pi0II}
\eea
which differs from the IOF one Eq.\eqref{Sphi} by extra SPP terms.

Actually, the SPP contributions in the anomaly terms of Eqs.\eqref{gapm} and \eqref{gappi0} render
$${\partial\over\partial\pi^0}{1\over V_4}\text {Tr}\;{G}(x,x)\neq {\partial\over\partial m}{1\over V_4}\text {Tr}\;{G}(x,x)i\gamma^5,$$
so no self-consistent thermodynamic potential can be defined for the "gap equations" in in-in scheme. This is not a terrible problem since the system is non-equilibrium {\it ab initio} in in-in scheme, which indicates that the thermodynamic potential might be not well defined. Nevertheless, a consistent thermodynamic potential can be found for the anomaly irrelevant terms with the regularized form
\begin{eqnarray}
V_{\rm EM}^{II}(m,\pi^0)
&=&{(m-m_0)^2+(\pi^0)^2\over 4G}+N_c\sum_{\rm f=u,d}\,\Re\left\{\int_{0}^{\infty}{\di s\over 8\pi^2}\;{e^{-m^{*2}s}\over s}\left[{q_{\rm f}B\over\tanh(q_{\rm f}Bs)}{q_{\rm f}E\over\tan(q_{\rm f}Es)}-{1\over s^2}-{s\,e^{-m^{*2}S_{{\rm f}E}}\over s+S_{{\rm f}E}}\times\right.\right.\nonumber\\
&&\left.\left.\left({q_{\rm f}B\over\tanh(q B(s+S_{{\rm f}E})}{q_{\rm f}E\over\tan(q_{\rm f}Es)}-{1\over s(s+\tilde{S}_{{\rm f}E}(1) )}\right)\right]-G_{\rm f\Lambda}^{0^+}(m^{*})+G_{\rm f\Lambda}^{\tilde{S}_{{\rm f}E}(1)}(m^{*})-(m^{*}\rightarrow0)\right\}.
\end{eqnarray}
Here, the $m^{*}$ independent but $B$ dependent terms are dropped for brevity compared to Eq.\eqref{VEM}, and the auxiliary function is defined as
\begin{eqnarray}
G_{\rm f\Lambda}^{S}(m^{*})&\equiv&\int_0^{m^{*}} M\di M~e^{-{M}^2S_{{\rm f}E}}F_{\Lambda}^{S}(M)=\int_0^{\Lambda}{p\di p\over 2\pi^{3/2}}{e^{p^2S_{{\rm f}E}}\left[{\rm Erf}\left({S}_{{\rm f}E}^{1/2}E_{\rm p}\right)-{\rm Erf}\left({S}_{{\rm f}E}^{1/2}p\right)\right]{D}\left(pS^{1/2}\right)\over {S}_{{\rm f}E}^{1/2}S^{1/2}}
\end{eqnarray}
with ${\rm Erf}(x)$ the error function. When chiral anomaly is not relevant to the ground state of the system, such as in pure electric field and beyond $I_2^c$ in PEM, the quantity $-V_{\rm EM}^{II}+V_{\rm EM}^{II}|_{E,B\rightarrow0}$ can be roughly understood as the in-in pressure. As should be, the pressure receives contributions from the persistent pair productions of real quarks and anti-quarks in the presence of electric field. 
\end{widetext}

\begin{figure}[!htb]
	\centering
	\includegraphics[width=0.41\textwidth]{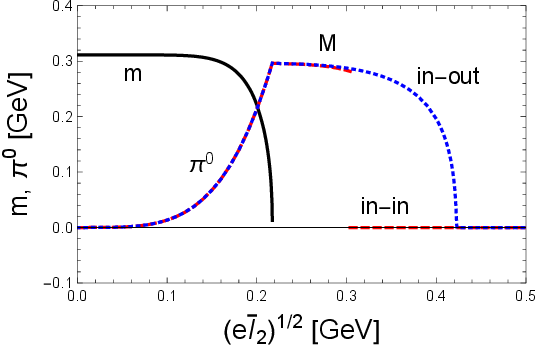}
	\caption{The illustrations of mass gap $m$ and pion condensate $\pi^0$ as functions of EM field $e\bar{I}_2$ in the in-out (blue dotted lines) and in-in (red dashed lines) schemes, respectively. The evolutions of $m$ are almost the same up to $\bar{I}_2^c$ thus are denoted by a single black solid line for both schemes. The plots are modified from Ref.~\cite{Cao:2019hku}.}\label{mpi0_2f}
\end{figure}

\begin{figure}[!htb]
	\centering
	\includegraphics[width=0.41\textwidth]{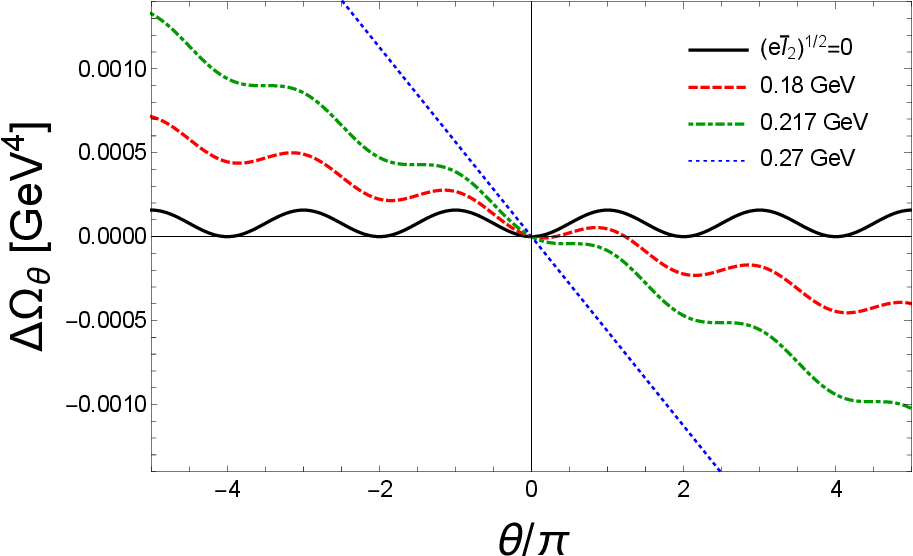}
	\caption{The $\theta$ dependence of the thermodynamic potential $\Delta\Omega_\theta=\Omega_\theta-\Omega_0$ for chosen EM fields $(e\bar{I}_2)^{1/2}=0, 0.18, 0.217$ and $0.27~{\rm GeV}$. The blue dotted line corresponds to the free case with $\Delta\Omega_\theta=-N_c{q^2I_2\over 4\pi^2}\theta$. The plots are from Ref.~\cite{Cao:2019hku}.}\label{omg_theta}
\end{figure}

Firstly, we consider the simple case $eE=eB=e\bar{I}_2$ and compare the results of in-out and in-in schemes in Fig.\ref{mpi0_2f}~\cite{Cao:2019hku}.
With $\pi^0$ condensate increasing with small $e\bar{I}_2$, the chiral rotation can be well checked~\cite{Cao:2015cka}. As we can  see, the difference is invisible up to the end of chiral rotation $\bar{I}_2^c$, but the feedback of Schwinger pair production in in-in scheme alters the second-order chiral restoration in in-out scheme to a first-order one at s smaller $e\bar{I}_2$. Actually, the study of the collective modes indicates that the presumed ground state might be unstable beyond the critical point $\bar{I}_2^c$~\cite{Cao:2019hku}. If we look at the $\theta$ dependence of the thermodynamic potential in Fig.\ref{omg_theta}, we can readily find the origin of the instability: it is the loss of local minima, usually separated by $2\pi$, that induces the system to a "chaotic" state where $\theta$ keeps changing. So beyond $\bar{I}_2^c$, one can only understand the curve as the changing of the magnitude $M$ with $e\bar{I}_2$ rather than the initially conceived $\pi^0$ condensate~\cite{Cao:2015cka}.

\begin{figure}[!htb]
	\begin{center}
		\includegraphics[width=8cm]{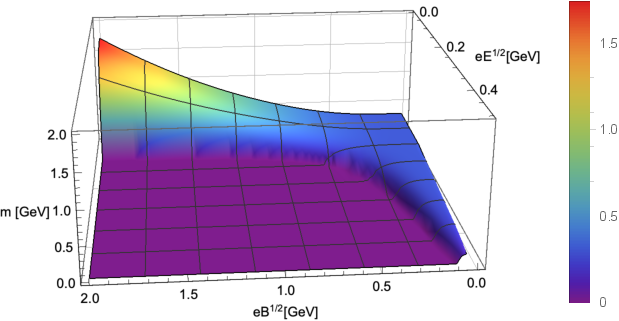}
		\includegraphics[width=8cm]{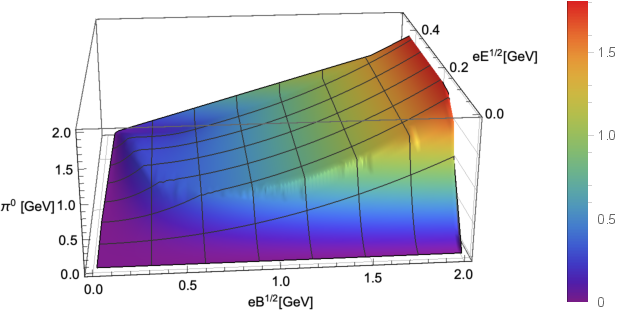}
		\caption{The mass gap $m$ and pion condensate $\pi^0$ as functions of the parallel electric field $eE$ and magnetic field $eB$. The plots are from Ref.~\cite{Wang:2017pje}.}\label{mpi0}
	\end{center}
\end{figure}

Secondly, it is also interesting to vary the electric and magnetic fields separately, see the in-out evaluations in Fig.\ref{mpi0}~\cite{Wang:2017pje}. Due to the competition between MCE and chiral rotation effect, we found a critical electric field ${eE_{\rm c}}=(86.4~{\rm MeV})^2$ for two-flavor NJL model, below which MCE dominates while above which chiral rotation effect dominates. Concretely, the dynamical mass $m$ increases with $eB$ for $eE<eE_{\rm c}$ but decreases with larger $eB$ for $eE>eE_{\rm c}$. 
 
\subsection{Three-flavor case in parallel EM field}\label{subsec:3PEM}

As had been naively demonstrated in the $U_A(1)$ symmetric two-flavor NJL model~\cite{Wang:2018gmj}, other neutral pseudoscalar meson like $\eta^0$ can also condense in PEM field when the corresponding interacting channel is turned on. The reason is simple: for a given flavor, the PEM field is able to induce condensation in the pseudoscalar channel through the triangle anomaly; then there should be condensations for various combinations of flavors, such as $\pi^0$ and $\eta^0$ in Ref.~\cite{Wang:2018gmj}. However, the $U_A(1)$ symmetric two-flavor NJL model is not so realistic at all: on one hand, strange quark can still be an important degree of freedom in QCD system, especially when EM field is strong; on the other hand, QCD $U_A(1)$ anomaly should be taken into account in NJL model in order to understand the meson properties properly~\cite{tHooft:1976rip,tHooft:1976snw}. Both problems can be automatically cured by adopting the three-flavor NJL model, where more low-lying collective modes are involved and the $U_A(1)$ anomaly has been taken into account through the 't Hooft term. In this model, the competition between QCD and QED anomalies might be very interesting. In the following, we derive the in-in formalism under the framework of three-flavor NJL model, and $\eta^8$ meson condensation would show up here~\cite{Cao:2020pjq}.

The corresponding Lagrangian density is given by~\cite{Klevansky:1992qe,Hatsuda:1994pi}:
\begin{eqnarray}
{\cal L}_{\rm NJL}\!\!&=&\!\!\bar\psi(i\slashed{D}-m_0)\psi+G\sum_{a=0}^8\left[(\bar\psi\lambda^a\psi)^2+(\bar\psi i\gamma_5\lambda^a\psi)^2\right]\nonumber\\
&&+{\cal L}_{\rm tH},\  \ {\cal L}_{\rm tH}=-K\sum_{s=\pm}{\det}~\bar\psi\Gamma^s\psi,\label{NJL3f}
\end{eqnarray}
where $\psi=(u,d,s)^T$ is now the three-flavor quark field and $ {\cal L}_{\rm tH}$ is the 't Hooft determinant with coupling $K$. Here, in the kinetic term, the current mass and charge matrices are respectively 
\bea
m_0&=&{\rm diag}(m_{\rm 0u},m_{\rm 0d},m_{\rm 0s}),\nonumber\\
Q&\equiv&{\rm diag}(q_{\rm u},q_{\rm d},q_{\rm s})={\rm diag}\left({2\over 3},-{1\over 3},-{1\over 3}\right)e
\eea
in the three-flavor case. In the four-quark interaction terms, more channels are involved in the flavor space with $\lambda^0=\sqrt{2/3}~{\rm diag}(1,1,1)$ and $\lambda^i~(i=1,\dots,8)$ the Gell-Mann matrices. Finally, in the 't Hooft term, the determinant is performed in flavor space thus renders this term six-quark interactions, and the vertices $\Gamma^\pm=1\pm\gamma_5$ correspond to right- and left-handed channels, respectively. 

Inspired from the study in Sec.\ref{subsec:2PEM}, we should set two kinds of condensates in PEM:
 \bea
 \pi_{\rm f}^5=\langle\bar\psi_{\rm f}i\gamma^5\psi_{\rm f}\rangle/N_c,\ \sigma_{\rm f}=\langle\bar\psi_{\rm f}\psi_{\rm f}\rangle/N_c\label{pi5s}
 \eea
 for ${\rm f}={\rm u,d,s}$. To facilitate the study, ${\cal L}_{\rm tH}$ can be reduced to an effective form of four-fermion interactions in Hartree approximation~\cite{Klevansky:1992qe}. When a quark-antiquark pair is contracted in each six-fermion interaction term, we find immediately
\begin{widetext}
\begin{eqnarray}
{\cal L}_{\rm tH}^4&=&-{K\over2}\sum_{s=\pm}\epsilon_{ijk}\epsilon_{imn}\langle\bar{\psi}^i\Gamma^s{\psi}^i\rangle(\bar{\psi}^j\Gamma^s{\psi}^m)(\bar{\psi}^k\Gamma^s{\psi}^n)\nonumber\\
&=&\!\!-N_c{K}\epsilon_{ijk}\epsilon_{imn}\left\{\sigma_i\left[(\bar{\psi}^j{\psi}^m)(\bar{\psi}^k{\psi}^n)
\!-\!(\bar{\psi}^ji\gamma^5{\psi}^m)(\bar{\psi}^ki\gamma^5{\psi}^n)\right]-2\pi_i^5(\bar{\psi}^ji\gamma^5{\psi}^m)(\bar{\psi}^k{\psi}^n)\right\}\nonumber\\
&=&-N_c{K\over6}\Big\{2\!\sum_{i=1}^3\sigma_{i}(\bar{\psi}\lambda^0\psi)^2\!-\!3\sigma_{\rm s}\!\sum_{i=1}^3(\bar{\psi}\lambda^i\psi)^2\!-\!3\sigma_{\rm d}\!\sum_{i=4}^5(\bar{\psi}\lambda^i\psi)^2
\!-\!3\sigma_{\rm u}\!\sum_{i=6}^7(\bar{\psi}\lambda^i\psi)^2\!-\!\!\sum_{i=1}^3\! \sigma_i\Big[{(\bar{\psi}\lambda^8\psi)^2
\over \sqrt{3}/2\lambda^8_{ii}}\!+\!\sqrt{6}\lambda^8_{ii}(\bar{\psi}\lambda^0\psi)\nonumber\\
&&(\bar{\psi}\lambda^8\psi)+\sqrt{6}\lambda^3_{ii}(\bar{\psi}\lambda^3\psi)(\bar{\psi}\lambda^0\psi-\sqrt{2}\bar{\psi}\lambda^8\psi)\Big]-(\lambda^a\rightarrow i\lambda^a\gamma^5)\Big\}\nonumber\\
&&+N_c{K\over3}\Big\{2\sum_{i=1}^3\pi^5_i(\bar{\psi}\lambda^0\psi)(\bar{\psi}i\lambda^0\gamma^5\psi)
\!-\!3\pi^5_{\rm s}\sum_{i=1}^3(\bar{\psi}\lambda^i\psi)(\bar{\psi}i\lambda^i\gamma^5\psi)
-3\pi^5_{\rm d}\sum_{i=4}^5(\bar{\psi}\lambda^i\psi)(\bar{\psi}i\lambda^i\gamma^5\psi)-3\pi^5_{\rm u}\sum_{i=6}^7
(\bar{\psi}\lambda^i\psi)\nonumber\\
&&(\bar{\psi}i\lambda^i\gamma^5\psi)\!-\!\sum_{i=1}^3\! \pi^5_i\Big[{(\bar{\psi}\lambda^8\psi)(\bar{\psi}i\lambda^8\gamma^5\psi)\over \sqrt{3}/2\lambda^8_{ii}}\!+\!{\sqrt{6}\over 2}\lambda^8_{ii}\Big((\bar{\psi}\lambda^0\psi)(\bar{\psi}i\lambda^8\gamma^5\psi)+(\bar{\psi}\lambda^8\psi)(\bar{\psi}i\lambda^0\gamma^5\psi)\Big)+{\sqrt{6}\over 2}\lambda^3_{ii}\Big((\bar{\psi}\lambda^3\psi)\nonumber\\
&&(\bar{\psi}i\lambda^0\gamma^5\psi-\sqrt{2}\bar{\psi}i\lambda^8\gamma^5\psi)+(\bar{\psi}i\lambda^3\gamma^5\psi)(\bar{\psi}\lambda^0\psi-\sqrt{2}\bar{\psi}\lambda^8\psi)\Big)\Big]\Big\},\label{LtH4}
\end{eqnarray}
where $\epsilon_{ijk}$ is the Levi-Civita symbol and the Einstein summation convention should be understood for all the flavor indices in the first two steps. Here and in the following, note the correspondence between the number index $i=1,2,3$ and the more explicit Latin index ${\rm f=u,d,s}$ -- the latter will be adopted whenever convenient. Substituting ${\cal L}_{\rm tH}$ by ${\cal L}_{\rm tH}^4$ in Eq.\eqref{NJL3f} , we obtain an effective Lagrangian density with only four-fermion interactions, that is,
\begin{eqnarray}
{\cal L}_{\rm NJL}^4\!\!&=&\!\!\bar\psi(i\slashed{D}-m_0)\psi+\sum_{a=0}^8\left\{G_{aa}^-(\bar\psi\lambda^a\psi)^2+G_{aa}^{+}(\bar\psi i\gamma_5\lambda^a\psi)^2+G_{aa}^5[(\bar\psi\lambda^a\psi)(\bar\psi i\lambda^a\gamma^5\psi)+(\bar\psi i\lambda^a\gamma^5\psi)(\bar\psi\lambda^a\psi)]\right\}+\nonumber\\
&&\!\!\!\sum_{a,b=0,3,8}^{a\neq b}\!\!\left\{G_{ab}^-(\bar\psi\lambda^a\psi)(\bar\psi\lambda^b\psi)\!+\!G_{ab}^+(\bar\psi i\lambda^a\gamma^5\psi)(\bar\psi i\lambda^b\gamma^5\psi)\!+\!G_{ab}^5[(\bar\psi \lambda^a\psi)(\bar\psi i\lambda^b\gamma^5\psi)\!+\!(\bar\psi i\lambda^a\gamma^5\psi)(\bar\psi \lambda^b\psi)]\right\}\!,\label{LNJL4}
\end{eqnarray}
where the elements of the symmetric scalar ($G^-$), pseudoscalar ($G^+$) and mixing ($G^5$) coupling matrices are respectively
\begin{eqnarray}\label{Gelements}
&&\!\!\!\!G_{00}^\mp=G\mp N_c{K\over3}\sum_{i=1}^3\sigma_i,~G_{11}^\mp=G_{22}^\mp=G_{33}^\mp=G\pm N_c{K\over2}\sigma_{\rm s},~G_{44}^\mp=G_{55}^\mp=G\pm N_c{K\over2}\sigma_{\rm d},~G_{66}^\mp=G_{77}^\mp=G\pm N_c{K\over2}\sigma_{\rm u},\nonumber\\
&&\!\!\!\!G_{88}^\mp=G\mp N_c{K\over6}(\sigma_{\rm s}-2\sigma_{\rm u}-2\sigma_{\rm d}),~G_{08}^\mp=\mp N_c{\sqrt{2}K\over12}(2\sigma_{\rm s}-\sigma_{\rm u}-\sigma_{\rm d}),~G_{38}^\mp=-\sqrt{2}G_{03}^\mp=\mp N_c{\sqrt{3}K\over6}(\sigma_{\rm u}-\sigma_{\rm d});\nonumber\\
&&\!\!\!\!G_{00}^5=N_c{K\over3}\sum_{i=1}^3\pi_i^5,~G_{11}^5=G_{22}^5=G_{33}^5=-N_c{K\over2}\pi_{\rm s}^5,~G_{44}^5=G_{55}^5=-N_c{K\over2}\pi_{\rm d}^5,~G_{66}^5=G_{77}^5=-N_c{K\over2}\pi_{\rm u}^5,\nonumber\\
&&\!\!\!\!G_{88}^5=N_c{K\over6}(\pi^5_{\rm s}-2\pi^5_{\rm u}-2\pi^5_{\rm d}),~G_{08}^5=N_c{\sqrt{2}K\over12}(2\pi^5_{\rm s}-\pi^5_{\rm u}-\pi^5_{\rm d}),~G_{38}^5=-\sqrt{2}G_{03}^5=N_c{\sqrt{3}K\over6}(\pi^5_{\rm u}-\pi^5_{\rm d}).
\end{eqnarray}
\end{widetext}
As the scalar-pseudoscalar mixing matrix $G^5$ is non-zero in the presence of pseudoscalar condensates, the isospin-parity eigenstates, scalars $\sigma,a_0,a_8$ and pseudoscalars $\pi^0,\eta_0,\eta_8$, will further mix with each other in the mass eigenstates. 

Armed with the reduced Lagrangian density Eq.\eqref{LNJL4}, all the subsequential analytic derivations can just parallel those given in Sec.\ref{subsec:2PEM} for the two-flavor case. By contracting a quark-antiquark pair further in the $\pi_{\rm f}^5$ and $\sigma_{\rm f}$ channels, the effective mass and pion condensate are respectively
\begin{eqnarray}
m_i&=&m_{0i}-4N_cG\sigma_i+2N_c^2K(\sigma_j\sigma_k\!-\!\pi^5_j\pi^5_k),\nonumber\\
\pi_i&=&-4N_cG\pi_i^5-2N_c^2K(\sigma_j\pi^5_k\!+\!\pi^5_j\sigma_k)\label{mpi}
\end{eqnarray}
 for each quark flavor with $i\neq j\neq k$. Then, six gap equations can be found by following the definitions of condensates in Eq.\eqref{pi5s}~\cite{Klevansky:1992qe}, that is,
\begin{eqnarray}
\sigma_i&=&\langle\bar{\psi}^i{\psi}^i\rangle/N_c=-{\rm tr} G_i,\label{sigmai}\\
\pi_i^5&=&\langle\bar{\psi}^ii\gamma^5{\psi}^i\rangle/N_c=-{\rm tr}~ i\gamma^5G_i\label{pii5}
\end{eqnarray}
where ${G}_i(x,x')=-(i\slashed{D}_i-m_i-i\,\pi_i\gamma_5)^{-1}$ is the quark propagator in PEM field and the trace should be taken over the spinor and coordinate spaces. 

As mentioned in Sec.\ref{subsec:2PEM}, in in-in scheme, the effective quark propagator can be revised from Eq.\eqref{GEM} as
\begin{widetext}
\begin{eqnarray}
\hat{G}_{\rm f}({p})
\!\!\!&=&\!\!\!i\int_{\Gamma_{{\rm f}\epsilon}}\!\!\di s\exp\left\{-i\,m_{\rm f}^{*2}s-{i\tanh(q_{\rm f}Es)\over q_{\rm f}E}({p}_4^2\!+\!p_3^2)-{i\tan(q_{\rm f}Bs)\over q_{\rm f}B}(p_1^2\!+\!p_2^2)\right\}\left[m_{\rm f}\!-\!i\gamma^5\pi_{\rm f}\!-\!\gamma^4(p_4\!-\!{i\tanh(q_{\rm f}Es)}p_3)\right.\nonumber\\
&&\left.-\gamma^3(p_3+{i\tanh(q_{\rm f}Es)}p_4)-\gamma^2(p_2+{\tan(q_{\rm f}Bs)}p_1)-\gamma^1(p_1-{\tan(q_{\rm f}Bs)}p_2)\right]\nonumber\\
&&\left[1+{i\gamma^5\tan(q_{\rm f}Bs)\tanh(q_{\rm f}Es)}
+{\gamma^1\gamma^2\tan(q_{\rm f}Bs)}+{i\gamma^4\gamma^3\tanh(q_{\rm f}Es)}\right],
\end{eqnarray}
\end{widetext}
where the chiral mass $m_{\rm f}^*=(m_{\rm f}^2+\pi_{\rm f}^2)^{1/2}$. Inserting the propagator into Eqs.\eqref{sigmai} and \eqref{pii5} and taking the integral variable transformations: $s\rightarrow-is$, the explicit forms of the gap equations are similar to Eqs.\eqref{gapm} and \eqref{gappi0}:
\begin{eqnarray}
\!\!\!\!\!\!\!\sigma_{\rm f}\!\!\!&=&\!\!\!-m_{\rm f}K_{\rm f}(m_{\rm f}^{*},E,B)\!+\!{\pi_{\rm f}q_{\rm f}^2I_2\over 4\pi^2m_{\rm f}^{*2}}\!\left(1\!-\!e^{-m_{\rm f}^{*2}S_{\rm fE}}\right)\!,\label{gapm3}\\
\!\!\!\!\!\!\!\pi^5_{\rm f}\!\!\!&=&\!\!\!-\pi_{\rm f}K_{\rm f}(m_{\rm f}^{*},E,B)\!-\!{m_{\rm f}q_{\rm f}^2I_2\over 4\pi^2m_{\rm f}^{*2}}\!\left(1\!-\!e^{-m_{\rm f}^{*2}S_{\rm fE}}\right)\!.\label{gappi3}
\end{eqnarray}
Here, the auxiliary function $K_{\rm f}$ has been defined in Eq.\eqref{Kfunc} and should be regularized to $K_{\rm f}^r$ as shown in Eq.\eqref{Kfuncr}. Since the mass gap ${m_{\rm f}}$ is positive before the end of chiral rotation,  Eqs.\eqref{mpi} and \eqref{gappi3} indicate that ${\pi_{\rm f}}$ always prefers the same sign as $I_2$ regardless of the quark charge when QED anomaly dominates over the QCD one. Then, getting rid of the regular terms proportional to $K_{\rm f}^r$ in Eqs.\eqref{gapm3} and \eqref{gappi3}, the anomaly terms give the model-independent results as
 \begin{eqnarray}
\sigma_{\rm f}{\pi_{\rm f}}-\pi^5_{\rm f}{m_{\rm f}}={q_{\rm f}^2I_2\over4\pi^{2}}\left(1-e^{-m_{\rm f}^{*2}S_{\rm fE}}\right),\label{Min}
\end{eqnarray}
which would consistently give Eq.\eqref{pi0II} by setting $K=0$.

Finally, the anomaly irrelevant part of the thermodynamic potential can be derived consistently as
\begin{widetext}
\begin{eqnarray}
V_{\rm EM}^r\!\!\!&=&\!\!\!2N_cG\!\!\sum_{{\rm f}=u,d,s}\!\!(\sigma_{\rm f}^2+(\pi_{\rm f}^5)^2)\!-\!2N_c^2K\left(2\sigma_{\rm u}\sigma_{\rm d}\sigma_{\rm s}\!-\!\epsilon_{ijk}^2\sigma_i\pi_j^5\pi_k^5\right)\!+\!N_c\!\!\sum_{{\rm f}=u,d,s}\!\!\Re\left\{\int_{0}^{\infty}\!\!{\di s\over 8\pi^2}\;{e^{-m_{\rm f}^{*2}s}\over s}\left[{q_{\rm f}B\over\tanh(q_{\rm f}Bs)}{q_{\rm f}E\over\tan(q_{\rm f}Es)}\right.\right.\nonumber\\
&&\!\!\!\!\!\!\!\!\!\!\!\!\left.\left.-{1\over s^2}-{s\,e^{-m_{\rm f}^{*2}S_{{\rm f}E}}\over s\!+\!S_{{\rm f}E}}\left({q_{\rm f}B\over\tanh(q B(s\!+\!S_{{\rm f}E})}{q_{\rm f}E\over\tan(q_{\rm f}Es)}-{1\over s(s\!+\!\tilde{S}_{{\rm f}E}(1) )}\right)\right]\!-\!G_{\rm f\Lambda}^{0^+}(m_{\rm f}^{*})\!+\!G_{\rm f\Lambda}^{\tilde{S}_{{\rm f}E}(1)}(m_{\rm f}^{*})\!-\!(m_{\rm f}^{*}\rightarrow0)\right\}.
\end{eqnarray}
\end{widetext}
 by combining the integrations over ${m_{\rm f}}$ of Eq.(\ref{gapm3}) and over ${\pi_{\rm f}}$ of Eq.(\ref{gappi3}) except the last anomaly terms. 

\begin{figure}[!htb]
	\centering
	\includegraphics[width=0.41\textwidth]{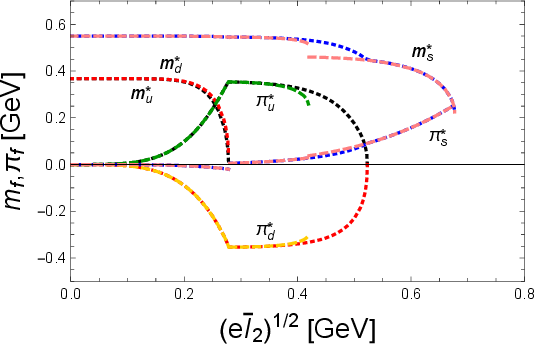}
	\caption{The illustrations of $m_{\rm f}^*$ and $\pi_{\rm f}^*$ as functions of EM field $e\bar{I}_2$ in the in-out (dotted lines) and in-in (dashed lines) schemes, respectively. The conventions roughly follow those of Fig.\ref{mpi0_2f} but the vanishing curves are neglected for simplicity here. }\label{mpi0_3f}
\end{figure}

In Fig.\ref{mpi0_3f}, we show the numerical results for $m_{\rm f}^*$ and $\pi_{\rm f}^*$ in both the in-out and in-in schemes. 
As we have predicted in Ref.~\cite{Cao:2020pjq}, the results are almost the same up to the point well above the end of first chiral rotation $e\bar{I}_2^{c_1}=(0.278~{\rm GeV})^2$, so the crossing feature of the meson masses in Ref.~\cite{Cao:2020pjq} is indeed reliable. Similar to the two-flavor case, the feedback of Schwinger pair production shifts the chiral restoration to an earlier first-order transition for the $u$ and $d$ quarks. Except that, there is not much difference between the results of in-out and in-in schemes in the three-flavor NJL model. As mentioned before, the signs of pion condensates $\pi_{\rm f}^*$ should be the same if QED anomaly dominates. But we find the sign of $\pi_{\rm d}^*$ and $\pi_{\rm s}^*$ to be opposite to that of $\pi_{\rm u}^*$ for small EM field here,  which actually indicates the dominance of the QCD anomaly. Then beyond $e\bar{I}_2^{c_1}$, $\pi_{\rm s}^*$ shifts to the sign implying the dominance of the QED anomaly for $s$ quark.

\subsection{Domain walls at finite baryon density}\label{subsec:DoW}
We have illustrated nonuniform $\pi^0$ condensate in the form of chiral density wave within NJL model in Sec.\ref{subsec:inhom}, here we briefly present an earlier study of $\pi^0$ domain wall within ChPT~\cite{Son:2007ny} . The $\sigma$ and $\pi^0$ domain walls take the following forms 
\bea
\!\!\!\!s=f_\pi\cos\theta(z),t_3=f_\pi\sin\theta(z),\ \tan{\theta(z)\over4}=e^{m_\pi z},\label{DW}
\eea
 which are similar to those for CDW phase except for fixed magnitude $f_\pi$ and nonlinear form of $\theta(z)$. As the $\pi^0$ domain wall was also found to be originated from the WZW anomaly term, it is properer to mention such phase right after the studies of PEM effect. In our opinion, this ansatz is pretty natural by comprehending the temporal vector potential for the electric field as the coordinate dependent chemical potential, see the beginning of Sec.\ref{sec:pseudo}. 

With $\theta(z)$ not a constant, Eq.\eqref{chirall0} reduces to
\begin{eqnarray}
{\cal L}_0(\theta)&=&{f^2_\pi\over2}\left[ -(\partial_z\theta)^2+2m_\pi^2 (\cos\theta-1)\right],
\end{eqnarray}
which then gives the EOM as
\bea
\partial_z^2\theta-m_\pi^2\sin\theta=0
\eea
by following Euler-Lagrangian equation. We can easily check that the $\theta(z)$ given in Eq.\eqref{DW} is indeed a solution of the EOM and the local pressure becomes explicitly
${\cal L}_0=-4{f^2_\pi}m_\pi^2$ for $|m_\pi z|\sim 0$. After integrating ${\cal L}_0(\theta(z))$ over $z$, the energy per unit area transverse to ${\bf B}$ was found to be
$E_0/S_\bot=8{f^2_\pi}m_\pi$~\cite{Son:2007ny}, larger than that in the homogeneous phase with $\theta=0$. 
By substituting the solution back to Eq.\eqref{chirall0}, the EOMs for the fluctuations $\pi^\pm$ become 
\bea
-{\bf \nabla}^2\pi^\pm+m_\pi^2\left[1-{6\over\cosh^2(m_\pi z)}\right]\pi^\pm={E^\pm}^2\pi^\pm.
\eea
There are two bound state solutions: zero-mode and tachyonic with energy
$${E^\pm}^2=(2n+1)eB-3\,m_\pi^2.$$
Thus, the $\pi^0$ domain wall is unstable for small magnetic field but would become metastable beyond the critical one $|eB_{\rm ms}|=3\,m_\pi^2$.

The story will change again when we take the WZW term Eq.\eqref{LWZW} into account. With the presence of baryon chemical potential, magnetic field and $\pi^0$ domain wall, the term can be rewritten as
\bea
{\cal L}_{\rm WZW}(\theta)={\mu_{\rm B}eB\over4\pi^2}\partial_z\theta(z)\label{LWZWz}
\eea
by compensating a total derivative. Then, by integrating Eq.\eqref{LWZWz} over $z$, the corresponding energy per unit area is $E_{\rm WZW}/S_\bot=-{\mu_{\rm B}eB/(2\pi)}$. Thus, the total energy per unit area can be obtained as
\bea
E/S_\bot\equiv(E_0-E_{\rm WZW})/S_\bot=8{f^2_\pi}m_\pi-{\mu_{\rm B}eB\over2\pi},
\eea
and the $\pi^0$ domain wall would  emerge to be globally favored ground state when the magnetic field is larger than the critical one: $|eB_{\rm c}|\equiv16\pi{f^2_\pi}m_\pi/\mu_{\rm B}$. For nuclear matter with $\mu_{\rm B}\lesssim m_N$, the critical value can be evaluated as $|eB_{\rm c}|\approx(0.25\,{\rm GeV})^2$, which is quite close to $eB_{\rm ms}$ and might be relevant to the cores of neutron stars. 

With the increasing of $\mu_{\rm B}$, the masses of $\eta$ and $\eta'$ would decrease quickly and their domain walls are also possible in color superconductor and color-flavor-locking phases, respectively. The critical values are estimated to be $eB_{\rm c}\approx0.36\,m_\pi^2$ for $\eta$ at $\mu_{\rm B}\sim 1\,{\rm GeV}$ and $eB_{\rm c}\approx0.025\,m_\pi^2$ for $\eta'$ at $\mu_{\rm B}\sim 1.5\, {\rm GeV}$~\cite{Son:2007ny}. Actually, $eB_{\rm c}$ can be reduced to that comparable to the observed magnitude of magnetars when the kaon condensation is taken into account in the color-flavor-locking phase.

\subsection{Section summary}
This section was devoted to exploring pseudoscalar, mainly $\pi^0$, condensates in parallel EM field or in magnetic field at finite density. In the former situation, constant pseudoscalar condensates were found in both two and three flavor cases (Fig.\ref{mpi0_2f} and Fig.\ref{mpi0_3f}) due to chiral anomaly, and model independent result was obtained  for $\pi^0$ condensate when the EM field was small. For larger EM field, the feedback of Schwinger mechanism becomes important and could be consistently taken into account by employing in-in formalism. The calculations showed that the $\chi$SR transition for $u$ and $d$ quarks occurred at smaller $I_2$ and altered from second order to first order (Fig.\ref{mpi0_2f} and Fig.\ref{mpi0_3f}). Incidentally, it is interesting to find the competition between magnetic catalysis effect and chiral rotation effect when one increased magnetic field at fixed electric fields (Fig.\ref{mpi0}). At finite density, inhomogeneous pseudoscalar condensates were expected to exist in the inner cores of neutron stars.

\section{Charged pion superfluidity}\label{sec:cpi}
The possibility of charged pion ($\pi^\pm$) superfluidity was first clearly and rigorously demonstrated in chiral perturbation theory by introducing isospin chemical potential $\mu_{\rm I}$ and the threshold was found to be $\mu_{\rm I}^c=m_\pi$~\cite{Son:2000xc}. Later, the studies with lattice QCD simulations~\cite{Kogut:2002zg,Brandt:2017oyy} and NJL model~\cite{He:2005nk} also well reproduced the results by adopting more fundamental degrees of freedom, quarks. In the meanwhile, the $\pi^\pm$ superfluidity was found to turn from the dilute pion Bose-Einstein condensation (BEC) phase to the Bardeen-Cooper superfluidity (BCS) phase with the increasing of $\mu_{\rm I}$ in NJL model~\cite{Sun:2007fc}. Once confinement is taken into account for real QCD, the BCS system can emerge as the intriguing quarksonic matter at moderate $\mu_{\rm I}$~\cite{Cao:2016ats}. Usually, NJL model is more convenient to explore other possibilities and such studies interestingly showed that the inhomogeneous Sarma phase would show up at low baryon density and the Larkin-Ovchinnikov-Fudde-Ferrell phase at high baryon density~\cite{He:2006tn}.

\subsection{Magnetic inhibition}\label{subsec:MI}
As mentioned in the introduction: ever since the discovery of inverse magnetic catalysis effect, the effect of magnetic field had been checked over other kinds of condensations besides the traditional $\sigma$ condensation. Charged pion condensation is one of them, here we'd like to show the magnetic inhibition effect by taking Ginzburg-Landau (GL) approximation within the two-flavor NJL model, see the Lagrangian density given in Eq.\eqref{LNJL2}. Assuming 
$$\langle\pi^+(x)\rangle=e^{ie\Phi(x,0)}\Delta,\ \langle\pi^-(x)\rangle=e^{-ie\Phi(x,0)}\Delta^*$$
 with $e^{\pm ie\Phi(x,y)}$ the Schwinger phases, the thermodynamic potential can be expressed in Minkowski space as
\begin{eqnarray}
V_{\rm B\mu_{\rm I}}\!(m,\Delta)\!\!=\!\!{(m\!-\!m_0)^2\!+\!|\Delta|^2\over 4G}\!+\!{i\over V_4}\text{Tr}\ln\!\!\left(\!\!\!\begin{array}{cc}
\ i\hat{G}_{\rm u}^{-1}\!&\!\!\!-i\gamma_5\Delta\\
-i\gamma_5\Delta^*\!&\!\!i\hat{G}_{\rm d}^{-1}
\end{array}\!\!\!\right)\!\!,\nonumber
\end{eqnarray}
where $V_4$ is the space-time volume and $\hat{G}_{\rm u/d}$ are the effective propagators getting rid of Schwinger phases from the full ones 
$$G_{\rm u/d}=i\left(i\slashed D_{\rm u/d}-m\pm {\mu_{\rm I}\over2}\gamma_0\right)^{-1}.$$ Note that the Schwinger phases are completely cancelled out in $V_{\mu_{\rm I}}$ through the transformations of the quark fields:
$$\psi_{\rm f}(y)\rightarrow e^{iq_{\rm f}\Phi(y,0)}\psi_{\rm f}(y),\bar\psi_{\rm f}(x)\rightarrow \bar\psi_{\rm f}(x)e^{-iq_{\rm f}\Phi(x,0)}.$$
So, the importance of introducing Schwinger phases to the charged condensates $\langle\pi^\pm(x)\rangle$ is that they guaranteer the gauge independence of the thermodynamic potential in external magnetic field~\cite{Cao:2015xja}. Recently, this manipulation had been justified by directly working in Euclidean space-time, where the pole mass of charged pion was extracted from the exponential law $\sim e^{-m_\pi\tau}$ of $\pi^\pm$ correlation function with the imaginary time $\tau$~\cite{Li:2020hlp}.

It can be checked that $V_{\mu_{\rm I}}$ depends on $\Delta$ and $\Delta^*$ only through a single variable $|\Delta|^2$, so we can set $\Delta$ to be real without loss of generality. And by following the consistent conclusion for a system with only finite $\mu_{\rm I}$~\cite{Son:2000xc,Kogut:2002zg,Brandt:2017oyy,He:2005nk}, we suppose the pion superfluidity transition to be still of second order in a magnetic field. Then, the Taylor expansion of $V_{\mu_{\rm I}}$ in terms of $\Delta$ around the critical point gives
\begin{eqnarray}
V_{\rm B\mu_{\rm I}} (m,\Delta)= V_{\rm B\mu_{\rm I}}(m)+{\cal A}\Delta^2+{{\cal B}\over 2}\Delta^4+o(\Delta^6)
\end{eqnarray}
up to $\Delta^4$. 
Here, $V_{\rm B\mu_{\rm I}}(m)$ is just the thermodynamic potential without pion condensation, refer to Eq.\eqref{Vm} for its formal expression, and the coefficients ${\cal A}$ and ${\cal B}$ in the GL approximation are respectively
\begin{eqnarray}
{\cal A}&=&{1\over 4G}+{i\over V_4}\text {Tr}\left[\hat{G}_{\rm u}(x-y)i\gamma_5\hat{G}_{\rm d}(y-x)i\gamma_5\right],\nonumber\\
{\cal B}&=&-{i\over V_4}\text {Tr}\left[\hat{G}_{\rm u}(x-y_1)i\gamma_5\hat{G}_{\rm d}(y_1-y_2)i\gamma_5 \hat{G}_{\rm u}(y_2-y_3)\right.\nonumber\\
&&\qquad\qquad\left. i\gamma_5\hat{G}_{\rm d}(y_3-x)i\gamma_5\right].
\label{ab}
\end{eqnarray}
To derive the explicit form of $V_{\mu_{\rm I}}$, the key is to obtain the explicit expressions of the quark propagators with finite $\mu_{\rm I}$ and $B$. The Schwinger formalism is more convenient for the evaluations of the expansion coefficients and the quark propagators can immediately be given as~\cite{Cao:2015xja}
\begin{eqnarray}
G_{\rm f}(x,y) &=&e^{i\,q_{\rm f}\int_y^x\bar A_{\rm f}^\mu \di x_\mu}\bar{G}_{\rm f}(x-y)\nonumber\\
&\equiv&e^{i\,q_{\rm f}\int_y^x A_{\rm f}^\mu \di x_\mu}\hat{G}_{\rm f}(x-y)\label{GmuI}
\end{eqnarray}
by defining effective vector potentials $\bar A_{\rm u/d}^\mu=(-\mu_{\rm I}/(2|q_{\rm u/d}|),0,0,0)+A^\mu$. Here, $\bar{G}_{\rm f}(x-y)
$ is the same as the effective propagator of Eq.\eqref{GEM} in vanishing $E$ limit.

In order to correctly take into account the effect of isospin chemical potential, we have to work in Euclidean energy-momentum space with the effective propagators modified from Eq.\eqref{prop_m} through the shiftings of energy~\cite{Cao:2015xja}:
 $$p_0\rightarrow ip_4\rightarrow i(p_4-{\cal S}(q_{\rm f})i\,\mu_{\rm I}/2).$$
 Substituting the explicit forms of $\hat{G}_{\rm f}(p)$ into $V_{\rm B\mu_{\rm I}}$, we find
\begin{eqnarray}
V_{\rm B\mu_{\rm I}}(m)\!\!\!&=&\!\!\!{(m-m_0)^2\over 4G}+{N_c\over 8\pi^2}\sum_{\rm f=u,d}\int_0^\infty{\di s\over s^3}\, e^{-sm^2}\times\nonumber\\
\!\!\!&&\!\!\!\!\!{q_{\rm f}Bs\over \tanh(q_{\rm f}Bs)}\vartheta_3\!\left(\!{\pi\over 2}\!-\!i{{\cal S}(q_{\rm f})\mu_{\rm I}\over 4T},e^{-{1\over 4sT^2}}\!\right),
\end{eqnarray}
where $\vartheta_3(z,q)$ is the third Jacobi theta function obtained by working out the summation over the Matsubara frequency, and
\begin{widetext}
\begin{eqnarray}
{\cal A}\!\!\!&=&\!\!\!{1\over 4G}\!-\!4N_cT\!\sum_n\!\int\! {\di^3{\bf k}\over(2\pi)^3}\!\int\! \di s\,\di t\, {\cal K}(s,t,\omega_n,{\bf k})\left\{\!\left(\!m^2\!+\!\omega_n^2\!+\!\left({\mu_{\rm I}\over 2}\right)^2\!+\!k_3^2\!\right)\left(1\!+\!f_{\rm u}(s)f_{\rm d}(t)\right)\!+\!{\bf k}_\bot^2\left(1\!-\!f_{\rm u}^2(s)\right)\left(1\!-\!f_{\rm d}^2(t)\right)\!\right\}\nonumber\\
{\cal B}\!\!\!&=&\!\!\!4N_cT\sum_n\int{d^3{\bf k}\over(2\pi)^3}\int \di s\, \di t\, \di s'\, \di\, t'\, {\cal K}(s,t,\omega_n,{\bf k}){\cal R}(s',t',\omega_n,{\bf k})\Bigg\{{1\over 2}\left[\left(\omega_n^{\rm u}\omega_n^{\rm d}+k_3^2+m^2\right)^2+\left(m\mu_{\rm I}\right)^2+\left(k_3\mu_{\rm I}\right)^2\right]\times\nonumber\\
&&\sum_{l=\pm}
  \left(1+lf_{\rm u}(s)\right)\left(1+lf_{\rm u}(s')\right)\left(1+lf_{\rm d}(t)\right)\left(1+lf_{\rm d}(t')\right)+{\bf k}_\bot^4\left(1-f_{\rm u}^2(s)\right)\left(1-f_{\rm u}^2(s')\right)\left(1-f_{\rm d}^2(t)\right)\left(1-f_{\rm d}^2(t')\right)\nonumber\\
&&
+{\bf k}_\bot^2\Big[4\left(\omega_n^{\rm u}\omega_n^{\rm d}+k_3^2+m^2\right)\left(1+f_{\rm u}(s)f_{\rm d}(t)\right)\left(1-f_{\rm u}^2(s')\right)\left(1-f_{\rm d}^2(t')\right)-\left(\left(\omega_n^{\rm u}\right)^2+k_3^2+m^2\right)\left(1-f_{\rm u}(s)f_{\rm u}(s')\right)\times\nonumber\\
&&\left(1-f_{\rm d}^2(t)\right)\left(1-f_{\rm d}^2(t')\right)
-\left(\left(\omega_n^{\rm d}\right)^2+k_3^2+m^2\right)\left(1-f_{\rm d}(t)f_{\rm d}(t')\right)\left(1-f_{\rm u}^2(s)\right)\left(1-f_{\rm u}^2(s')\right)\Big]\Bigg\}\label{CoeB}
\end{eqnarray}
with $f_{\rm u}(s)=\tanh B_{\rm u}^s$, $f_{\rm d}(t)=\tanh B_{\rm d}^t$, transverse momentum ${\bf k}_\bot^2=k_1^2+k_2^2$ and the effective frequencies $\omega_n^{\rm u}=\omega_n-i\mu_{\rm I}/2$ and $\omega_n^{\rm d}=\omega_n+i\mu_{\rm I}/2$. Here, we have defined an integral kernel as
\begin{equation}
 {\cal K}(s,t,\omega_n,{\bf k})=e^{-s\left[m^2+\left(\omega_n-i\mu_{\rm I}/2\right)^2+k_3^2+{\bf k}_\bot^2f_{\rm u}(s)/B_{\rm u}^s\right]-t\left[m^2+\left(\omega_n+i\mu_{\rm I}/2\right)^2+k_3^2+{\bf k}_\bot^2f_{\rm d}(t)/B_{\rm d}^t\right]},
 \end{equation}
for brevity. 

According to the GL theory, the condition ${\cal A}=0$ determines the critical point of transition to pion superfluidity for a given magnetic field. To carry out numerical calculations, it is enough to regularize $V_{\rm B\mu_{\rm I}}^r(m)$ and ${\cal A}$ by adopting the well-behaved "vacuum regularization" scheme. In this way, the leading term of the thermodynamic potential becomes
\bea
V_{\rm B\mu_{\rm I}}^r(m)={(m-m_0)^2\over 4G}+V_{\rm B}(m)+{N_c\over 8\pi^2}\sum_{\rm f=u,d}\int_0^\infty{\di s\over s^3}\, e^{-sm^2}{q_{\rm f}Bs\over \tanh(q_{\rm f}Bs)}\left[\vartheta_3\!\left(\!{\pi\over 2}\!-\!i{{\cal S}(q_{\rm f})\mu_{\rm I}\over 4T},e^{-{1\over 4sT^2}}\!\right)-1\right]
\eea
with $V_{\rm B\mu_{\rm I}}(m)$ given by Eq.\eqref{VB}, then the gap equation follows from $\partial_mV_{\rm B\mu_{\rm I}}^r(m)=0$ as
\bea 
0&=&{m-m_0\over 2G}-4N_c\int^\Lambda{\di^3p\over(2\pi)^3}{m\over (p^2+m^2)^{1\over2}}-{N_c\over4\pi^2}m\sum_{\rm f=u,d}\int_0^\infty{\di s\over s^2}e^{-m^2s}\left[{q_{\rm f}Bs\over\tanh(q_{\rm f}Bs)}-1\right]\nonumber\\
&&-{N_c\over 4\pi^2}m\sum_{\rm f=u,d}\int_0^\infty{\di s\over s^2}\, e^{-sm^2}{q_{\rm f}Bs\over \tanh(q_{\rm f}Bs)}\left[\vartheta_3\!\left(\!{\pi\over 2}\!-\!i{{\cal S}(q_{\rm f})\mu_{\rm I}\over 4T},e^{-{1\over 4sT^2}}\!\right)-1\right].
\eea
And the regularized quadratic coefficient ${\cal A}^r$ is composed of a vacuum part ${\cal A}_0$ and two magnetic field dependent parts ${\cal A}_{B1}$ and ${\cal A}_{B2}$, that is,
${\cal A}^r={\cal A}_0+{\cal A}_{\rm B}+{\cal A}_{\rm \mu_{\rm I}}$. The explicit expressions of the components are as the following~\cite{Cao:2015xja}:
\bea
{\cal A}_0 \!\!\!&=&\!\!\!{1\over4G}\!-\!{N_c\Lambda^2\over 2\pi^2}\Bigg[\sqrt{1\!+\!\tilde{m}^2}\!-\!\left(\tilde{m}^2\!-\!{1\over 2}\tilde{\mu}_I^2\right)\ln\left({1\over\tilde{m}}
\!+\!\sqrt{1\!+\!{1\over\tilde{m}^2}}\right)\!-\!{\tilde{\mu}_I}\sqrt{\tilde{m}^2\!-\!\left({\tilde{\mu}_I\over2}\right)^2}\arctan{\tilde{\mu}_I/2\over\sqrt{\left(1\!+\!\tilde{m}^2\right)\left(\tilde{m}^2\!-\!\left({\tilde{\mu}_I\over2}\right)^2\right)}}\nonumber\\
&&-\sum_{t=\pm}\int_0^\infty k^2 dk{1\over E_t(k)}{2\over 1+e^{E_t(k)/T}}\Bigg],\nonumber\\
{\cal A}_{\rm B}\!\!\!&=&\!\!\!-{N_cT\over 4\pi^{3/2}}\sum_n\int_0^\infty {ds\over\sqrt s}\int_{-1}^1 dv e^{-s\left(m^2+\left(\omega_n-iv{\mu_{\rm I}\over 2}\right)^2-\left({\mu_{\rm I}\over 2}\right)^2(1-v^2)\right)}\Bigg[
{1\over s}\left({1\over \left(f_{\rm u}\left({1+v\over2}s\right)/B_{\rm u}^s+f_{\rm d}\left({1-v\over 2}s\right)/B_{\rm d}^s\right)^2}-1\right)\nonumber\\
&&+\left({m^2+\omega_n^2+\left({\mu_{\rm I}\over 2}\right)^2+{1\over 2s}\over f_{\rm u}\left({1+v\over2}s\right)/B_{\rm u}^s+f_{\rm d}\left({1-v\over 2}s\right)/B_{\rm d}^s}-1\right)\Bigg],\nonumber\\
{\cal A}_{\rm \mu_{\rm I}}\!\!\!&=&\!\!\!-{N_cT\over 4\pi^{3/2}}\sum_n\int_0^\infty {ds\over\sqrt s}\int_{-1}^1 dv e^{-s\left(m^2+\left(\omega_n-iv{\mu_{\rm I}\over 2}\right)^2-\left({\mu_{\rm I}\over 2}\right)^2(1-v^2)\right)}\Bigg[
{1\over s}{f_{\rm u}^2\left({1+v\over 2}s\right)f_{\rm d}^2\left({1-v\over 2}s\right)-f_{\rm u}^2\left({1+v\over 2}s\right)-f_{\rm d}^2\left({1-v\over 2}s\right)\over \left(f_{\rm u}\left({1+v\over 2}s\right)/B_{\rm u}^s+f_{\rm d}\left({1-v\over 2}s\right)/B_{\rm d}^s\right)^2}\nonumber\\
&&+{\left(m^2+\omega_n^2+\left({\mu_{\rm I}\over 2}\right)^2+{1\over 2s}\right)f_{\rm u}\left({1+v\over 2}s\right)f_{\rm d}\left({1-v\over 2}s\right)\over f_{\rm u}\left({1+v\over2}s\right)/B_{\rm u}^s+f_{\rm d}\left({1-v\over 2}s\right)/B_{\rm d}^t}\Bigg]\label{ABmu}
\eea
with $\tilde{m}={m/\Lambda}, \tilde{\mu}_I={\mu_{\rm I}/ \Lambda}$ and $E_t(k)=\sqrt{k^2+m^2}+t\,\mu_{\rm I}/2$. We remark that there is an implicit condition $\mu_{\rm I}/2<m$ in these equations, which guarantees the convergence of the integrations over the ultraviolet region of $s$. As the critical isospin chemical potential $\mu_{\rm I}^c=m_\pi$ is much smaller than the dynamical quark mass $m$ at $B=0$~\cite{He:2005nk}, it is enough to demonstrate the magnetic effect on $\pi^\pm$ superfluidity by confining ourselves to the case $\mu_{\rm I}/2<m$ here. Of course, the study can be extended to the case $m<\mu_{\rm I}/2$ if we adopt presentations of the quark propagators with the real proper time or alternative with Ritus formalism.

Actually, the derivation of the quartic coefficient ${\cal B}$ is very tedious and its expression shown in Eq.\eqref{CoeB} is quite complicated. On one hand, if the transition to $\pi^\pm$ superfluidity is continuous, ${\cal B}$ is not relevant to the determination of critical point; on the other hand, the sign of ${\cal B}$ would affect the transition order thus the validity of the GL approximation. In the following, we will study the sign of  ${\cal B}$ by just focusing on the integrand as regularization would not shift that. With the help of the exchanging symmetry between $s (t)$ and $s' (t')$ in the integrations, the following inequality can be found
\begin{eqnarray}
&&\!\!\!\!\!4\left(1\!\!+\!\!f_{\rm u}(s)f_{\rm d}(t)\right)\!\left(1\!\!-\!\!f_{\rm u}^2(s')\right)\!\left(1\!\!-\!\!f_{\rm d}^2(t')\right)\!-\!\left(1\!\!-\!\!f_{\rm u}(s)f_{\rm u}(s')\right)\!\left(1\!\!-\!\!f_{\rm d}^2(t)\right)\!\left(1\!\!-\!\!f_{\rm d}^2(t')\right)
\!-\!\left(1\!\!-\!\!f_{\rm d}(t)f_{\rm d}(t')\right)\!\left(1\!\!-\!\!f_{\rm u}^2(s')\right)\!\left(1\!\!-\!\!f_{\rm u}^2(s)\right)\nonumber\\
&\ge& \left(1-f_{\rm u}^2(s')\right)\left(1-f_{\rm d}^2(t')\right)
\left[4\left(1+f_{\rm u}(s)f_{\rm d}(t)\right)-\left(1+f_{\rm u}^2(s)\right)\left(1-f_{\rm d}^2(t)\right)-\left(1+f_{\rm d}^2(t)\right)\left(1-f_{\rm u}^2(s)\right)\right]\nonumber\\
&=&2\left(1+f_{\rm u}(s)f_{\rm d}(t)\right)^2\left(1-f_{\rm u}^2(s')\right)\left(1-f_{\rm d}^2(t')\right)\ge 0.
\end{eqnarray}
Then, the condition $\mu_{\rm I}/2<m$ and the fact $\Im\,{\cal K}/\Re\,{\cal K}\ll 1$ imply the quartic coefficient ${\cal B}$ to be positive definite at the critical point. Thus, even in the presence of finite magnetic field, the transition from normal to uniform pion superfluidity phase is consistent with that of second order within NJL model. However, one should remember that uniform $\pi^\pm$ superfluidity is not allowed in magnetic field, because $\pi^\pm$ superfluidity is also superconductor and Meissner effect forbids the penetration of magnetic field. It is much better to understand the critical point as the unstable point where $\pi^\pm$ superfluidity can probably organize itself as vortical lattices, corresponding to the Type-II superconductor, see Ref.~\cite{Chernodub:2010qx}.
\end{widetext}

\begin{figure}[!htb]
\centering
\includegraphics[width=0.4\textwidth]{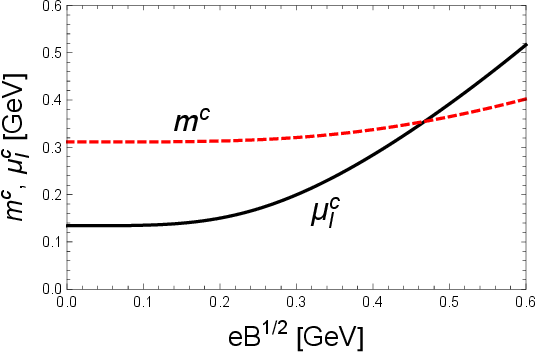}
\caption{The critical isospin chemical potential $\mu_{\rm I}^c$ (black solid) and the corresponding quark dynamical mass $m^c$ (red dashed) as functions of the magnetic field $eB$. }
\label{muI_B}
\end{figure}

In the following, we keep changing the magnetic field and try to find the critical isospin chemical potential $\mu_{\rm I}^c$ according to the condition ${\cal A}=0$, see the results in Fig.\ref{muI_B}. As we can tell, $\mu_{\rm I}^c$ increases with $eB$, thus there is magnetic inhibition effect to $\pi^\pm$ superfluidity and the dHvA oscillation does not demonstrate itself at zero temperature. However, we'd like to mention that there is some window at finite temperature where dHvA oscillation can be found~\cite{Cao:2015xja}.

\subsection{Possibility in parallel magnetic field and rotation}\label{subsec:PMRpi}

\subsubsection{Intuition from Klein-Gordon theory}\label{subsec:KGT}
The possibility of $\pi^\pm$ superfluidity in parallel magnetic field and rotation (${\bf B}\parallel\boldsymbol{\Omega}$) was first proposed within the Klein-Gordon theory for pions~\cite{Liu:2017spl}. The basic logic is that: the orbital angular momenta (OAMs) of $\pi^\pm$ tend to align along opposite directions in external magnetic field, then the rotation-OAM coupling would cause the splitting between $\pi^\pm$ spectra, just like that with $\mu_{\rm I}$. Take the $n$-th Landau level for example, the OAMs of $\pi^\pm$ are constrained by $\pm l\in[-n,{N}-n]\ (n\in\mathbb{N})$ in order that the eigenfunction exists and the correct degeneracy is found. So there is symmetry between the OAM $l$ for $\pi^+$ and the OAM $-l$ for $\pi^-$, and the corresponding energies with rotation effect are~\cite{Liu:2017spl} 
\bea
E_{\pi^\pm}(n,p_3,l)=\sqrt{(2n+1)eB+p_3^2+m_\pi^2}\mp\Omega l
\eea
One can immediately tell the $\mu_{\rm I}$-like role of $\Omega l$, thus it is not surprising that $\pi^\pm$ superfluidity can happen in ${\bf B}\parallel\boldsymbol{\Omega}$.

That is only a very simple demonstration. As mentioned in Sec.\ref{subsec:RMI}, the rotational system can not be very large for the sake of causality and the boundary condition must be applied. So, the issue has to be explored more carefully by adopting the Lagrangian density
\bea
{\cal L}=\left|\left(D_\mu-i\,\Omega\hat{L}_z\delta_{\mu 0}\right){\pi}^+\right|^2-{m_\pi^2\,{\pi}^+{\pi}^-}
\eea
in Minkowski space. The explicit equation of motion for $\pi^+$ follows from Euler-Lagrangian equation as
\bea
\left[\left(D_0-i\,\Omega\hat{L}_z\right)^2-{\bf D}_i^2+m_\pi^2\right]{\pi}^+=0,
\eea
which then can be solved in cylindrical coordinate system (see Sec.\ref{subsec:RMI}) as~\cite{Liu:2017spl,Cao:2020pmm}
$${\pi}^+(x)={\cal N}_{\rm nl}e^{-i(Et-p_3x_3)}{e^{i\, l\theta}}~\tilde{r}^le^{-{\tilde{r}^2\over 2}}{}_1F_1(-\lambda_l^n,|l|+1,\tilde{r}^2).$$
Here, ${\cal N}_{\rm nl}$ is a normalization factor and $_1F_1$ is a hypergeometrical function satisfying the boundary condition: $${}_1F_1(-\lambda_l^n,|l|+1,\tilde{r}^2)|_{r\rightarrow R}=0$$ where $R$ is the radius of the cylinder and the transverse energy follows the convention: $0<\lambda_l^0<\lambda_l^1<\cdots$. Then for positive $l$, the dispersion of $\pi^+$ is given by
$$E_{\pi^+}=\sqrt{(2\lambda_l^n+1)eB+p_3^2+m_\pi^2}-l\Omega,$$
and that of $\pi^-$ follows as
$$E_{\pi^-}=\sqrt{(2\lambda_l^n+1)eB+p_3^2+m_\pi^2}+l\Omega.$$

In the limit $B\rightarrow0$, the transverse energy gap $(2\lambda_l^n+1)eB$ would alter to another sets of eigenvalues always greater than $|l\Omega|$; thus the presence of both $B$ and $\Omega$ is important for the occurrence of $\pi^\pm$ superfluidity. In Ref.~\cite{Liu:2017spl}, two examples are compared:
\bea
&&N=25,\,\ l=20,\, E_{\pi^+}=\sqrt{1.86eB+m_\pi^2}-20\Omega;\nonumber\\
&&N=100,\,l=84,\,E_{\pi^+}=\sqrt{1.36eB+m_\pi^2}-84\Omega;\nonumber
\eea
thus the critical angular velocity $\Omega_c$ given by $E_{\pi^+}=0$ decreases with $eB$ for a given $R$.

\subsubsection{Ambiguity in NJL model}\label{subsec:NJLcpi}
In the Klein-Gordon theory, pions were treated as point particles without any internal structures; but in reality, pions are of finite sizes and composed of elementary quark degrees of freedom. In this sense, nontrivial spin dynamics of quarks might play an important role in spinless charged pion system. Take $\pi^+=\bar{d}i\gamma^5u$ for example~\cite{Cao:2019ctl}: Applying a magnetic field will align both the spins of $u$ and anti-$d$ quarks along its direction due to their positive charges. If we further turn on 
a parallel rotation in the system, the situation would not change. Then, we would expect the total spin of $u$ and anti-$d$ quarks to be $1$ because of the polarization effect, which contradicts with the pseudoscalar nature of $\pi^+$. Hence, according to this spin breaking picture, we might conclude that $\pi^\pm$ superfluidity is not favored in ${\bf B}\parallel\boldsymbol{\Omega}$ at all, similar to that in pure magnetic field~\cite{Cao:2015xja}, see Sec.\ref{subsec:MI}.   

Following a similar procedure as that in Sec.\ref{subsec:MI}, we still use the Ginzburg-Landau approximation to explore the possibility of $\pi^\pm$ superfluidity in ${\bf B}\parallel\boldsymbol{\Omega}$. The leading order contribution has already been well discussed in Sec.\ref{subsec:RMI} with the regularized thermodynamic potential and gap equation given in Eqs.\eqref{Omeg_BOmg1} and \eqref{Gm_Omg}, respectively. 
In this section, we mainly focus on deriving the quadratic coefficient $${\cal A}={1\over4G}+{\cal A}_{\rm FL}$$ for $\pi^+$ with the help of the well constructed Feynman Green's functions for quarks Eqs.\eqref{propagator+} and \eqref{propagator-}. Here, the complicated polarization function ${\cal A}_{FL}$ can be evaluated through the $u-d$ quark mixing loop:
\begin{eqnarray}
	{\cal A}_{\rm FL}={i\over V_4}{\rm Tr}\left[S_{\rm F}^{\rm u}(x,y)i\gamma^5S_{\rm F}^{\rm d}(y,x)i\gamma^5e^{-ie\Phi}\right],
\end{eqnarray}
where $e^{-ie\Phi}$ defines the compensatory Schwinger phases of $\langle\pi^\pm\rangle$ in ${\bf B}\parallel\boldsymbol{\Omega}$ and the trace should be taken over all internal spaces and the coordinate space. Carrying out the traces in the internal spaces, we get
\begin{widetext}
	\begin{eqnarray}
	{\cal A}_{FL}&=&{-2N_ci\over S}\sum_{\rm n=0}^\infty\sum_l\sum_{\rm n'=0}^\infty\sum_{l'}\sum_{r,r'}\sum_{\theta,\theta'}\int_{-\infty}^{\infty}{dp_0\over2\pi}\int_{-\infty}^{\infty}{dp_3\over2\pi}
	{e^{-i\Phi}\over\left[\left({p}_0^{l+}\right)^2-E_{\rm un}^2\right]\left[\left({p}_0^{l'-}\right)^2-E_{\rm dn'}^2\right]}\nonumber\\
	&&\Bigg\{\left[\chi_{\rm n,l}(\theta,r)\chi_{\rm n,l}^*(\theta',r')\chi_{\rm n'-1,l'-1}^{-}(\theta',r')\chi_{\rm n'-1,l'-1}^{-*}(\theta,r)+\chi_{\rm n-1,l+1}(\theta,r)\chi_{\rm n-1,l+1}^*(\theta',r')\chi_{\rm n',l'}^{-}(\theta',r')\chi_{\rm n',l'}^{-*}(\theta,r)\right]\nonumber\\
	&&\ \ \ \times\left({p}_0^{l+}{p}_0^{l'-}-p_3^2-m^2\right)+2\sqrt{(2nq_{\rm u}B)2n'|q_{\rm d}B|}\chi_{\rm n,l}(\theta,r)\chi_{\rm n-1,l+1}^*(\theta',r')\chi_{\rm n',l'}^{-}(\theta',r')\chi_{\rm n'-1,l'-1}^{-*}(\theta,r)\Bigg\},\label{AFL1}
	\end{eqnarray}
\end{widetext}
where we have used the denotations $$\sum_{r,r'}=\int_0^\infty rdr\int_0^\infty r'dr',\ \sum_{\theta,\theta'}=\int_0^{2\pi}d\theta\int_0^{2\pi}d\theta'$$ for simplicity. 

There are two ansatzes for the form of $\Phi$ in the market: one the same as that in the Minkowski space and the other in curved space with the the involved integral path along geodesic line. The explicit forms are respectively~\cite{Cao:2019ctl,Chen:2019tcp}:
\bea
\Phi_{\rm M}&=&{eB\over2}rr'\sin(\theta-\theta'),\label{SPM}\\
\Phi_{\rm C}&=&{eB\over2}rr'\sin[\theta-\theta'+\Omega(t-t')]\label{SPC}
\eea
with the difference only from the variable of the function $\sin(x)$. The choice of $\Phi_{\rm C}$ is quite sophisticatedly explained by introducing the curved space metric and Synge’s world function, but it is a pity that such phase can only {\it correctly} emerge from the $\pi^\pm$ propagator after we take a nonphysical energy shift: ${p}_0\rightarrow p_0-l\Omega$, refer to Eq.\eqref{Spi}. In our opinion, even though rotation effect is introduced through rotating frame, the Schwinger phase is not necessarily defined in curved space as the magnetic field is defined in Minkowski space. Actually,
 the $\pi^\pm$ propagator can also be constructed from the eigenfunctions as:
 \bea
 S_{\rm \pi^+}(x,x')\!\!\!&=&\!\!\!\sum_{\rm n=0}^\infty\!\sum_l\!\int\!{\di p_0\di p_3\over(2\pi)^2}{i~e^{-ip_0(t-t^\prime)+ip_3(z-z^\prime)}\over\left({p}_0\!+\!l\,\Omega\right)^2\!-\!E_{\pi^\pm}^2(n,p_3,0)}\nonumber\\
&&\chi_{\rm n,l}(\theta,r)\chi_{\rm n,l}^*(\theta',r').\label{Spi}
\eea
Here, $l\,\Omega$ shows up like $\mu_{\rm I}$ in the denominator~\cite{He:2005nk}, and the $1/T$ periodicity is guaranteed for the imaginary time $\tau-\tau'=i(t-t')$ after continuing from the real one $t-t'$ and adopting the boson Matsubara frequency $\omega_n=2\pi n T$. Then, it is natural to compensate the Schwinger phase of $\langle\pi^\pm\rangle$ as that given in Sec.\ref{subsec:MI}, that is, with $\Phi_{\rm M}$; otherwise, both the $\mu_{\rm I}$-like role of $l\,\Omega$ and the $1/T$ periodicity would be lost by shifting ${p}_0$ to $p_0-l\Omega$.
Of course, both Schwinger phases give the same gauge invariant result in the limit $\Omega\rightarrow0$ and $R\rightarrow\infty$.
	
In the following, we work out the more complicated polarization function ${\cal A}_{\rm FL}^{\rm M}$ with the phase $\Phi_{\rm M}$, and the result for $\Phi_{\rm C}$ can be immediately adjusted from ${\cal A}_{\rm FL}^{\rm M}$ by modifying the internal implementation from $p_0=p_0'$ to ${p}_0^{l+}={p'}_0^{l'-}$. The integrals over the polar angles can be completed in Eq.\eqref{AFL1} to give
\begin{widetext}
	\begin{eqnarray}
	{\cal A}_{\rm FL}^{\rm M}&=&{-2N_ci\over S}\sum_{\rm n=0}^\infty\sum_{l=0}\sum_{\rm n'=0}^\infty\sum_{l'=0}\int_{-\infty}^{\infty}{dp_0\over2\pi}\int_{-\infty}^{\infty}{dp_3\over2\pi}
	{{n!n'!\over l!l'!}\left({q_{\rm u}B\over 2}\right)^{l-n+1}\left({|q_{\rm d}B|\over 2}\right)^{l'-n'+1}\over\left[\left({p}_0^{(l-n)+}\right)^2-E_{\rm un}^2\right]\left[\left({p}_0^{(n'-l')-}\right)^2-E_{\rm dn'}^2\right]}\nonumber\\
	&&\Bigg\{H_{\rm nl,n'l'}(q_{\rm u}B,|q_{\rm d}B|)+\left[{|q_{\rm d}B|\over n'}G_{\rm nl,n'l'}(q_{\rm u}B,|q_{\rm d}B|)+{q_{\rm u}B\over n}G_{\rm n'l',nl}(|q_{\rm d}B|,q_{\rm u}B)\right]\nonumber\\
	&&\ \ \times\left[\left({p}_0^{(l-n)+}\right)^2-E_{\rm un}^2+\left({p}_0^{(n'-l')-}\right)^2-E_{\rm dn'}^2-\Omega_{\rm nl,n'l'}^2\right]\Bigg\},
	\end{eqnarray}
	where the auxiliary functions are defined as
	\begin{eqnarray}
	G_{\rm nl,n'l'}(q_{\rm u}B,|q_{\rm d}B|)\!\!\!&\equiv&\!\!\!\int_0^\infty{\di r \di r'}~J_{\rm k}\left({eB\over2}rr'\right)(rr')^{k+1}e^{-eB(r^2+{r'}^2)/4}\prod_{x=r,r'}{F}_{\rm nl,n'l'}(q_{\rm u}B,|q_{\rm d}B|;x),\\
	H_{\rm nl,n'l'}(q_{\rm u}B,|q_{\rm d}B|)\!\!\!&\equiv&\!\!\!2\int_0^\infty\!\!\!{\di r \di r'}J_{\rm k}\left({eB\over2}rr'\right)(rr')^{k+1}e^{-eB(r^2+{r'}^2)/4}\left\{\prod_{x=r,r'}\Big[{|q_{\rm d}B|}{F}_{\rm nl,n'l'}(q_{\rm u}B,|q_{\rm d}B|;x)+\right.\nonumber\\
	&&\!\!\!\!\!\!\!\!\!\!\!\!\!\!\!\!\!\!\!\!\!\!\!\!\!\!\!\!\!\!\!\!\!\!\!\!\!\!\!\!\!\left.{q_{\rm u}B}{F}_{\rm n'l',nl}(|q_{\rm d}B|,q_{\rm u}B;x)\Big]\!+\!nn'{q_{\rm u}B}{|q_{\rm d}B|}\!\!\prod_{x=r,r'}\!\!\left({{F}_{\rm nl,n'l'}(q_{\rm u}B,|q_{\rm d}B|;x)\over n'}\!+\!{{F}_{\rm n'l',nl}(|q_{\rm d}B|,q_{\rm u}B;x)\over n}\right)\right\}
	\end{eqnarray}
	with $k\equiv l+l'-n-n'+1$ and
\bea
F_{\rm nl,n'l'}(q_{\rm u}B,|q_{\rm d}B|;x)\equiv L_{\rm n}^{l-n}\left({q_{\rm u}B ~x^2\over2}\right)L_{\rm n'-1}^{l'-n'+1}\left({|q_{\rm d}B| ~x^2\over2}\right).
\eea
	
Finally, we transfer to imaginary-time formalism and sum over the fermion Mastubara frequency to get
\begin{eqnarray}\label{AFLM}
{\cal A}_{\rm FL}^{\rm M}
&=&-{N_c\over 2S}\sum_{\rm n=0}^\infty\sum_{l=0}\sum_{\rm n'=0}^\infty\sum_{l'=0}\sum_{s=\pm}\int_{-\infty}^{\infty}{dp_3\over(2\pi)}{\tanh\left({E_{\rm un}-s~\Omega(l-n+{1\over2})\over2T}\right){n!n'!\over l!l'!}\left({q_{\rm u}B\over 2}\right)^{l-n+1}\left({|q_{\rm d}B|\over 2}\right)^{l'-n'+1}\over E_{\rm un}\left[\left(E_{\rm un}-s\,k\Omega\right)^2-E_{\rm dn'}^2\right]}\nonumber\\
&&\Bigg\{\Bigg[{|q_{\rm d}B|\over n'}G_{\rm nl,n'l'}(q_{\rm u}B,|q_{\rm d}B|)+{q_{\rm u}B\over n}G_{\rm n'l',nl}(|q_{\rm d}B|,q_{\rm u}B)\Bigg]\left[\left(E_{\rm un}-s\,k\Omega\right)^2-E_{\rm dn'}^2-(k\Omega)^2\right]\nonumber\\
&&+H_{\rm nl,n'l'}(q_{\rm u}B,|q_{\rm d}B|)\Bigg\}+\left(E_{\rm un}\leftrightarrow E_{\rm dn'},nl\leftrightarrow n'l',q_{\rm u}\leftrightarrow |q_{\rm d}|\right).
\end{eqnarray}
 And the explicit form with $\Phi_{\rm C}$ can be modified from Eq.\eqref{AFLM} by taking the substitution $k\rightarrow0$ as	
\begin{eqnarray}\label{AFLC}
{\cal A}_{\rm FL}^{\rm C}
&=&-{N_c\over 2S}\sum_{\rm n=0}^\infty\sum_{l=0}\sum_{\rm n'=0}^\infty\sum_{l'=0}\sum_{s=\pm}\int_{-\infty}^{\infty}{dp_3\over(2\pi)}{1\over E_{\rm un}}{\tanh\left({E_{\rm un}-s~\Omega(l-n+{1\over2})\over2T}\right){n!n'!\over l!l'!}\left({q_{\rm u}B\over 2}\right)^{l-n+1}\left({|q_{\rm d}B|\over 2}\right)^{l'-n'+1}}\nonumber\\
&&\Bigg\{\Bigg[{|q_{\rm d}B|\over n'}G_{\rm nl,n'l'}(q_{\rm u}B,|q_{\rm d}B|)+{q_{\rm u}B\over n}G_{\rm n'l',nl}(|q_{\rm d}B|,q_{\rm u}B)\Bigg]+{1\over\left(E_{\rm un}\right)^2-E_{\rm dn'}^2}H_{\rm nl,n'l'}(q_{\rm u}B,|q_{\rm d}B|)\Bigg\}\nonumber\\
&&+\left(E_{\rm un}\leftrightarrow E_{\rm dn'},nl\leftrightarrow n'l',q_{\rm u}\leftrightarrow |q_{\rm d}|\right).
\end{eqnarray}
So, the angular velocity $\Omega$ would induce $\mu_{\rm I}$-like effect in Eq.\eqref{AFLM} through the $\Omega$-dependent terms in the denominators, while it can only induce $\mu_B$-like effect in Eq.\eqref{AFLC} due to the absence of such terms. The correctness of these forms has been checked by calculating several low Landau level contributions in the limit $\Omega\rightarrow 0$, see Appendix.\ref{sec:check}.

Again, by following the ``vacuum regularization" scheme as in Sec.\ref{subsec:MI}, the quadratic coefficient ${\cal A}$ can be decomposed into three parts: ${\cal A}^r={\cal A}_{0}+{\cal A}_{\rm B}+{\cal A}_\Omega$ with ${\cal A}_{0}$ and ${\cal A}_{\rm B}$ the same as those given in Eq.\eqref{ABmu}. The rotation and temperature effects are reserved in ${\cal A}_\Omega$ which is explicitly
\begin{eqnarray}\label{AOM}
{\cal A}_\Omega^{\rm M}\!\!\!&=&\!\!\!-{N_c\over 2S}\sum_{\rm n=0}^\infty\sum_{l=0}^{N_{\rm u}}\sum_{\rm n'=0}^\infty\sum_{l'=0}^{N_{\rm d}}\!{{n!n'!\over l!l'!}\!\!\left({q_{\rm u}B\over 2}\right)^{\!l\!-\!n\!+\!1}\!\!\!\left({|q_{\rm d}B|\over 2}\right)^{\!l'\!-\!n'\!+\!1}}\!\left\{\Bigg[{|q_{\rm d}B|\over n'}G_{\rm nl,n'l'}(q_{\rm u}B,|q_{\rm d}B|)\!+\!{q_{\rm u}B\over n}G_{\rm n'l',nl}(|q_{\rm d}B|,q_{\rm u}B)\Bigg]\right.\nonumber\\
&&\sum_{s=\pm}\int_{-\infty}^{\infty}{dp_3\over(2\pi)}\left[ {\tanh\left({E_{\rm un}-s\,\Omega(l-n+{1\over2})\over2T}\right)-1\over E_{\rm un}}-{(k\Omega)^2\tanh\left({E_{\rm un}-s\,\Omega(l-n+{1\over2})\over2T}\right)\over E_{\rm un}\left[\left(E_{\rm un}-s\,k\Omega\right)^2-E_{\rm dn'}^2\right]}\right]+H_{\rm nl,n'l'}(q_{\rm u}B,|q_{\rm d}B|)\nonumber\\
&&\sum_{s=\pm}\left.\int_{-\infty}^{\infty}{dp_3\over(2\pi)}{1\over E_{\rm un} }\left[{\tanh\left({E_{\rm un}-s\,\Omega(l-n+{1\over2})\over2T}\right)\over\left(E_{\rm un}-s\,k\Omega\right)^2-E_{\rm dn'}^2}-{1\over\left(E_{\rm un}\right)^2-E_{\rm dn'}^2}\right]\right\}+\left(E_{\rm un}\leftrightarrow E_{\rm dn'},nl\leftrightarrow n'l',q_{\rm u}\leftrightarrow |q_{\rm d}|\right)
\end{eqnarray}
\end{widetext}	
for $\Phi_{\rm M}$, and the one for $\Phi_{\rm C}$ can be obtained from Eq.\eqref{AOM} by taking the limit $k\rightarrow0$.
Eventually, the coefficient ${\cal A}^r$ can be evaluated after solving the value of dynamical mass $m$ from the gap equation Eq.\eqref{Gm_Omg}. To facilitate the numerical calculations, the functions
 $G_{\rm nl,n'l'}(q_{\rm u}B,|q_{\rm d}B|)$ and $H_{\rm nl,n'l'}(q_{\rm u}B,|q_{\rm d}B|)$ were analytically worked out up to the $10$-th Landau levels ($n_{\rm max}=n'_{\rm max}=10$) by using {\it Mathematica}. we have checked that such Landau level truncations give very accurate values of ${\cal A}^r$ for the strong magnetic fields considered here: $eB=0.5\,{\rm GeV}^2$.  

\begin{figure}[!htb]
\centering
\includegraphics[width=0.42\textwidth]{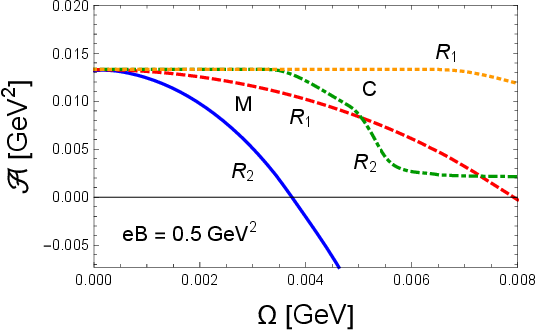}
\caption{With the Schwinger phases in Minkowski (M) and curved (C) spaces, the quadratic coefficients ${\cal A}$ are evaluated as functions of the angular momentum $\Omega$ for given magnetic field $eB=0.5~{\rm GeV}^2$. The system sizes  $R_1$ and $R_2$ are the same as those chosen in Fig.\ref{m_omg}.}\label{A_omg}
\end{figure}

Following the study in Sec.\ref{subsec:RMI}, two system sizes are chosen for comparison and the numerical results are illustrated in Fig.\ref{A_omg} for the Schwinger phases in both the Minkowski and curved spaces. We regret that the numerical calculations are not so correct in our previous work Ref.~\cite{Cao:2019ctl}, but the qualitative conclusion is still true. According to the evaluations with the Schwinger phase in Minkowski space, the quadratic coefficient ${\cal A}$ decreases to negative for both system sizes thus $\pi^\pm$ superfluidity is favored in PEM. However, as pointed out in Ref.~\cite{Chen:2019tcp}, we find that $\pi^\pm$ superfluidity is not possible if we choose the Schwinger phase in curved space -- that is the "ambiguity".  The results of three-flavor NJL model will be presented in Sec.\ref{subsec:PMRrho}.

\section{Charged rho superconductivity}\label{sec:crho}
The initial idea of charged rho superconductor comes from the decreasing of some $\rho^\pm$ meson mass with magnetic field in the point particle picture~\cite{Chernodub:2010qx}, similar to that of $W^\pm$ gauge bosons. For an elementary vector boson without internal structure, the lowest dispersion is the one with spin along the same direction of magnetic field, that is,
\bea
E_{\rm v}^2(n,p_3)=(2n-1)eB+p_3^2+m_{\rm v}^2,\ (n\in\mathbb{N})
\eea
in $3+1$ dimensions. Then, for the lowest Landau level $n=0$ and $p_3=0$, we can easily tell that the energy square vanishes at $eB_{\rm c}=m_{\rm v}^2$ and the system becomes unstable beyond the critical point. In this sense, the critical magnetic field $eB_{\rm c}=m_\rho^2$ for QCD beyond which the ground state becomes {\it vacuum superconductor} (VS)~\cite{Chernodub:2010qx}. Actually, the detailed study within the modified Weinberg model found vortical structures of $\rho^\pm$, thus the VS is of type-II consistent with Meissner effect~\cite{Chernodub:2010qx}. For vector bosons with degeneracy larger than $1$, the kinetic term is usually introduced through non-Abelian strength tensors in the Lagrangian. Similar to $W^\pm$ and $Z$ bosons in quantum electroweak dynamics, the mesons $\rho^\pm$ and $\rho^0$ play the roles of $SU(2)$ gauge bosons in low energy QCD. So, due to the interactions between $\rho^\pm$ and $\rho^0$ in the strength tensors, the $\rho^0$ vortical condensate shows up through the derivative of $|\langle\rho^\pm\rangle|^2$, that is, $\rho^0_\mu={ig_\rho\over-\partial_\bot^2+m_\rho^2}\partial_\mu|\langle\rho^\pm\rangle|^2$~\cite{Chernodub:2010qx}. 

However, the story is really different when the internal quark structures of $\rho^\pm$ are taken into account. It was argued that such VS violates Vafa-Witten theorem and the $\rho^\pm$ meson mass was generally proved to be larger than the $\pi^0$ mass in external magnetic field~\cite{Hidaka:2012mz}. As mentioned in the introduction, the lattice QCD simulations from different groups never found the $\rho^\pm$ meson mass to vanish at large magnetic field thus disfavored VS~\cite{Hidaka:2012mz,Luschevskaya:2016epp,Bali:2017ian}. But with the chemical potential like effect induced by rotation, the Vafa-Witten theorem is not valid anymore, so the $\rho^\pm$ condensation is not forbidden in principle, similar to the $\pi^\pm$ condensation at finite $\mu_I$. In parallel magnetic field and rotation, the lowest dispersion of $\rho^\pm$ is modified to~\cite{Cao:2020pmm}
\bea
E_{\rho^\pm}=\sqrt{(2\lambda_l^n-1)eB+p_3^2+m_\rho^2}\mp (l+1)\Omega,
\eea
hence the critical angular velocity is given by
 \bea
 \Omega_{\rm c}={1\over |l+1|}\sqrt{(2\lambda_l^0-1)eB+m_\rho^2}
 \eea
 for given $eB$ and $l$. On the other hand, as $\pi^\pm$ mass increases with $eB$ but $\rho^\pm$ mass decreases, one would expect that $\rho^\pm$ condensation might favor over $\pi^\pm$ condensation for $$2eB> m_\rho^2-m_\pi^2$$ in the point particle picture. Note that the threshold $eB\gtrsim m_\rho/\sqrt{2}$ is consistent with Vafa-Witten theorem.

\subsection{Two-flavor case in pure magnetic field}\label{subsec:2fM}
In the simple two-flavor NJL model, $\rho^\pm$ mass was indeed found to decrease to zero with the magnetic field thus favors {\it vacuum superconductivity} beyond the critical $B$~\cite{Chernodub:2011mc,Liu:2014uwa}. Here, for the sake of completeness, we'd like to start with showing the results of two-flavor NJL model by comparing the masses of $\pi^0$ and $\rho^\pm$~\cite{Cao:2019res}. In mean field approximation, the thermodynamic potential and gap equation are the same as those given in Sec.\ref{ssubsec:MCE}, and then the dispersions of mesons can be consistently studied in random phase approximation (RPA). In the original NJL model, the vector interaction channels are absent between quarks and antiquarks but must be included in the Lagrangian in order to study the properties of $\rho$ mesons. In low energy nuclear physics, the vector-isoscalar interaction is important to reproduce the saturation properties of nuclear matter, see the Walecka model in Ref.~\cite{Kapusta2006} for example. Similarly, the vector-isovector interactions should play a role in asymmetric nuclear matter, see the review Ref.~\cite{Li:2008gp}. One also notes that the vector interactions can be obtained by taking Fierz transformations of the scalar interactions in the initial version of NJL model~\cite{Klevansky:1989vi}. Then, the extended two-flavor NJL model can be written as~\cite{Cao:2019res}
\begin{eqnarray}\label{njl2}
{\cal L}&=&\bar \psi\left(i\slashed D-m_0\right)\psi+G\left[\left(\bar\psi\psi\right)^2+\left(\bar\psi i\gamma_5\boldsymbol\tau\psi\right)^2\right]\nonumber\\
&&-G_{\rm v}\left[\left(\bar\psi\gamma^\mu\tau^a\psi\right)^2+\left(\bar\psi i\gamma^\mu\gamma_5\tau^a\psi\right)^2\right],
\end{eqnarray}
where the new four-fermion vector coupling constant $G_{\rm v}$ should be determined by the $\rho$ meson mass in vacuum in principle. However, since $m_\rho$ is usually larger than the cutoff $\Lambda$, numerical results could be sensitive to any ad hoc choice of $G_{\rm v}$.

In RPA, the inverse meson propagators are in the following forms~\cite{Klevansky:1992qe}:
\bea
\!\!\!\!\!\!\!\!D^{-1}_{SS}(y,x)\!\!\!&=&\!\!\!-{e^{-iq_s\!\int_{x}^y\!A\cdot\di x}\over2G}\!+\!{i\over V_4}{\rm Tr}\, {\cal G}\Gamma_{S^*} {\cal G}\Gamma_{S},\label{Sprp}\\
\!\!\!\!\!\!\!\!D^{-1}_{\bar{V}_{\mu}\bar{V}_{\nu}}(y,x)\!\!\!&=&\!\!\!{e^{-iq_{\rm v}\!\int_{x}^y\!A\cdot\di x}g_{{\mu}{\nu}}\over2G_{\rm v}}\!+\!{i\over V_4}{\rm Tr}\, {\cal G}\Gamma_{{\bar{V}_{\mu}}^*} {\cal G}\Gamma_{\bar{V}_{\nu}},\label{Vprp}
\eea
where the exponents are Schwinger phases, ${\cal G}={\rm diag}(G_{\rm u},G_{\rm d})$ is the fermion propagator in flavor space, and the coupling vertices $\Gamma_{S/S^*}$ and $\Gamma_{\bar{V}_{\mu}/\bar{V}_{\mu}^*}$ are for the scalar/pseudoscalar and vector/pseudovector channels, respectively. The explicit forms of all the relevant coupling vertices are~\cite{Cao:2019res}
\bea
&&\!\!\!\!\!\Gamma_{\sigma/\sigma^*}=-1,~\Gamma_{\pi^0/{\pi^0}^*}=-i\gamma^5\tau_3,~\Gamma_{\pi_\pm}=-i\gamma^5\tau_\pm,\nonumber\\
&&\!\!\!\!\!\Gamma_{\bar{\omega}_\mu/\bar{\omega}_\mu^*}=\bar{\gamma}_\mu^\pm,~\Gamma_{\bar{\rho}_{0\mu}/\bar{\rho}_{0\mu}^*}=\bar{\gamma}_\mu^\pm\tau_3,~\Gamma_{\bar{\rho}_{\pm\mu}}=\bar{\gamma}_\mu^\pm\tau_\pm,\label{Gammas}
\eea
where we have defined $\bar{\gamma}^{\pm}_\mu=(\gamma_0,{\gamma_1\pm i\gamma_2\over \sqrt{2}},{\gamma_1\mp i\gamma_2\over \sqrt{2}},\gamma_3)$ and the spin eigenstate $\bar{V}_{\mu}/\bar{V}_{\mu}^*=({V}_{0},{{V}_{1}\mp i{V}_{2}\over \sqrt 2},{{V}_{1}\pm i{V}_{2}\over \sqrt 2},V_3)$ with the spatial components $\bar{V}_{1}, \bar{V}_{2}$ and $\bar{V}_{3}$ corresponding to $S_z=1,-1$ and $0$ along ${\bf B}$. Since $D^{-1}_{\bar{V}_{\mu}\bar{V}_{\nu}}$ vanishes at zero effective momentum for the mixing terms with $\mu\neq\nu$, the spin eigenstates are much more convenient for the exploration of pole masses in magnetic field. As illustrated in Ref.~\cite{Li:2020hlp}, the Schwinger phases can be safely dropped for the evaluations of pole masses without violating gauge invariance. Then, meson masses can be more conveniently evaluated in energy-momentum space by requiring $D^{-1}(p_0,{\bf p}={\bf 0})=0$, where the involved inverse meson propagators are
\bea
\!\!\!\!\!\!\!\!\!\!\!\!D^{-1}_{SS}(p)\!\!\!&\equiv&\!\!\!{1\over2G}+\Pi_{SS}(p)\nonumber\\
\!\!\!&=&\!\!\!\int\!\di^4x~e^{-ip\cdot (y-x)}e^{iq_s\!\int_{x}^y\!A\cdot\di x}D^{-1}_{SS}(y,x),\label{SprpP}\\
\!\!\!\!\!\!\!\!\!\!\!\!D^{-1}_{\bar{V}_{\mu}\bar{V}_{\mu}}(p)\!\!\!&\equiv&\!\!\!{1\over2G_{\rm v}}+\Pi_{\bar{V}_{\mu}\bar{V}_{\mu}}(p)\nonumber\\
\!\!\!&=&\!\!\!\int\!\!\di^4x~e^{-ip\cdot (y-x)}e^{iq_{\rm v}\!\int_{x}^y\!A\cdot\di x}D^{-1}_{\bar{V}_{\mu}\bar{V}_{\mu}}(y,x).\label{VprpP}
\eea
To derive the explicit forms of the polarization functions $\Pi(p)$, it is equivalent to substitute $G_{\rm f}$ in Eqs.\eqref{Sprp} and \eqref{Vprp} by the corresponding effective propagator $\hat{G}_{\rm f}({p})$ defined in Eq.\eqref{prop_m}.

\subsubsection{Intuition in lowest Landau level approximation}\label{LLL}

Previously, it was found that both $\pi^0$ and $\bar{\rho}^+_1$ meson masses decrease with weak magnetic field~\cite{Chernodub:2011mc,Liu:2014uwa,Wang:2017vtn,Mao:2018dqe,Liu:2018zag,Avancini:2016fgq,Shushpanov:1997sf,Xu:2020sui}, which may indicate the system unstable to neutral pion ($\pi^0$) superfluidity or vacuum superconductivity at large $B$. To get an intuitive comprehension, we adopt the lowest Landau level (LLL) approximation and the effective quark propagator is~\cite{Miransky:2015ava}
\bea
\hat{G}_{\rm f}^{LLL}({k})\!=\!-i\,e^{{-{\bf k}_\bot^2\over|q_{\rm f}B|}}{m\!-\!k_4\gamma^4\!-\!k_3\gamma^3\over k_4^2\!+\!k_3^2\!+\!m^2}[1\!+\!{\cal S}(q_{\rm f}B)i\gamma^1\gamma^2]\nonumber\\
\eea
with the transverse momenta ${\bf k}_\bot=(k_1,k_2)$. Then, the corresponding explicit forms of the effective inverse propagators of $\pi^0$ and $\bar{\rho}^+_1$ are respectively
\begin{widetext}
\bea
{D}^{-1}_{\pi^0\pi^0}(p)&=&-{1\over2G}+N_c\sum_{\rm f=u,d}\int{\di^4k\over (2\pi)^4}{\rm tr}~S_{\rm f}^{LLL}({k}+p)i\gamma^5S_{\rm f}^{LLL}({k})i\gamma^5\nonumber\\
&=&-{1\over2G}+8N_c\sum_{\rm f=u,d}\int{\di^4k\over (2\pi)^4}{e^{-{{\bf k}_\bot^2+({\bf k}_\bot+{\bf p}_\bot)^2\over|q_{\rm f}B|}}[m^2\!+\!k_4(k_4+p_4)\!+\!k_3(k_3+p_3)]\over (k_4^2\!+\!k_3^2\!+\!m^2)[(k_4+p_4)^2\!+\!(k_3+p_3)^2\!+\!m^2]},\label{NPLLL}\\
{D}^{-1}_{\bar{\rho}^+_1\bar{\rho}^+_1}(p)&=&-{1\over2G_{\rm v}}+{2N_c}\int{\di^4k\over (2\pi)^4}{\rm tr}~S_{\rm d}^{LLL}({k}+p)\Gamma_{\bar{\rho}^-_1}S_{\rm u}^{LLL}({k})\Gamma_{\bar{\rho}^+_1}\nonumber\\
&=&-{1\over2G_{\rm v}}+32N_c\int{\di^4k\over (2\pi)^4}{e^{-{{\bf k}_\bot^2\over|q_{\rm u}B|}-{({\bf k}_\bot+{\bf p}_\bot)^2\over|q_{\rm d}B|}}[m^2\!+\!k_4(k_4+p_4)\!+\!k_3(k_3+p_3)]\over (k_4^2\!+\!k_3^2\!+\!m^2)[(k_4+p_4)^2\!+\!(k_3+p_3)^2\!+\!m^2]}\label{RPLLL}
\eea
\end{widetext}
with the trace "${\rm tr}$" only taken over spinor space. If we assume $q_{\rm u}=-q_{\rm d}$, the equality of the second terms in Eq.(\ref{NPLLL}) and Eq.(\ref{RPLLL}) can be immediately recognized except for a factor $2$, so it is fair to compare the masses of $\pi^0$ and $\bar{\rho}^+_1$ in external magnetic field.

In the limit $p\rightarrow0$, after completing integrations over the transverse momenta ${\bf k}_\bot$ and inserting the values of $q_{\rm u}$ and $q_{\rm d}$, the effective inverse propagators become
\bea
-{D}^{-1}_{\pi^0\pi^0}(0)
\!\!\!&=&\!\!\!{1\over2G}-{N_c\over \pi}\int{\di^2k\over (2\pi)^2}{eB\over k^2+m^2},\\\label{qcp}
-{D}^{-1}_{\bar{\rho}^+_1\bar{\rho}^+_1}(0)
\!\!\!&=&\!\!\!{1\over2G_{\rm v}}-{16N_c\over9\pi}\int{\di^2k\over (2\pi)^2}{eB\over k^2+m^2}.\label{qcr}
\eea
Actually, they are just the quadratic GL expansion coefficients (GLC$_2$s) around small $\langle\pi^0\rangle$ and $\langle\bar{\rho}^+_1\rangle$ in the LLL approximation, refer to the full one around small $\langle\pi^\pm\rangle$ in Sec.\ref{subsec:MI}. Note that the effect of not so strong  magnetic field should only be taken seriously for the qualitative aspects, because the second terms do not reproduce the correct results in the limit $B\rightarrow0$ at all. In this respect, some useful conclusions can already be drawn even without introducing any regularization, similar to the proof of the positivity of ${\cal B}$ in Sec.\ref{subsec:MI}.
\begin{itemize}
\item[(1)] In the small $B$ region where $m$ is almost a constant, both the GLC$_2$s decrease with magnetic field thus seem to favor decreasing masses in order to maintain the pole condition $D^{-1}(p_0,{\bf p}=0)=0$, see Ref.~\cite{Chernodub:2011mc,Liu:2014uwa,Wang:2017vtn,Mao:2018dqe,Liu:2018zag,Avancini:2016fgq,Shushpanov:1997sf,Xu:2020sui}.
\item[(2)] The magnetic field effect on $\bar{\rho}^+_1$ meson is stronger than that on $\pi^0$ meson, since the linear coefficient of $eB$ in Eq.\eqref{qcr} is larger than that in Eq.\eqref{qcp}. This is consistent with the steeper reduction of $\bar{\rho}^+_1$ mass found in Ref.~\cite{Liu:2014uwa,Liu:2018zag,Bali:2017ian}.	
\item[(3)] In LLL approximation, the gap equation takes the form
\bea
{m-m_0\over2G}-m{N_c\over \pi}\int{\di^2k\over (2\pi)^2}{eB\over k^2+m^2}=0,
\eea
where the MCE can be easily told. Then, we have 
$$-{D}^{-1}_{\pi^0\pi^0}(0)={m_0\over2mG}>0,$$ and $\pi^0$ superfluidity is disfavored, see Ref.~\cite{Liu:2018zag}. However, for the chosen model parameters,
 $$-{D}^{-1}_{\bar{\rho}^+_1\bar{\rho}^+_1}(0)\approx{1\over2G_{\rm v}}-{16\over18G}<0,$$ thus vacuum superconductivity is favored for strong enough magnetic field.
 \item[(4)] As $u$ and $d$ quarks possess different electric charges, their masses would definitely split in strong magnetic field~\cite{Wang:2018gmj}. For $\pi^0$, $u$ and $d$ quarks contribute through polarization loops of pure flavors, see Eq.\eqref{NPLLL}; while for $\bar{\rho}^+_1$, they contribute through a polarization loop of flavor-mixing, see Eq.\eqref{RPLLL}. Thus, the mass splitting effect is expect to be more important for the evaluation of $\bar{\rho}^+_1$ mass and has to be taken care of carefully before the possibility of vacuum superconductivity is concluded.
\end{itemize}

\subsubsection{Full Landau levels formalism and numerical results}\label{FLL}

By utilizing the full propagators of quarks Eq.\eqref{prop_m}, the effective inverse propagators Eq.\eqref{SprpP} and Eq.\eqref{VprpP} become explicitly
\begin{widetext}
\bea
-{D}^{-1}_{\pi^0\pi^0}(p)
&=&{1\over2G}-4N_c\sum_{\rm f=u,d}\int{\di^4k\over (2\pi)^4}\int{\di s \di s'}e^{-s\left[m^2+(k_4+p_4)^2+(k_3+p_3)^2+({\bf k+p})_\bot^2 {\tanh B_{\rm f}^s\over B_{\rm f}^s}\right]}e^{-s'\left[m^2+k_4^2+k_3^2+{\bf k}_\bot^2 {\tanh B_{\rm f}^{s'}\over B_{\rm f}^{s'}}\right]}\nonumber\\
&&\left[(m^2\!+\!({\bf k}_\parallel\!+\!{\bf p}_\parallel)\cdot{\bf k}_\parallel)(1\!+\!\tanh B_{\rm f}^s\tanh B_{\rm f}^{s'})\!+\!({\bf k}_\bot\!+\!{\bf p}_\bot)\cdot{\bf k}_\bot(1\!-\!\tanh^2 B_{\rm f}^s)(1\!-\!\tanh^2 B_{\rm f}^{s'})\right],\label{NPFLL}\\
-{D}^{-1}_{\bar{\rho}^+_1\bar{\rho}^+_1}(p)
&=&{1\over2G_{\rm v}}-8N_c\int{\di^4k\over (2\pi)^4}\int{\di s \di s'}e^{-s\left[m^2+(k_4+p_4)^2+(k_3+p_3)^2+({\bf k+p})_\bot^2 {\tanh B_{\rm u}^s\over B_{\rm u}^s}\right]}e^{-s'\left[m^2+k_4^2+k_3^2+{\bf k}_\bot^2 {\tanh B_{\rm d}^{s'}\over B_{\rm d}^{s'}}\right]}\nonumber\\
&&(m^2\!+\!({\bf k}_\parallel\!+\!{\bf p}_\parallel)\cdot{\bf k}_\parallel)(1\!+\!\tanh B_{\rm u}^s)(1-\tanh B_{\rm d}^{s'}).\label{RPFLL}
	\eea
with ${\bf k}_\parallel\equiv(k_4,k_3)$ the longitudinal momenta. In large $B$ limit, $\tanh B_{\rm u/d}^s\rightarrow\pm1$ according to the signs of $q_{\rm u}$ and $q_{\rm d}$, thus the LLL results Eq.\eqref{NPLLL} and Eq.\eqref{RPLLL} can be well reproduced.
For ${\bf p}={\bf 0}$, by carrying out the integrations over internal energy-momentum and transforming the proper time variables, we simply obtain
\bea
\!\!\!-{D}^{-1}_{\pi^0\pi^0}(p_4)
\!\!\!&=&\!\!\!{1\over2G}-{N_c\over8\pi^2}\sum_{\rm f=u,d}\int{\di s}\int_{-1}^1 {\di u}~e^{-s\left(m^2+{{1-u^2\over4}}p_4^2\right)}\left[\left(m^2\!+\!{1\over s}-{{1-u^2\over4}}p_4^2\right){q_{\rm f}B\over\tanh B_{\rm f}^{s}}\!+\!{(q_{\rm f}B)^2\over\sinh^2B_{\rm f}^{s}}\right],\label{NPFLL1}\\
\!\!\!-{D}^{-1}_{\bar{\rho}^+_1\bar{\rho}^+_1}(p_4)\!\!\!&=&\!\!\!{1\over2G_{\rm v}}-{N_c\over4\pi^2}\int\!{\di s\over s}\!\int_{-1}^1\! {\di u}~{e^{-s\left(m^2+u^+u^-p_4^2\right)}}\left(m^2\!+\!{1\over s}-u^+u^-p_4^2\right){\left[1\!+\!\tanh{B_{\rm u}^s}^+\right]\!\!\left[1\!-\!\tanh{B_{\rm d}^s}^-\right]\over {\tanh{B_{\rm u}^s}^+/B_{\rm u}^s}\!+\!{\tanh{B_{\rm d}^s}^-/B_{\rm d}^{s}}}\label{RPFLL1}
\eea
\end{widetext}
where $u^\pm={1\pm u\over2}, {B_{\rm u}^s}^+={B_{\rm u}^s}u^+$ and ${B_{\rm d}^s}^-={B_{\rm d}^s}u^-$ are defined for brevity. 

Similar to the implementation in Sec.\ref{subsec:MI} , the effective inverse propagator of $\pi^0$ can be regularized in the same way as the quadratic coefficient for $\pi^\pm$, that is,
\bea
-{D}^{-1}_{\pi^0\pi^0}
\!\!\!&=&\!\!\!{1\over2G}+\Delta\Pi_{\pi^0\pi^0}(p_4)-8N_c\int^{\rm reg}{\di^4k\over (2\pi)^4}\nonumber\\
&&{k_4(k_4+p_4)\!+\!E_{\bf k}^2\over (k_4^2\!+\!E_{\bf k}^2)[(k_4+p_4)^2\!+\!E_{\bf k}^2]}\label{NPFLLr}
\eea
with $\Delta\Pi_{\pi^0\pi^0}=\Pi_{\pi^0\pi^0}-(B\rightarrow0)$. However, the divergence of $B$-odd terms in Eq.\eqref{RPFLL1} can not be canceled out with vacuum regularization, so it is subtle to deal with the case of vector $\bar{\rho}^+_1$ meson. As can be checked, the signs of such terms are proportional to the spin component $S_z$ of $\bar{\rho}^+$ meson, their origin must be the $S_z-B$ coupling. If we expand Eq.\eqref{RPFLL1} to linear term on $B$, that is,~\cite{Cao:2019res}
\bea
\!\!\!\!\!\!\!\!\!\Pi_{\bar{\rho}^+_1\bar{\rho}^+_1}^{o(B^2)}(p_4)\!\!\!&=&\!\!\!-{N_c\over4\pi^2}\int{\di s\over s}\int_{-1}^1 {\di u}~{e^{-s\left(m^2+{{1-u^2\over4}}p_4^2\right)}}\nonumber\\
&&\left(m^2\!+\!{1\over s}-{{1-u^2\over4}}p_4^2\right)\left(1+{eBs\over 2}\right),
\eea
such divergence can be completely absorbed by the coefficient of $S_z-B$ coupling term, which is itself $B$-independent.
Eventually, we can perform a modified version of vacuum regularization to the effective inverse propagator of $\bar{\rho}^+_1$ and find
\bea\label{RPFLLr}
-{D}^{-1}_{\bar{\rho}^+_1\bar{\rho}^+_1}
\!\!\!&=&\!\!\!{1\over2G_{\rm v}}\!+\!\Delta\Pi_{\bar{\rho}^+_1\bar{\rho}^+_1}\!-\!8N_c\!\int^{\rm reg}\!\!\!\!{\di^4k\over (2\pi)^4}\!\!\left(1\!+\!{eB\over k_4^2\!+\!E_{\bf k}^2}\right)\nonumber\\
&&{m^2\!+\!k_4(k_4+p_4)\!+\!k_3^2\over (k_4^2\!+\!E_{\bf k}^2)[(k_4+p_4)^2\!+\!E_{\bf k}^2]},
\eea
where we have defined $\Delta\Pi_{\bar{\rho}^+_1\bar{\rho}^+_1}=\Pi_{\bar{\rho}^+_1\bar{\rho}^+_1}-\Pi_{\bar{\rho}^+_1\bar{\rho}^+_1}^{o(B^2)}$ instead of the scheme for $\Delta\Pi_{\pi^0\pi^0}$.

As mentioned before, the advantage of vacuum regularization is that no artifacts would be encountered for the $B$-dependent parts even with $(eB)^{1/2}$ much larger than the regularization cutoff $\Lambda$. This is obvious for $\boldsymbol{\pi}$ mesons, and for $\bar{\rho}^+_1$ meson, the absorption of the $B$-dependent divergence into the $B$-independent coefficient renders the regularized formula suitable to explore the effect of large magnetic field. It is much better to understand the coefficient as the spin magnetic moment, which can be alternatively calculated through quark loops for infinitesimal $B$. To explore the meson spectra, the analytic continuation $p_4\rightarrow i p_0$ should be performed in Eq.\eqref{NPFLLr} and Eq.\eqref{RPFLLr}, but one should keep in mind that the proper time integral is only ultraviolet finite for $p_0<2m$, similar to the condition for $\mu_I$ in Sec.\ref{subsec:MI}. As had been pointed out in Sec.\ref{subsec:MI}, the Ritus formalism or real proper time presentation can be adopted to avoid such technical problem. In Appendix.~\ref{equality}, the equality between the Ritus and imaginary proper-time presentations is well checked for $p_0<2m$.

Even though we can deal with the artificial divergence through mathematical methods, the case with $p_0\geq2m$ would still cause nonphysical consequences. In Appendix.~\ref{invalidity}, we  first compare meson propagators at $B=0$ for different regularization schemes, and then show the invalidity of NJL model to investigate the mass of physical $\bar{\rho}^+_1$ meson in the presence of magnetic field. The incapability of NJL model is due to the lack of confinement effect there, and one should note that the extensive PNJL model is only different from NJL model at finite temperature thus cannot help the situation~\cite{Ferreira:2013tba}. But most recently, Polyakov loop dependence of the couplings was adopted to produce confinement at zero temperature within PNJL model~\cite{Mattos:2021alf}. For such consideration and with the purpose of qualitative comprehension, it is helpful to assume $\rho$ meson mass to be $\lesssim2m$ in the vacuum. Then, by regularizing the $B$-independent divergences with the three-momentum cutoff, the effective inverse meson propagators are explicitly
\bea
\!\!\!\!\!\!\!-{D}^{-1}_{\pi^0\pi^0}
\!\!\!&=&\!\!\!{1\over2G}+\Delta\Pi_{\pi^0\pi^0}-N_c\int_0^\Lambda {k^2dk\over\pi^2}\frac{8E_k}{4 E_k^2+p_4^2},\label{NPFLLr1}\\
-{D}^{-1}_{\bar{\rho}^+_1\bar{\rho}^+_1}
\!\!\!&=&\!\!\!{1\over2G_{\rm v}}+\Delta\Pi_{\bar{\rho}^+_1\bar{\rho}^+_1}-N_c\int_0^\Lambda {k^2dk\over \pi^2E_k}\left[\frac{8(m^2\!+\!{2\over3}k^2)}{4 E_k^2+p_4^2}\right.\nonumber\\
&&\left.+\frac{8E_k^2({2m^2\!+\!k^2})\!-\!{2\over3}k^2p_4^2}{E_k^2(4 E_k^2+p_4^2)^2}eB\right].
\label{RPFLLr1}
\eea
Here, the integrations over the energy $k_0$ have been carried out in the last terms, which are consistent with those obtained in Ref.~\cite{Klevansky:1992qe,Wang:2018gmj,Brauner:2016lkh}. 

\begin{figure}[!htb]
	\begin{center}
		\includegraphics[width=8cm]{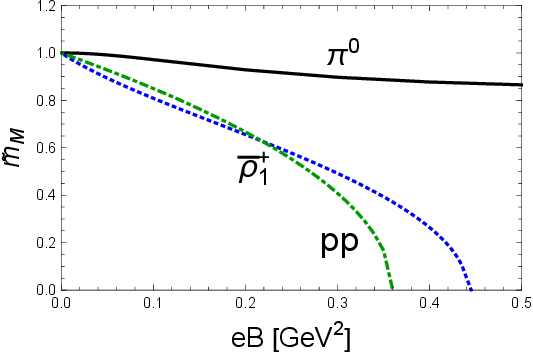}
		\caption{The self-consistent masses of $\pi^0$ (black solid line) and $\bar{\rho}^+_1$ (blue dotted line) mesons as functions of magnetic field $eB$ in two-flavor NJL model. For comparison, point particle mass for $\bar{\rho}^+_1$ (green dot-dashed line) are also included. All meson masses $m_M$ are normalized by their vacuum values. The plots are modified from Ref.~\cite{Cao:2019res}.}\label{mass_2f}
	\end{center}
\end{figure}

Finally, we calculate the masses of $\pi^0$ and $\bar{\rho}^+_1$ mesons self-consistently and show them together in Fig.\ref{mass_2f}. Though $m_{\pi^0}$ decreases slowly with $eB$ and is always finite, $m_{\bar{\rho}^+_1}$ indeed decreases to zero in the two-flavor case~\cite{Wang:2017vtn,Mao:2018dqe,Liu:2018zag,Avancini:2016fgq,Shushpanov:1997sf,Xu:2020sui} which is qualitatively consistent with the point particle result.

\subsection{Three-flavor case in pure magnetic field}\label{subsec:3fM}

As mentioned in the end of Sec.\ref{LLL}, the magnetic field inevitably induces mass splitting between $u$ and $d$ quarks. Under the mean field approximation, that is only allowed when we introduce the iso-vector scalar interactions to NJL model. In this respect, it is more suitable to adopt the three-flavor NJL model for realistic study and the Lagrangian density can be extended from Eq.\eqref{NJL3f} to~\cite{Klimt:1989pm}
\begin{eqnarray}\label{NJL3v}
{\cal L}_{\rm NJL}\!\!\!\!\!&=&\!\!\!\!\!\bar\psi(i\slashed{D}-m_0)\psi+G\sum_{a=0}^8[(\bar\psi\lambda^a\psi)^2+(\bar\psi i\gamma_5\lambda^a\psi)^2]\nonumber\\
&&+{\cal L}_{\rm tH}-G_{\rm v}\left[\left(\bar\psi\gamma^\mu\tau^a\psi\right)^2+\left(\bar\psi i\gamma^\mu\gamma_5\tau^a\psi\right)^2\right]\nonumber\\
&{\cal L}_{\rm tH}&\!\!\!=-K\sum_{s=\pm}{\rm Det}\bar\psi\Gamma^s\psi
\end{eqnarray} 
by adding the four-fermion vector channels. Compared to the iso-scalar interaction $(\bar\psi\psi)^2$ in the two-flavor case, the extra iso-vector term $(\bar\psi\lambda^{3}\psi)^2$ here allowes the dynamical masses of $u$ and $d$ quarks to be different. Now, similar to the two-flavor case, only nonzero chiral condensations, $\sigma_{\rm i}\equiv\langle\bar{\psi}^i{\psi}^i\rangle$ with $i$ flavor index, are considered. Then, by following the same procedure as deriving Eq.\eqref{LNJL4}, ${\cal L}_{\rm tH}$ can be reduced to an effective form of four-fermion interactions~\cite{Klevansky:1992qe}, which then gives the Lagrangian as
\begin{widetext}
\begin{eqnarray}\label{NJL3vr}
\!\!\!{\cal L}_{\rm NJL}^4=\bar\psi(i\slashed{D}-m_0)\psi+\!\!\sum_{a,b=0}^8\!\left[G_{ab}^-(\bar\psi\lambda^a\psi)(\bar\psi\lambda^b\psi)\!+\!G_{ab}^+(\bar\psi i\gamma_5\lambda^a\psi)(\bar\psi i\gamma_5\lambda^b\psi)\right]\!-\!G_{\rm v}\!\!\left[\left(\bar\psi\gamma^\mu\tau^a\psi\right)^2\!+\!\left(\bar\psi i\gamma^\mu\gamma_5\tau^a\psi\right)^2\right]
\end{eqnarray}
with $G^\pm$ listed in Eq.\eqref{Gelements}. Armed with the reduced Lagrangian density, all the evaluations can be carried out parallel to the two-flavor case presented in Sec.\ref{subsec:2fM}.

First of all, the pure magnetic field case can be considered as the vanishing electric field limit of the parallel EM field case, so the gap equations and thermodynamic potentials can be directly obtained from those given in Sec.\ref{subsec:3PEM} by taking the limits: $E,\pi_{\rm f}^5\rightarrow0$. The regularized forms are respectively
\begin{eqnarray}
V_{\rm B3f}^r&=&2G\sum_{{\rm f}=u,d,s}\sigma_{\rm f}^2-4K\prod_{{\rm f}=u,d,s}\sigma_{\rm f}-N_c\sum_{{\rm f}=u,d,s}\left\{{{m_{\rm f}}\Lambda^3\over8\pi^2}\Big[\Big(2+\tilde{m}_{\rm f}^{2}\Big)\Big({1+\tilde{m}_{\rm f}^{2}}\Big)^{1\over2}+\tilde{m}_{\rm f}^{3}\ln\Big(
\Big({1\!+\!\tilde{m}_{\rm f}^{2}}\Big)^{1\over2}\!-\!1\Big)\tilde{m}_{\rm f}^{-1}\Big]\right.\nonumber\\
&&\left.-{1\over8\pi^2}\int_0^\infty {ds\over s^3}e^{-m_{\rm f}^2s}\left({q_{\rm f}Bs
	\over\tanh(q_{\rm f}Bs)}-1\right)-(m_{\rm f}\rightarrow 0)\right\},\label{Omeg_B3f}\\
	-\sigma_{\rm f}
&=&N_c{{m_{\rm f}}\Lambda^2\over2\pi^2}\Big[\Big({1\!+\!\tilde{m}_{\rm f}^{2}}\Big)^{1\over2}\!+\!\tilde{m}_{\rm f}^{2}\ln\Big(
\Big({1\!+\!\tilde{m}_{\rm f}^{2}}\Big)^{1\over2}\!-\!1\Big)\tilde{m}_{\rm f}^{-1}\Big]+N_c{m_{\rm f}\over4\pi^2}\int_0^\infty {ds\over s^2}e^{-m_{\rm f}^2s}\left({q_{\rm f}Bs
	\over\tanh(q_{\rm f}Bs)}-1\right)\label{mgap}
\end{eqnarray}
with the reduced masses $\tilde{m}_{\rm f}={m_{\rm f}/\Lambda}$.
\end{widetext}

Finally, we will pay attention to the collective modes, mainly the neutral pseudoscalar and charged vector modes. For the neutral pseudoscalar sector, it is helpful to define the polarization loops with only one flavor:
	\begin{eqnarray}
	\Pi_{\rm f}=-N_c\int{\di^4k\over (2\pi)^4}\text{tr}~S_{\rm f}({k}+p)i\gamma^5S_{\rm f}({k})i\gamma^5,
	\end{eqnarray}
whose regularized forms can be directly read out from those given in Eq.\eqref{NPFLLr1} but now with different mass ${m}_{\rm f}$ for each flavor.
Then, by setting the diagonal matrix $\Pi^+_0\equiv{\rm diag}(\Pi_{\rm u},\Pi_{\rm d},\Pi_{\rm s})$ in flavor space,
the polarization functions for the $SU(3)$ flavor eigenstates can be directly given by $$\Pi^{+}_{ij}={\rm tr_f}~\lambda^i\Pi^+_0\lambda^j$$
with $i,j=0,3,8$ corresponding to $\eta_0, \pi^0$ and $\eta_8$ channels, respectively. As diagonal matrices commute with each other, we find $\Pi^{+}_{ij}=\Pi^{+}_{ji}$ and only $6$ independent functions are involved here, that is~\cite{Cao:2019res}, 
\bea
	\!\!\!\!\!\!\!\!\!&&\Pi^{+}_{00}={2\over3}\sum_{\rm f=u,d,s}\Pi_{\rm f},~\Pi^{+}_{03}=\sqrt{2\over3}(\Pi_{\rm u}-\Pi_{\rm d}),\nonumber\\
	\!\!\!\!\!\!\!\!\!&&\Pi^{+}_{08}={\sqrt{2}\over3}(\Pi_{\rm u}+\Pi_{\rm d}-2\Pi_{\rm s}),~\Pi^{+}_{33}=\Pi_{\rm u}+\Pi_{\rm d},\nonumber\\
	\!\!\!\!\!\!\!\!\!&&\Pi^{+}_{38}=\sqrt{1\over3}(\Pi_{\rm u}\!-\!\Pi_{\rm d}),~\Pi^{+}_{88}={1\over3}(\Pi_{\rm u}\!+\!\Pi_{\rm d}\!+\!4\Pi_{\rm s}).
\eea
Thus, by recalling the coupling matrix given in Eq.\eqref{Gelements}, the effective inverse propagator matrix becomes
	\begin{eqnarray}\label{NPSM}
	-D^{-1}_{ij}(p)={1\over2}({G}^{+})^{-1}_{ij}+\Pi^{+}_{ij}
	\end{eqnarray}
for the neutral pseudoscalar sector. Besides the $\eta_0-\eta_8$ mixing induced by heavier $s$ quark and $U_{\rm A}(1)$ anomaly in the vacuum~\cite{tHooft:1976snw}, magnetic field further develops mixing among $\eta_0, \pi^0$ and $\eta_8$ since all the non-diagonal elements $\Pi^{+}_{i(\neq)j}\neq 0$ now. Then, the pole masses of the neutral pseudoscalar mesons can be solved numerically according to the condition $\det D^{-1}_{ij}(p_0,{\bf p}={\bf 0})=0$~\cite{Klimt:1989pm} and there should be three independent solutions.
	
Turn to the vector mode $\bar{\rho}^+_1$, the different masses between $u$ and $d$ quarks would alter the effective inverse propagator Eq.\eqref{RPFLL} to
\bea
-{D}^{-1}_{\bar{\rho}^+_1\bar{\rho}^+_1}(p_4)={1\over2G_{\rm v}}+\Pi_{\bar{\rho}^+_1\bar{\rho}^+_1}^*(p_4)
\eea
with the polarization function~\cite{Cao:2019res}
\begin{widetext}
\bea	
\Pi_{\bar{\rho}^+_1\bar{\rho}^+_1}^*(p_4)
=\!-\!{N_c\over4\pi^2}\!\!\int\!{\di s\over s}\!\!\int_{-1}^1\!\! {\di u}~{e^{-s\left[m_{\rm u}^2u^+\!+m_{\rm d}^2u^-\!+u^+u^-p_4^2\right]}}\!\!\left(m_{\rm u}m_{\rm d}\!+\!{1\over s}-u^+u^-p_4^2\right)\!\!{\left[1\!+\!\tanh{B_{\rm u}^s}^+\right]\!\!\!\left[1\!-\!\tanh{B_{\rm d}^s}^-\right]\over {\tanh{B_{\rm u}^s}^+\over B_{\rm u}^s}+{\tanh{B_{\rm d}^s}^-\over B_{\rm d}^{s}}}.
\eea
By following a similar procedure as in Sec.\ref{FLL}, it can be regularized to
\bea\label{RP3f}	
-{D}^{-1}_{\bar{\rho}^+_1\bar{\rho}^+_1}(p_4)
&=&{1\over2G_{\rm v}}+\Delta\Pi_{\bar{\rho}^+_1\bar{\rho}^+_1}^*-N_c\int_0^\Lambda\!\! {2k^2dk\over\pi^2}\frac{(E_{\rm u}E_{\rm d}\!+\!{m_{\rm u}}{m_{\rm d}}\!+\!{1\over3}k^2)(E_{\rm u}\!+\!E_{\rm d})}{E_{\rm u}E_{\rm d}[(E_{\rm u}\!+\!E_{\rm d})^2\!+\!p_4^2]}-{N_c}\int_0^\Lambda {k^2dk\over\pi^2}\left\{{{q_{\rm u}B}\over(E_{\rm u}\!+\!E_{\rm d})^2\!+\!p_4^2}\right.\nonumber\\
&&\left.\left[\left(\frac{E_{\rm u}E_{\rm d}\!+\!{m_{\rm u}}{m_{\rm d}}\!+\!{1\over3}k^2}{E_{\rm u}^3}\!+\!{1\over E_{\rm u}}\!+\!{1\over E_{\rm d}}\right)\!-\!\frac{[p_4^2\!+\!(m_{\rm u}\!-\!m_{\rm d})^2\!+\!{4\over3}k^2](E_{\rm u}\!+\!E_{\rm d})^2}{E_{\rm u}^2E_{\rm d}[(E_{\rm u}+E_{\rm d})^2+p_4^2]}\right]\!-\!(u\leftrightarrow d)\right\},
\eea
\end{widetext}
where $\Delta\Pi_{\bar{\rho}^+_1\bar{\rho}^+_1}^*(p_4)=\Pi_{\bar{\rho}^+_1\bar{\rho}^+_1}^*(p_4)-\Pi_{\bar{\rho}^+_1\bar{\rho}^+_1}^{o(B^2)*}(p_4)$ and the quark dispersions are $E_{\rm f}=\sqrt{m_{\rm f}^2+k^2}$. One can check that Eq.\eqref{RPFLLr1} can be well reproduced if we take $m_{\rm u}=m_{\rm d}=m$ in Eq.\eqref{RP3f}. As we have shown that the qualitative conclusions do not depend on the choices of regularization schemes~\cite{Cao:2019res}, it is enough to adopt the simple three-momentum cutoff scheme to deal with the $B$-independent divergences here.

\begin{figure}[!htb]
	\begin{center}
		\includegraphics[width=8cm]{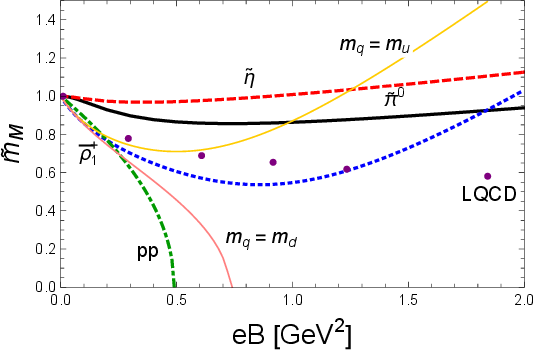}
		\caption{The self-consistent masses of $\tilde{\pi}^0$ (black solid line), $\tilde\eta$ (red dashed line) and $\bar{\rho}^+_1$ (blue dotted line) mesons as functions of magnetic field $eB$. Also shown are the $\bar{\rho}^+_1$ masses in point particle ansatz (green dot-dashed line), from LQCD simulations (purple points) and with both composite quark masses equally $m_{\rm u}$ (thin yellow solid line) or  $m_{\rm d}$ (thin pink dashed line). All masses are normalized by their vacuum values. The plots are from Ref.~\cite{Cao:2019res}.}\label{mM_3f}
	\end{center}
\end{figure}

Now, we are ready to study the properties of collective modes in magnetic field through Eq.\eqref{NPSM} and Eq.\eqref{RP3f} after solving the gap equations in Eq.\eqref{mgap} self-consistently. As the vacuum mass of $u$ and $d$ quarks in this case ($0.368~{\rm GeV}$) is larger than that in two-flavor case ($0.313~{\rm GeV}$), the vector coupling is fixed to $G_{\rm v}\Lambda^2=2.527$ by fitting to a vacuum mass $m_\rho^v=0.7~{\rm GeV}$ closer to the physical value. The main results of the meson masses are illustrated in Fig.\ref{mM_3f}, where $\tilde{\pi}^0$ and $\tilde\eta$ are the effective neutral pseudoscalars corresponding to ${\pi}^0$ and $\eta$ mesons at $B=0$.
It is easy to tell that all the consistently obtained masses show a similar feature with each other: first decreasie and then increase with $eB$, though the variation of the $\bar{\rho}^+_1$ mass is stronger. The latter is consistent with the flavor-mixing polarization loop shown in Sec.\ref{LLL} and the enhancement at larger $B$ is due to the splitting MCE among the quarks~\cite{Cao:2019res}. To help the understanding, the $\bar{\rho}^+_1$ mass with both composite quark masses equal to $m_{\rm u}$ or $m_{\rm d}$ is also depicted in Fig.\ref{mM_3f}. With the $\bar{\rho}^+_1$ mass decreasing to zero when $m_{\rm d}$ is adopted, we can easily discover that the great enhancement of $m_{\rm u}$~\cite{Cao:2019res} avoids further reduction at larger $eB$. For comparison, we also demonstrate the normalized $\bar{\rho}^+_1$ mass with point particle ansatz and from LQCD simulations~\cite{Bali:2017ian}. An important finding is that our results are semi-quantitatively consistent with that of LQCD at relatively weak magnetic field region, where the curvatures with respect to $eB$ are both positive. In contrary, they are negative for both the point particle formula andtwo-flavor NJL model (except for the weakest $B$), see Fig.\ref{mM_3f}.  Thus, in any sense,  the realistic three-flavor NJL model is much better than the two-flavor one for exploring the properties of vector mesons.

\subsection{Possibility in parallel magnetic field and rotation}\label{subsec:PMRrho}
According to last section, we adopt the three-flavor NJL model to explore the $\rho^\pm$ superconductivity in parallel magnetic field and rotation (${\bf B}\parallel\boldsymbol{\Omega}$). And the Lagrangian density can be modified from Eq.\eqref{NJL3v} by adding the rotation effect through the temporal covariant derivative. In this case, it is more convenient to adopt the symmetric gauge, then the longitudinal and transverse covariant derivatives are respectively~\cite{Cao:2020pmm} 
\bea
&&D_0=\partial_t-i\Omega \left(\hat{L}_z+\hat{S}_z\right),~D_3=\partial_z,\\
&&D_1=\partial_x+iQ\,By/2,~D_2=\partial_y-iQ\,Bx/2.
\eea
If only $\sigma_{\rm f}$ condensations are considered, the modification of the temporal covariant derivative does not affect the reduction procedure of the Lagrangian Eq.\eqref{NJL3v} at all, so the reduced Lagrangian with only four-quark interactions keeps the same form as Eq.\eqref{NJL3vr}. Based on Eq.\eqref{NJL3vr}, we are going to check if the system would undergo any instability to charged rho ($\rho^\pm$) superconductor in ${\bf B}\parallel\boldsymbol{\Omega}$.

Similar to the case of $\pi^\pm$ superfluid discussed in Sec.\ref{subsec:NJLcpi}, we assume the possible transition to $\rho^\pm$ superconductor to be of continuous for simplicity. Then, the thermodynamic potential can be expressed in Ginzburg-Landau (GL) approximation as
\begin{eqnarray}\label{thermo}
V_{\rm B\Omega}^r(\sigma_{\rm f},\Delta)=V_{\rm B\Omega}^r(\sigma_{\rm f})+{\cal A}~\Delta^2+{\cal B}~\Delta^4+\dots,
\end{eqnarray}
where the leading order term can be extended from Eqs.\eqref{Omeg_BOmgT} and \eqref{Omeg_B3f} to
\begin{widetext}
\bea
V_{\rm B\Omega}^r(\sigma_{\rm f})=V_{\rm B3f}^r-{N_c}T \sum_{t=\pm}\sum_{\rm f=u,d,s}\sum_{\rm n=0}^{n_{\rm max}}{1\over S}\sum_{\rm l=0}^{N_{\rm f}}\alpha_n\int_{-\infty}^\infty{\di p_3\over\pi}\ln\left(1\!+\!e^{-(E_{{\rm f}n}(p_3)+t\,\Omega_{\rm nl})/T}\right),
\eea
and the gap equations follow as
\bea
-\sigma_{\rm f}
&=&N_c{{m_{\rm f}}\Lambda^2\over2\pi^2}\Big[\Big({1\!+\!\tilde{m}_{\rm f}^{2}}\Big)^{1\over2}\!+\!\tilde{m}_{\rm f}^{2}\ln\Big(
\Big({1\!+\!\tilde{m}_{\rm f}^{2}}\Big)^{1\over2}\!-\!1\Big)\tilde{m}_{\rm f}^{-1}\Big]+N_c{m_{\rm f}\over4\pi^2}\int_0^\infty {ds\over s^2}e^{-m_{\rm f}^2s}\left({q_{\rm f}Bs
	\over\tanh(q_{\rm f}Bs)}-1\right)\nonumber\\
&&-N_cm_{\rm f}\sum_{t=\pm}\sum_{\rm n=0}^{n_{\rm max}}{1\over S} \sum_{\rm l=0}^{{N}^{{\rm f}}}\int_{-\infty}^{\infty}{dp_3\over\pi}{\alpha_n\over E_{{\rm f}n}(p_3)}
{1\over 1\!+\!e^{(E_{{\rm f}n}(p_3)+t\,\Omega_{\rm nl})/T}}.
\eea
\end{widetext}
Of course, $\Delta$ can be understood as any kinds of mesonic condensates and the expansion coefficients can be evaluated according to their interaction vertices with quarks, refer to Sec.\ref{subsec:MI}. For simplicity, we assume the transition to be solely determined by the quadratic coefficient ${\cal A}$~\cite{Cao:2019ctl}: If ${\cal A}<0$, the meson condensation is favored; and if ${\cal A}_a<{\cal A}_b<0$, we take meson $a$ condensation more preferred than meson $b$. As demonstrated in Sec.\ref{sec:cpi}, the coefficient is given by
$${\cal A}={1\over4G}+\Pi$$
with the bare polarization function defined as
\begin{equation}\label{PF}
\Pi={i\over V_4}{\rm Tr}\left[S(x,y)\Gamma_{M*}S(y,x)\Gamma_{M}e^{-i\Phi_M}\right].
\end{equation}

For $\rho^+$ with its spin along the direction of ${\bf B}$,  $\Gamma_{\bar{\rho}^+_{1}}=\bar{\gamma}_1^+\tau_+$, and the explicit evaluations of the polarization functions can follow that of $\pi^\pm$ given in Sec.\ref{subsec:NJLcpi}. By substituting the fermion propagators Eqs.\eqref{propagator+} and \eqref{propagator-} into the polarization function Eq.\eqref{PF}, we find explicitly
\begin{widetext}
	\begin{eqnarray}
	\Pi_{\bar{\rho}^+_1}&=&{-i\over S}\sum_{\rm n=0}^{n_{\rm max}}\sum_l\sum_{\rm n'=0}^{n_{\rm max}}\sum_{\rm l'}\int_{-\infty}^{\infty}{dp_0\over2\pi}\int_{-\infty}^{\infty}{dp_3\over2\pi}
	~{\rm Tr}\Bigg\{\left[{\cal P}_\uparrow\chi_{\rm n,l}^+(\theta,r)\chi_{\rm n,l}^{+*}(\theta',r')+{\cal P}_\downarrow\chi_{\rm n-1,l+1}^{+}(\theta,r)\chi_{\rm n-1,l+1}^{+*}(\theta',r')\right]\nonumber\\
	&&\ \ \ \ \times\left(\gamma^0{p}_0^{l+}-\gamma^3p_3+m_{\rm u}\right)-\left[{\cal P}_\uparrow\chi_{\rm n,l}^{+}(\theta,r)\chi_{\rm n-1,l+1}^{+*}(\theta',r')+{\cal P}_\downarrow\chi_{\rm n-1,l+1}^{+}(\theta,r)\chi_{\rm n,l}^{+*}(\theta',r')\right]
	\gamma^2\sqrt{2n q_{\rm u}B}\Bigg\}\nonumber\\
	&&\times\Bigg\{-\left[{\cal P}_\downarrow\chi_{\rm n'-1,l'-1}^{-}(\theta',r')\chi_{\rm n'-1,l'-1}^{-*}(\theta,r)+{\cal P}_\uparrow\chi_{\rm n',l'}^{-}(\theta',r')\chi_{\rm n',l'}^{-*}(\theta,r)\right]\left(\gamma^0{p}_0^{l'-}-\gamma^3p_3-m_{\rm d}\right){\gamma^1+i\gamma^2\over \sqrt{2}}\nonumber\\
	&&\ \ \ \ \  -\left[{\cal P}_\downarrow\chi_{\rm n'-1,l'-1}^{-}(\theta',r')\chi_{\rm n',l'}^{-*}(\theta,r)+{\cal P}_\uparrow\chi_{\rm n',l'}^{-}(\theta',r')\chi_{\rm n'-1,l'-1}^{-*}(\theta,r)\right]\gamma^2\sqrt{2n'|q_{\rm d}B|}{\gamma^1-i\gamma^2\over \sqrt{2}}\Bigg\}{\gamma^1-i\gamma^2\over \sqrt{2}}\nonumber\\
	&&\times{e^{-i\Phi}\over\left[\left({p}_0^{l+}\right)^2-E_{\rm un}^2\right]\left[\left({p}_0^{l'-}\right)^2-E_{\rm dn'}^2\right]}
	\end{eqnarray}
	with the trace over the spinor and space-time spaces. By working out the trace over the spinor space, the expression is greatly simplified, that is~\cite{Cao:2020pmm}, 
	\begin{eqnarray}\label{A1}
	\Pi_{\bar{\rho}^+_1}&=&{-4N_ci\over S}\sum_{\rm n=0}^{n_{\rm max}}\sum_l\sum_{\rm n'=0}^{n_{\rm max}}\sum_{\rm l'}\sum_{r,r'}\sum_{\theta,\theta'}\int_{-\infty}^{\infty}{dp_0\over2\pi}\int_{-\infty}^{\infty}{dp_3\over2\pi}
	{e^{-i\Phi}\over\left[\left({p}_0^{l+}\right)^2\!-\!E_{\rm un}^2\right]\left[\left({p}_0^{l'-}\right)^2\!-\!E_{\rm dn'}^2\right]}\left({p}_0^{l+}{p}_0^{l'-}-p_3^2-m_{\rm u}m_{\rm d}\right)\nonumber\\
	&&\chi_{\rm n,l}^{+}(\theta,r)\chi_{\rm n,l}^{+*}(\theta',r')\chi_{\rm n',l'}^{-}(\theta',r')\chi_{\rm n',l'}^{-*}(\theta,r).
	\end{eqnarray}
Compared to that of charged pion given in Eq.\eqref{AFL1}, there is only one kind of combination of $\chi^{+}\chi^{+*}$ and $\chi^{-}\chi^{-*}$ here; except for that, the numerator is independent of Landau levels which is consistent with that obtained in Appendix.~\ref{equality}. Incidentally, it can be checked that the polarization function of the axial vector $\bar{a}^+_1$ can be adjusted from Eq.\eqref{A1} by taking the substitution $-m_{\rm u}m_{\rm d}\rightarrow+m_{\rm u}m_{\rm d}$. As we had shown in Ref.~\cite{Cao:2019res} that the mass term contributes negatively to $\Pi_{\bar{\rho}^+_1}$, we would expect $m_{\bar{a}^+_1}>m_{\bar{\rho}^+_1}$  in the presence of pure magnetic field. This is consistent with the observation that $\rho$ mesons are the lightest (axial) vectors in the vacuum and it turns out that the $\bar{a}^+_1$ superconductor is not possible in this case. However, due to the chemical potential like effect induced by $\Omega$, the mass term can be positive for $\bar{\rho}^+_1$ in ${\bf B}\parallel\boldsymbol{\Omega}$, so we have to check the possibility of $\bar{a}^+_1$ superconductor here.
	
We work out the case with Schwinger phase in Minkowski space first, that is, by adopting $\Phi_{\rm M}$ shown in Eq.\eqref{SPM}. Then, inserting into Eq.\eqref{A1} the explicit forms of $\chi^\pm$ as shown in Eqs.\eqref{chi+} and \eqref{chi-} and carrying out the integrations over the polar angles, we obtain
	\begin{eqnarray}\label{A2}
	\Pi_{\bar{\rho}^+_1}^{\rm M}&=&{-16N_ci\over S}\sum_{\rm n,n'=0}^{n_{\rm max}}\sum_{\rm l=0}^{\mathcal{N}_{\rm u}}\sum_{\rm l'=0}^{\mathcal{N}_{\rm d}}\sum_{r,r'}\!\int_{-\infty}^{\infty}\!{dp_0\over2\pi}\int_{-\infty}^{\infty}\!{dp_3\over2\pi}
	{e^{-{eB\over 4}(r^2\!+{r'}^2)}\left({p}_0^{(l-n)+}{p}_0^{(n'-l')-}\!-\!p_3^2\!-\!m_{\rm u}m_{\rm d}\right)\over\left[\left({p}_0^{(l-n)+}\right)^2-E_{\rm un}^2\right]\left[\left({p}_0^{(n'-l')-}\right)^2-E_{\rm dn'}^2\right]}{n!n'!\over l!l'!}\left({q_{\rm u}B\over 2}\right)^{l-n+1}\nonumber\\
	&&\left({|q_{\rm d}B|\over 2}\right)^{l'-n'+1}J_{\rm l+l'-n-n'}\left({eB\over2}rr'\right)(rr')^{l+l'-n-n'}{F}_{\rm nl,n'l'}(q_{\rm u}B,|q_{\rm d}B|;r,r'),
	\end{eqnarray}
	where the auxiliary function is defined as
	\begin{eqnarray}
	F_{\rm nl,n'l'}(q_{\rm u}B,|q_{\rm d}B|;r,r')\equiv \prod_{x=r,r'}L_n^{l-n}\left({q_{\rm u}B ~x^2\over2}\right)L_{\rm n'}^{l'-n'}\left({|q_{\rm d}B| ~x^2\over2}\right).
	\end{eqnarray}	
Here, we find the contribution of LLL with $n=n'=0$ is finite due to the special combining structure of $\chi^{+}\chi^{+*}$ and $\chi^{-}\chi^{-*}$ in Eq.\eqref{A1}. For charged pion, we can see in Sec.\ref{subsec:NJLcpi} that such combination doesn't exist, so the LLL won't contribute. Physically, this is related to the fact that the spins of of $u$ and $\bar{d}$ quarks are along the same direction in LLL thus cannot form a state with total spin zero.
	
For future convenience, we define two reduced radii: $\bar{r}=(eB/2)^{1/2} r$ and $\bar{r}'=(eB/2)^{1/2} r'$, then Eq.\eqref{A2} becomes
	\begin{eqnarray}\label{A3}
	&&\!\!\!\!\!\!	\Pi_{\bar{\rho}^+_1}^{\rm M}={-16N_ci\over S}\sum_{\rm n,n'=0}^{n_{\rm max}}\sum_{\rm l=0}^{\mathcal{N}_{\rm u}}\sum_{\rm l'=0}^{\mathcal{N}_{\rm d}}\int_{-\infty}^{\infty}\!{dp_0\over2\pi}\int_{-\infty}^{\infty}\!{dp_3\over2\pi}
	{\left({p}_0^{(l-n)+}{p}_0^{(n'-l')-}\!\!-\!p_3^2\!-\!m_{\rm u}m_{\rm d}\right)\!	{\cal F}_{\rm nl,n'l'}(\tilde{q}_{\rm u},|\tilde{q}_{\rm d}|)\over\left[\left({p}_0^{(l-n)+}\right)^2-E_{\rm un}^2\right]\left[\left({p}_0^{(n'-l')-}\right)^2-E_{\rm dn'}^2\right]},\\
	&&\!\!\!\!\!\!\!\!	{\cal F}_{\rm nl,n'l'}(\tilde{q}_{\rm u},|\tilde{q}_{\rm d}|)\!=\!{n!n'!\over l!l'!}\tilde{q}_{\rm u}^{l-n+1}{|\tilde{q}_{\rm d}|}^{l'-n'+1}\sum_{\bar{r},\bar{r}'}e^{-{\bar{r}^2+(\bar{r}')^2\over 2}}J_{\rm l+l'-n-n'}\left(\bar{r}\bar{r}'\right)(\bar{r}\bar{r}')^{l+l'-n-n'}{F}_{\rm nl,n'l'}(2\tilde{q}_{\rm u},2|\tilde{q}_{\rm d}|;\bar{r},\bar{r}').
	\end{eqnarray}
To proceed, we transform the LLs independent numerator as the following
	\bea
	&&{p}_0^{(l-n)+}{p}_0^{(n'-l')-}-p_3^2-m_{\rm u}m_{\rm d}\nonumber\\
	&\rightarrow&
	{1\over2}\left[\left({p}_0^{(l-n)+}\right)^2-E_{\rm un}^2+\left({p}_0^{(n'-l')-}\right)^2-E_{\rm dn'}^2-(k\,\Omega)^2+(m_{\rm u}-m_{\rm d})^2+2n\,q_{\rm u}B+2n'|q_{\rm d}B|\right]
	\eea
	with $k$ the same as that defined in Sec.\ref{subsec:NJLcpi}, which then alters the polarization function to
	\bea
	\Pi_{\bar{\rho}^+_1}^{\rm M}&=&{-8N_ci\over S}\sum_{\rm n,n'=0}^{n_{\rm max}}\sum_{\rm l=0}^{\mathcal{N}_{\rm u}}\sum_{\rm l'=0}^{\mathcal{N}_{\rm d}}	{\cal F}_{\rm nl,n'l'}(\tilde{q}_{\rm u},|\tilde{q}_{\rm d}|)\int_{-\infty}^{\infty}\!{dp_0\over2\pi}\int_{-\infty}^{\infty}\!{dp_3\over2\pi}\left\{
	{1\over\left({p}_0^{(l-n)+}\right)^2-E_{\rm un}^2}+{1\over\left({p}_0^{(n'-l')-}\right)^2-E_{\rm dn'}^2}\right.\nonumber\\
	&&\left.+{-\Omega_{\rm nl,n'l'}^2+(m_{\rm u}-m_{\rm d})^2+2n\,q_{\rm u}B+2n'|q_{\rm d}B|\over\left[\left({p}_0^{(l-n)+}\right)^2-E_{\rm un}^2\right]\left[\left({p}_0^{(n'-l')-}\right)^2-E_{\rm dn'}^2\right]}\right\}.
	\eea
Now, we're ready to work within the finite temperature field theory by taking the transformations: $p_0\rightarrow i\omega_m$ and $-i\int_{-\infty}^{\infty}\!{dp_0\over2\pi}\rightarrow T\sum_{m=-\infty}^\infty$. Then, by carrying out the summation over the fermion Matsubara frequency $\omega_m=(2m+1)\pi T$, we get
	\bea
	\Pi_{\bar{\rho}^+_1}^{\rm M}&=&{-2N_c\over S}\sum_{\rm n,n'=0}^{n_{\rm max}}\sum_{\rm l=0}^{\mathcal{N}_{\rm u}}\sum_{\rm l'=0}^{\mathcal{N}_{\rm d}}	{\cal F}_{\rm nl,n'l'}(\tilde{q}_{\rm u},|\tilde{q}_{\rm d}|)\sum_{s=\pm}\int_{-\infty}^{\infty}{dp_3\over(2\pi)}\left\{\tanh\left({E_{\rm un}-s\,\Omega_{\rm nl,{1\over2}0}\over2T}\right){1\over E_{\rm un}}\Bigg[ 1+\right.\nonumber\\
	&&\left.{-\Omega_{\rm nl,n'l'}^2+(m_{\rm u}-m_{\rm d})^2+2n\,q_{\rm u}B+2n'|q_{\rm d}B|\over\left(E_{\rm un}-s\,\Omega_{\rm nl,n'l'}\right)^2-E_{\rm dn'}^2}\Bigg]+\left(E_{\rm dn'}\leftrightarrow E_{\rm un},nl\leftrightarrow n'l',q_{\rm u}\leftrightarrow |q_{\rm d}|\right)\right\}.
	\eea
Similar to that of charged pion in Sec.\ref{subsec:NJLcpi},	the temperature and rotation dependent parts should be convergent and can be completely separated out by 
	$\Pi_{\bar{\rho}^+_1}^{\rm M}-\Pi_{\bar{\rho}^+_1}^{\rm M}\big|_{\Omega\rightarrow0, T\rightarrow0}$. We immediately recognize the subtracted term to be the polarization function in pure magnetic field, thus the regularized quadratic coefficient can be simply given by 
\bea
{\cal A}_{\rm M}^r=-{1\over 2}{D}^{-1}_{\bar{\rho}^+_1\bar{\rho}^+_1}(0)+\Pi_{\bar{\rho}^+_1}^{\rm M}-\Pi_{\bar{\rho}^+_1}^{\rm M}\big|_{\Omega\rightarrow0, T\rightarrow0}
\eea
with the help of Eq.\eqref{RP3f}.

Next, we choose the Schwinger phase in curved space, that is, with the corresponding exponent $\Phi_{\rm C}$ shown in Eq.\eqref{SPC}. Compared to $\Phi_{\rm M}$, the difference is only from the angle shift: $\theta-\theta' \rightarrow \theta-\theta'-\Omega(t-t')$, which then modifies the relation between $u$ and $\bar{d}$ quark energies in the loop. In that sense, the polarization function  with $\Phi_{\rm C}$ can be directly corrected from Eq.\eqref{A3} by changing ${p}_0^{(l-n)+}$ to ${p}_0^{(n'-l')+}$, that is,
\begin{eqnarray}\label{A4}
&&\Pi_{\bar{\rho}^+_1}^{\rm C}={-16N_ci\over S}\sum_{\rm n,n'=0}^{n_{\rm max}}\sum_{\rm l=0}^{\mathcal{N}_{\rm u}}\sum_{\rm l'=0}^{\mathcal{N}_{\rm d}}\int_{-\infty}^{\infty}\!{dp_0\over2\pi}\int_{-\infty}^{\infty}\!{dp_3\over2\pi}
{\left({p}_0^{(n'-l')+}{p}_0^{(n'-l')-}\!\!-\!p_3^2\!-\!m_{\rm u}m_{\rm d}\right)	{\cal F}_{\rm nl,n'l'}(\tilde{q}_{\rm u},|\tilde{q}_{\rm d}|)\over\left[\left({p}_0^{(n'-l')+}\right)^2-E_{\rm un}^2\right]\left[\left({p}_0^{(n'-l')-}\right)^2-E_{\rm dn'}^2\right]}.
\end{eqnarray}
Then, within the finite temperature field theory, the polarization function can be evaluated explicitly as
\bea
\Pi_{\bar{\rho}^+_1}^{\rm C}\!\!\!&=&\!\!\!{-2N_c\over S}\sum_{\rm n,n'=0}^{n_{\rm max}}\sum_{\rm l=0}^{\mathcal{N}_{\rm u}}\sum_{\rm l'=0}^{\mathcal{N}_{\rm d}}	{\cal F}_{\rm nl,n'l'}(\tilde{q}_{\rm u},|\tilde{q}_{\rm d}|)\sum_{s=\pm}\int_{-\infty}^{\infty}{dp_3\over(2\pi)}\Bigg\{\tanh\left({E_{\rm un}\!+\!s\,\Omega_{{3\over2}0,n'l'}\over2T}\right){1\over E_{\rm un}}+\tanh\left({E_{\rm dn'}-s\,\Omega_{{1\over2}0,n'l'}\over2T}\right)\nonumber\\
&&\!\!\!{1\over E_{\rm dn'}}\!+\!\left[{-\Omega^2\!+\!(m_{\rm u}\!-\!m_{\rm d})^2\!+\!2n\,q_{\rm u}B\!+\!2n'|q_{\rm d}B|}\right]\left[{1\over E_{\rm un}}{\tanh\left({E_{\rm un}\!+\!s\,\Omega_{{3\over2}0,n'l'}\over2T}\right)\over\left(E_{\rm un}-s\,\Omega\right)^2-E_{\rm dn'}^2}+{1\over E_{\rm dn'}}{\tanh\left({E_{\rm dn'}-s\,\Omega_{{1\over2}0,n'l'}\over2T}\right)\over\left(E_{\rm dn'}-s\,\Omega\right)^2-E_{\rm un}^2}\right]\Bigg\}.\nonumber\\
\eea
In contrary to that of charged pion in Sec.\ref{subsec:NJLcpi},  the appearances of $\Omega$ in the denominators indicate that  the angular velocity can still  induce an effect of isospin chemical potential to $\bar{\rho}^+_1$ in this case. Finally, following the same logic as the previous section, the quadratic coefficient can be regularized as
\bea
{\cal A}_{\rm C}^r=-{1\over 2}{D}^{-1}_{\bar{\rho}^+_1\bar{\rho}^+_1}(0)+\Pi_{\bar{\rho}^+_1}^{\rm C}-\Pi_{\bar{\rho}^+_1}^{\rm C}\big|_{\Omega\rightarrow0, T\rightarrow0}.
\eea

\end{widetext}

\begin{figure}[!htb]
	\centering
	\includegraphics[width=0.42\textwidth]{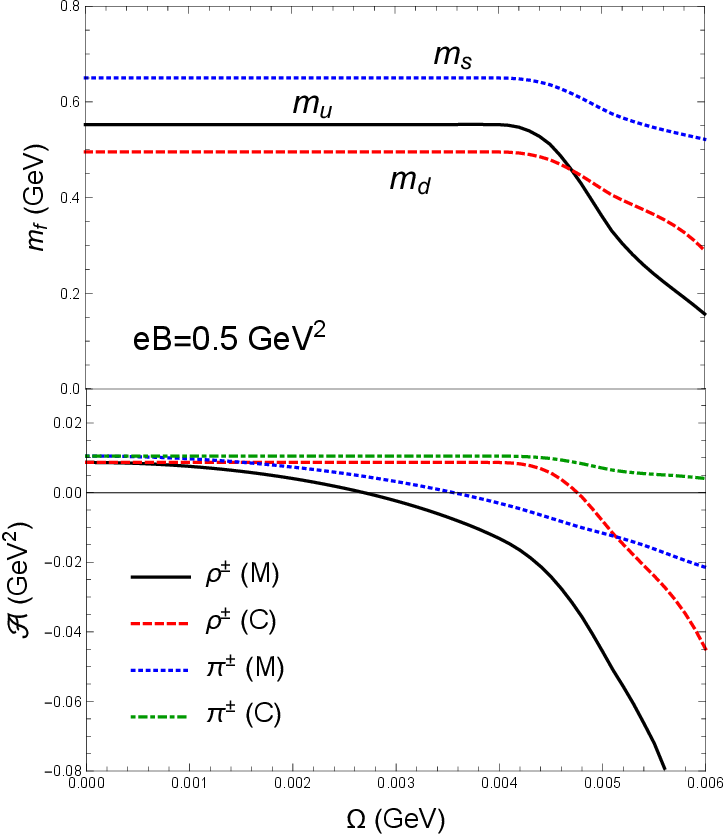}
	\caption{The dynamical quark masses $m_{\rm f}$ (upper panel) and quadratic GL expansion coefficients ${\cal A}$ (lower panel) as functions of the angular velocity $\Omega$ at the magnetic field $eB=0.5~{\rm GeV}^2$. In the lower panel, the coefficients for $\rho^\pm$ condensation are compared to those for $\pi^\pm$ with Schwinger phase in Minkowski (M) and curved (C) spaces. The plots are from Ref.~\cite{Cao:2020pmm}.}\label{Am_05}
\end{figure}

In Fig.\ref{Am_05}, we show the dynamical quark masses $m_{\rm f}$ together with the quadratic coefficients ${\cal A}$ for both $\pi^\pm$ superfluidity and $\rho^\pm$ superconductivity. For the given magnetic field, $m_{\rm u}$ decreases with $\Omega$ much more quickly than the other flavors and eventually $m_{\rm u}<m_{\rm d}$ which somehow can be understood as a manifestation of dHvA oscillation. The quadratic coefficients ${\cal A}$ for $\pi^\pm$ superfluidity are qualitatively consistent with those found in Sec.\ref{subsec:NJLcpi}, that is, the ambiguity continues in the three-flavor NJL model. However, there is no ambiguity as to the question whether $\rho^\pm$ superconductivity can happen or not in ${\bf B}\parallel\boldsymbol{\Omega}$ -- the answer is consistently yes for the Schwinger phases basic in Minkowski and curved spaces. For Schwinger phase in curved space, the difference between $\pi^\pm$ and $\bar{\rho}^+_1$ mesons is that the spin-rotation coupling can still induce $\mu_{\rm I}$ effect to $\bar{\rho}^+_1$. Comparing the critical $\Omega$'s , $\rho^\pm$ superconductivity is more favored than $\pi^\pm$ superfluidity for the chosen magnetic field, thus $\rho^\pm$ superconductivity is indeed possible in ${\bf B}\parallel\boldsymbol{\Omega}$. It is interesting to note that a similar situation was encountered at finite $\mu_{\rm I}$ where $\rho^\pm$ superconductivity was also found to be more favored when $\Omega$ is large enough~\cite{Zhang:2018ome}. 
 
 There are lengthy derivations in this section. To summarize, we adopted the Ginzburg-Landau approximation to explore the possibility of charged rho condensation within the two- and three-flavor NJL models. In Sec.\ref{subsec:3fM}, we denied the charged rho condensation in pure magnetic field due to the dynamical mass splitting between the $u$ and $d$ quark components of $\rho^\pm$ mesons. While in Sec.\ref{subsec:PMRrho}, we proposed that the $\rho^\pm$ superconductor could happen in a system with parallel magnetic field and rotation, because a series of isospin chemical potentials were effectively generated and some could exceed the pole mass of $\rho^\pm$ mesons.

\section{Color superconductivity}\label{sec:CS}
\subsection{Color superconductivity in a rotated magnetic field}\label{subsec:CSRM}
As mentioned in the introduction, the de Haas–van Alphen oscillation can be found in the color superconductor when strong magnetic field is present~\cite{Noronha:2007wg,Fukushima:2007fc,Fayazbakhsh:2010bh,Cao:2015xja}. Usually, constant diquark condensates are not allowed in external magnetic field, as the color superconductor is also the type-I electric superconductor which would expel magnetic field due to the Meissner effect~\cite{Schmitt:2003xq,Feng:2009vt}. However, the interplay between the EM and gluon fields opens a gate to study the constant diquark condensates in magnetic field. Actually, similar to the Higgs mechanism in quantum electroweak dynamics, the {\it rotated} photon remains massless in quantum "electro-strong" dynamics thus the {\it rotated} magnetic field can penetrate into the color superconductor~\cite{Alford:1999pb}. 

In the two-flavor color superconductor (2SC) and color-flavor locking (CFL) phases, all diquark condensates become effectively neutral to the {\it rotated} Abelian gauge field after rearranging the electric and color charges of involved quarks. To see that explicitely, the electric and $8$-th color matrices are
\bea
Q&=&{1\over 3}{\rm diag}\left(2,-1,-1\right),\\
T_8&=&{1\over \sqrt{3}}{\rm diag}\left(1,1,-2\right)
\eea
    in (u,d,s) flavor space and (r,g,b) color space, respectively; then the effective charge matrices of quarks are defined to be proportional to~\cite{Noronha:2007wg,Fukushima:2007fc}
\bea
\tilde{Q}&=&{1\over 2}\left(Q\otimes {\bf I}_{\rm c}+a\,{\bf I}_{\rm f}\otimes T_8\right)
\eea
with $a=-{1\over2\sqrt{3}}$ and $a={1\over\sqrt{3}}$ for 2SC and CFL phases, respectively. So, for 2SC phase, the relevant diquarks $u_{\rm r}d_{\rm g}$ and $u_{\rm g}d_{\rm r}$ are both neutral; for CFL phase, the relevant diquarks $u_{\rm r}s_{\rm b},u_{\rm g}d_{\rm b}, u_{\rm b}s_{\rm r}, u_{\rm b}d_{\rm g},s_{\rm g}d_{\rm r}$ and $s_{\rm r}d_{\rm g}$ are all neutral.

In order to find the corresponding gauge field $\tilde{A}_\mu$, we can simply look at the vector potential of quarks in flavor and color spaces:
\bea
e\,{A}_\mu\, Q \otimes {\bf I}_{\rm c}+g\, {G}_\mu^8\, {\bf I}_{\rm f}\otimes T_8.
\eea
By assuming the effective orthogonal gauge fields to be of the form
\bea
\tilde{A}_\mu&=&\cos\theta_{\rm es}{A}_\mu+\sin\theta_{\rm es}{G}_\mu^8,\\
 \tilde{G}_\mu^8&=&\cos\theta_{\rm es}{G}_\mu^8-\sin\theta_{\rm es}{A}_\mu,
\eea
 the vector potential can be rewritten as
\bea
&&\!\!\!\!\!\!\!\!\!\!\!\!\!\!\!(e\,\cos\theta_{\rm es}\,{Q}\otimes {\bf I}_{\rm c}+g\, \sin\theta_{\rm es}\,{\bf I}_{\rm f}\otimes T_8)\,\tilde{A}_\mu+\nonumber\\
&&(-e\,\sin\theta_{\rm es}\,{Q}\otimes {\bf I}_{\rm c}+g\, \cos\theta_{\rm es}\,{\bf I}_{\rm f}\otimes T_8)\,\tilde{G}_\mu^8.
\eea
Then, the "electro-strong" mixing angle $\theta_{\rm es}$ is immediately found to satisfy
$$\cos\theta_{\rm es}={g\over\sqrt{(a\,e)^2+g^2}},\ \sin\theta_{\rm es}={a\,e\over\sqrt{(a\,e)^2+g^2}}$$
in order to give the effective charge matrix $\tilde{Q}$. The relations are quite similar to those for the Weinberg angle in quantum electroweak dynamics. Thus, the vector potential can be represented as
\bea
\cos\theta_{\rm es}({e}\,\tilde{A}_\mu\,\tilde{Q}+{g}\,\tilde{G}_\mu^8\,\tilde{T}_8)
\eea
with the effective gluon interaction matrix
$$\tilde{T}_8={\bf I}_{\rm f}\otimes T_8-a\left({e\over g}\right)^2{Q}\otimes {\bf I}_{\rm c}.$$

Eventually, the kinetic term of quarks can be simply modified from the usual one~\cite{Alford:1998mk} to
\bea
\bar{\psi}[i\gamma^\mu(\partial_\mu-i\,{e}\,\cos\theta_{\rm es}\,\tilde{A}_\mu\,\tilde{Q})+\mu\gamma^0-m]\psi\label{KR}
\eea
by keeping only the massless gauge field $\tilde{A}_\mu$ in color superconductor. As we know, the coupling constants of QED and QCD are
$${e^2\over 4\pi}={1\over 137},\ {g^2\over 4\pi}\sim1$$
in natural unit, the mixing angle $\theta_{\rm es}$ is very small: $-1/40$ for 2SC and $1/20$ for CFL, respectively. Then, the effect of QED is only reduced by 
$$\cos^2\theta_{\rm es}-1\approx\theta_{\rm es}^2$$
according to the gauge coupling term in Eq.\eqref{KR}, hence it is fair enough to approximate the {\it rotated} magnetic field by the external one. Incidentally, the gluon condensation was found to be $$2\langle g^2{\bf B}_a^{2}\rangle\sim (0.88\,  {\rm GeV})^4$$
in the vacuum~\cite{DiGiacomo:1981lcx} and not sensitive to deconfinement transition~\cite{Lee:1989qj}, then we expect $\langle g^2({\bf B}_{8}^2\rangle\sim (0.44\, {\rm GeV})^4$ and $|e\sin\theta_{\rm es}{\bf B}_{8}|\sim|\theta_{\rm es}|(0.13\, {\rm GeV})^2$. For the range of $eB$ we are interested in, the chromomagnetic field, $\sim (20\, {\rm MeV})^2$ for 2SC and $\sim(29\, {\rm MeV})^2$ for CFL, is small thus can be safely neglected.

In the following, we take the simpler two flavor case for example to show how the chiral and diquark condensates are affected by the external magnetic field at large $\mu$~\cite{Cao:2015xja}. By turning on the diquark interaction channels, the effective Lagrangian density can be taken as~\cite{Fayazbakhsh:2010bh}
\begin{eqnarray}
{\cal L}\!\!&=&\!\!\bar\psi\left[i\slashed{\tilde D}-m_0+\mu\gamma_0\right]\psi+G\left[\left(\bar\psi\psi\right)^2+\left(\bar\psi i\gamma_5\boldsymbol \tau\psi\right)^2\right]\nonumber\\
&&+G_{\rm d}\left(i\bar\psi_C\varepsilon\epsilon_{\rm b}\gamma_5\psi\right)\left(i\bar\psi\varepsilon\epsilon_{\rm b}\gamma_5\psi_C\right)
\end{eqnarray}
without loss of generality. Here, $\psi_C=C\bar\psi^T$ and $\bar\psi_C=\psi^TC$ are charge-conjugate spinors with $C=i\gamma_2\gamma_0$, $\varepsilon_{ij}$ and $(\epsilon_{\rm b})_{\alpha\beta}=\epsilon_{\alpha\beta b}$ are respectively the antisymmetric matrices in flavor and color spaces with indexes $i,j=(u,d)$ and $\alpha,\beta=(r,g,b)$, and the coupling constants $G$ and $G_{\rm d}$ are related to each other by the Fierz transformation $G_{\rm d}=3G/4$~\cite{Huang:2002zd}. The covariant derivative is similar to that in usual EM field:
$${\tilde D}_\mu=\partial_\mu-i\,{e}{A}_\mu\,\tilde{Q},$$
but now the effective charge also depends on the color, that is,
$$\tilde{Q}_{\rm i\alpha}=\left({1\over 2},{1\over 2},1,-{1\over 2},-{1\over 2},0\right).$$

In mean field approximation, we assume
\begin{eqnarray}
\sigma&=&-2G\langle\bar\psi\psi\rangle\equiv m-m_0,\nonumber\\
\Delta_c&=&-2G_{\rm d} \langle i\,\bar\psi_C\varepsilon\epsilon_{\rm b}\gamma_5\psi\rangle,\nonumber\\
\Delta_c^*&=&-2G_{\rm d} \langle i\,\bar\psi\varepsilon\epsilon_{\rm b}\gamma_5\psi_C\rangle,
\end{eqnarray}
then the Lagrangian density can be formally given as
\bea
{\cal L}&=&-{(m-m_0)^2\over 4G}-{|\Delta_c|^2\over 4G_{\rm d}}+\bar\psi\left(i\slashed{\tilde D}-m+\mu\gamma_0\right)\psi\nonumber\\
&&-{1\over2}\left(\Delta_ci\bar\psi\varepsilon\epsilon_{\rm b}\gamma_5\psi_C+\Delta_c^*
i\bar\psi_C\varepsilon\epsilon_{\rm b}\gamma_5\psi\right).
\eea
Here, we can present the $1/2$ charged sector $u_{\rm r,g}$ and $d_{\rm r,g}$ in the more convenient Nambu-Gorkov space as:
\bea
{1\over2}\bar\Psi_{\rm rg}\left(\begin{array}{cc}
i\,G_+^{-1}&-i\Delta_c\tau_2\lambda_2\gamma_5P_{-q}\\ -i\Delta_c\tau_2\lambda_2\gamma_5P_q & i\,G_-^{-1}\end{array}\right)\Psi_{\rm rg},
\eea
where $\lambda_2$ is the second Pauli matrix in color space, the extended quark fields are defined as
 $$\bar\Psi_{\rm rg}=(\bar{u}_{\rm r},\bar{u}_{\rm g},  {d}_{\rm C r} , {d}_{\rm C g}),\  \Psi_{\rm rg}=\left(\begin{array}{c}u_{\rm r}\\ u_{\rm g} \\ \bar{d}_{\rm C r} \\ \bar{d}_{\rm C g}\end{array}\right),$$ 
 and the propagators of $u$ quark and $\bar{d}_{\rm C}$ anti-quark are respectively
 $$i\,G_\pm^{-1}(x)=i\gamma^\mu(\partial_\mu\mp i\,{e\over2}{A}_\mu)-m\pm \mu\gamma_0.$$
 Then, by integrating out all the bilinear spinors and taking into account the double counting in the Nambu-Gorkov space, we can finally obtain the thermodynamic potential as
 \begin{widetext}
 \bea
V_{\rm 2SC} &=&{(m\!-\!m_0)^2\over 4G}\!+\!{|\Delta_c|^2\over 4G_{\rm d}}-\sum_{j=\pm}\bigg\{\int {d^3{\bf p}\over(2\pi)^3}\left[{E^0}\!+\!2T\ln\left(1\!+\!e^{-\left(E^0\!+\!j\mu\right)/T}\right)\right]\!+\!\sum_{n,p_3}\left[{E^1}\!+\!2T\ln\left(1\!+\!e^{-\left(E^n\!+\!j\mu\right)/T}\right)\right]\nonumber\\
&&+2\sum_{n,p_3}\left[E_{j}^{1/2}\!+\!2T\ln\left(1\!+\!e^{-E_{j}^{1/2}/T}\right)\right]\bigg\}.
 \eea
Here, we use the abbreviation
$$\sum_{n,p_3}\equiv{eB\over 2\pi}\sum_{n=0}(1-\delta_{n0}/2)\int{dp_3\over 2\pi}$$
and the involved dispersions are
\begin{eqnarray}
E^0(p)=\sqrt{p^2+m^2},\ E^k(p_3)=\sqrt{{2kneB+p_3^2}+m^2},\
E_{j}^{1/2}(p_3)=\sqrt{\left[(E^{1/2}(p_3)\!+\!j\,\mu\right]^2\!+\!|\Delta_c|^2}.
\end{eqnarray}

Finally, restricting ourselves to zero temperature, we perform Pauli-Villars regularization to get 
 \bea
V_{\rm 2SC}^{\rm r} &=&{(m-m_0)^2\over 4G}+{|\Delta_c|^2\over 4G_{\rm d}}-\sum_{j=\pm}\sum_{i=0}^3(-1)^i C_3^i\Bigg\{\int{d^3{\bf p}\over(2\pi)^3}\left[{E_i^0}\theta(E_i^0-\mu)+{\mu}\theta(\mu-E_i^0)\right]\nonumber\\
&&+\sum_{n,p_3}\left[{E_i^1}\theta(E_i^1-\mu)+{\mu}\theta(\mu-E_i^1)+2E_{i,j}^{1/2}\right]\Bigg\},
\eea
 where the regularized dispersions are 
 $$E_i^{0,k}=\sqrt{(E^{0,k})^2+i\,\Lambda^2},\ E_{i,j}^{1/2}=\sqrt{\left(E_i^{1/2}\!+\!j\,\mu\right)^2\!+\!|\Delta_c|^2}.$$
 Without loss of generality, we assume $\Delta_c=\Delta_c^*$, and the gap equations can be given through $\partial_{m}V^{\rm r}_{\rm 2SC}=\partial_{\Delta_c}V^{\rm r}_{\rm 2SC}=0$ as
 \begin{eqnarray}
 {m-m_0\over2G}&=&m\sum_{j=\pm}\sum_{i=0}^3(-1)^i C_3^i\bigg\{\int{d^3{\bf p}\over (2\pi)^3}{1\over E_i^0}\theta(E_i^0-\mu)+\sum_{n,p_3}\bigg[{1\over E_i^1}\theta(E_i^1-\mu)+2{E_i^{1/2}+j\mu\over E_i^{1/2}E_{i,j}^{1/2}}\bigg]\bigg\},\nonumber\\
{\Delta_c\over 2G_{\rm d}}&=&\Delta_c\sum_{j=\pm}\sum_{i=0}^3(-1)^i C_3^i\sum_{n,p_3}{2\over E_{i,j}^{1/2}}.
\label{gaps}
\end{eqnarray}
\end{widetext}

\begin{figure}[!htb]
	\centering
	\includegraphics[width=0.45\textwidth]{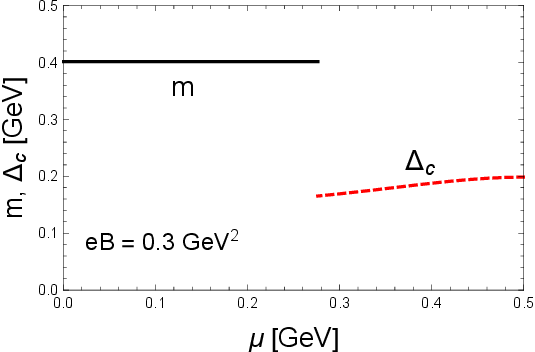}
	\caption{The mass $m$ and diquark condensate $\Delta_c$ as functions of baryon chemical potential $\mu$ for a given rotated magnetic field $eB=0.3~{\rm GeV}^2$.}\label{mDelta}
\end{figure}

\begin{figure}[!htb]
	\centering
	\includegraphics[width=0.45\textwidth]{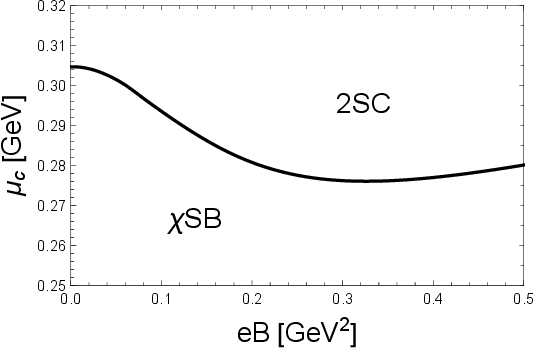}
	\caption{The critical baryon chemical potential $\mu_{\rm c}$ as a function of the rotated magnetic field $eB$ for the $\chi$SB-2SC transition.}\label{cmuB_2SC}
\end{figure}

In chiral limit $m_0=0$, we consistently solve the gap equations in Eq.\eqref{gaps} and find only three kinds of solutions is possible: $m=\Delta_c=0; m\neq0,\Delta_c=0$ and $m=0,\Delta_c\neq0$. At a given rotated magnetic field, the order parameters are shown as functions of the baryon chemical potential in Fig.\ref{mDelta}: for small $\mu$, the solution $m\neq0,\Delta_c=0$ is the ground state; while for large $\mu$, the solution $m=0,\Delta_c\neq0$ dominates.
Then, by changing the magnetic field, we can easily work out the phase boundary between $\chi$SB and 2SC phases, see Fig.\ref{cmuB_2SC}. Here, we can again identify the dHvA oscillation. Compared to the phase boundary in Fig.\ref{critical_CDW}, we can conclude that $\chi$SB-2SC transition is more favored at small $eB$ where the discreteness of Landau levels is not so important. At large $eB$, the $\chi$SB phase is quite different from that in Sec.\ref{subsec:inhom} as the "electric charges" are rotated here, so it is not justified to conclude that $\chi$SB-CDW transition is favored. To settle that problem, we have to carry out the calculation of the thermodynamic potential for CDW phase in the rotated magnetic field.

\section{Summary and perspectives}\label{sec:summ}
In this review, mainly four kinds of phase transitions were explored in the circumstances where magnetic fields are present: chiral symmetry breaking and restoration, neutral pseudoscalar superfluidity, charged pion superfluidity and charged rho superconductivity. Among these transitions, chiral symmetry breaking and restoration attracts the most attention in high energy nuclear physics and was thus studied with great details in Sec.\ref{sec:chiral}. In particular, we revisited the unsolved problems of the inverse magnetic catalysis effect and the competition between the chiral density wave and solitonic modulation phases. By adopting reasonable schemes, it was shown that some satisfactory conclusions can be drawn: the coupling running with $eB$ at zero temperature is capable of consistently explaining the decreasing of $\pi^0$ mass and the IMCE found in LQCD; the CDW phase is favored over the SM phase for the baryon chemical potential well away from the cutoff. On the other hand, based on the ansatz of either homogeneous or inhomogeneous  chiral condensate, the de Haas-van Alphen oscillation was found in the $\mu_{\rm B}-eB$ phase diagram. But the oscillation is missing in the $\Omega-eB$ phase diagram even though the effect of $\Omega$ was found to be $\mu_{\rm B}$-like in Ref.~\cite{Chen:2015hfc}. The reason is that there are a series of angular momentum dependent "$\mu_{\rm B}$'s" which are so close to each other that deep dip cannot be developed for the dHvA oscillation. Incidentally, the study of color superconductivity in Sec.\ref{subsec:CSRM} also found the dHvA oscillation in the phase boundary. As mentioned, further work is still demanded to clarify the true $\mu_{\rm B}-eB$ phase diagram when CDW, SM and 2SC phases are all considered on the same basis. 

The neutral pion superfluidity in parallel EM field was rediscovered by us several years ago and has become an important candidate for the ground state of QCD system. This review covered our previous studies in the two- and three-flavor NJL models and found that other neutral pseudoscalar condensates, such as $\eta, \eta'$, also exist in parallel EM fields. Besides, the in-in and in-out (or Schwinger) schemes were compared with each other, and the feedback of  the Schwinger pair production was expected to alter the chiral restoration from second order to first order. Nevertheless, the chiral rotation is almost the same for both schemes in the $u$ and $d$ quark sector. Furthermore, charged pion and rho condensates were shown to be impossible in a pure magnetic field but can occur in parallel magnetic field and rotation. According to our explorations in Sec.\ref{subsec:PMRpi} and Sec.\ref{subsec:PMRrho}, $\pi^\pm$ superfluidity might be favored at small magnetic field and rotation though with ambiguity,  while $\rho^\pm$ superconductivity might be preferred at large magnetic field and rotation. In the future, we hope to work out the true phase diagram for $\chi$SB and $\chi$SR, $\pi^\pm$ superfluidity and $\rho^\pm$ superconductivity in ${\bf B}\parallel\boldsymbol{\Omega}$ by consistently taking the type-II superconductor and boundary condition into account. Additionally, the running coupling effect has to be examined in order to obtain more realistic results and intriguing phenomena might emerge. 

\section*{Acknowledgments}
G.C. would like to thank his wife Yu Zhou for her great supports when he was preparing the review. G.C. also appreciates Li Yan for his kind invitation to visit the Institute of Modern Physics at Fudan University and Pengfei Zhuang for allowing him to take part in the project at Tsinghua University. G.C. is funded by the National Natural Science Foundation of China with Grant No. 11805290. 
\newpage
\appendix

\section{The check of polarization functions in pure magnetic field}\label{sec:check}
To check the correctness of Eq.\eqref{AFLM} and \eqref{AFLC}, we consider the nontrivial contributions from low Landau levels in pure magnetic field. First, we evaluate the contributions from $n=0,n'=1$ and $n=1,n'=0$. It turns out that the corresponding auxiliary functions are
	\begin{eqnarray}
	\!\!\!\!\!\!\!\!\!\!\!\!\!&&G_{\rm 0\,l(l'),1\,l'(l)}(q_{\rm u}B,|q_{\rm d}B|)={(l\!+\!l')!\over eB}\left({2\over eB}\right)^{\!l\!+\!l'\!+\!1},\nonumber\\
	\!\!\!\!\!\!\!\!\!\!\!\!\!&&H_{\rm 0\,l,1\,l'}(q_{\rm u}B,|q_{\rm d}B|)=2{(l\!+\!l')!\over eB}\left({2\over eB}\right)^{\!l\!+\!l'\!+\!1}\!\! |q_{\rm d}B|^2,\nonumber\\
	\!\!\!\!\!\!\!\!\!\!\!\!\!&&H_{\rm 1\,l,0\,l'}(q_{\rm u}B,|q_{\rm d}B|)=2{(l\!+\!l')!\over eB}\left({2\over eB}\right)^{\!l\!+\!l'\!+\!1} \!\!(q_{\rm u}B)^2,
	\end{eqnarray}
	then we obtain from Eq.\eqref{AFLM} and \eqref{AFLC} that the $\Omega\rightarrow0$ limit is
	\begin{eqnarray}\label{A0C}
	{\cal A}_1\!\!\!&=&\!\!\!{2N_ci\over \pi}\left(\tilde{q}_{\rm u}\tilde{q}_{\rm d}eB\right)\int_{-\infty}^{\infty}{dp_0dp_3\over(2\pi)^2}
	\left[{1\over{p}_0^2\!-\!E_{\rm d1}^2}+{1\over{p}_0^2\!-\!E_{\rm u1}^2}\right]\nonumber\\
	\!\!\!&=&\!\!\!{N_c\over 2\pi}\left(\tilde{q}_{\rm u}\tilde{q}_{\rm d}eB\right)\int_{-\infty}^{\infty}{dp_3\over(2\pi)}\left[{1\over E_{\rm d1}}+{1\over E_{\rm u1}}\right].
	\end{eqnarray}
Next, we calculate the contribution from $n=n'=1$. Utilizing the following results
	\begin{widetext}
	\begin{eqnarray}
	G_{\rm 1\,l,1\,l'}(q_{\rm u}B,|q_{\rm d}B|)&=&{(l+l'-1)!\over (eB)^3}\left({2\over eB}\right)^{l+l'}\left[\left(|q_{\rm d}B|l-q_{\rm u}B l'\right)^2-(l+l')(q_{\rm u}B)^2\right],\nonumber\\
	H_{\rm 1\,l,1\,l'}(q_{\rm u}B,|q_{\rm d}B|)&=&2|q_{\rm d}B|eBG_{\rm 1\,l,1\,l'}(q_{\rm u}B,|q_{\rm d}B|)+2q_{\rm u}BeBG_{\rm 1\,l',1\,l}(|q_{\rm d}B|,q_{\rm u}B)\nonumber\\
	&&-8|q_{\rm d}B|q_{\rm u}B{(l+l'-1)!\over (eB)^3}\left({2\over eB}\right)^{l+l'}\left[\left(|q_{\rm d}B|l-q_{\rm u}B l'\right)^2+(l+l')|q_{\rm d}B|q_{\rm u}B\right],
	\end{eqnarray}
	the $\Omega\rightarrow0$ limit can be obtained from Eq.\eqref{AFLM} and \eqref{AFLC} as
	\begin{eqnarray}\label{A1C}
	{\cal A}_2
	&{=}&{-2N_ci\over \pi}{q_{\rm u}B|q_{\rm d}|B\over eB}\int_{-\infty}^{\infty}{dp_0\over2\pi}\int_{-\infty}^{\infty}{dp_3\over2\pi}
	{1\over\left[{p}_0^2-E_{\rm u1}^2\right]\left[{p}_0^2-E_{\rm d1}^2\right]}\left\{\left({p}_0^2-p_3^2-m^2-4eB{q_{\rm u}\over e}{|q_{\rm d}|\over e}\right)\right.\nonumber\\
	&&\left.-\left({p}_0^2-p_3^2-m^2+4eB{q_{\rm u}\over e}{|q_{\rm d}|\over e}\right)\right\}\nonumber\\
	&=&-{16N_c\over \pi}\left(\tilde{q}_{\rm u}\tilde{q}_{\rm d}eB\right)^2\int_{-\infty}^{\infty}{dp_4\over2\pi}\int_{-\infty}^{\infty}{dp_3\over2\pi}{1\over\left[{p}_4^2+E_{\rm u1}^2\right]\left[{p}_4^2+E_{\rm d1}^2\right]}.
	\end{eqnarray}
	
Now, we try to reevaluate ${\cal A}_1$ and ${\cal A}_2$ by directly using the effective fermion propagator in Euclidean space~\cite{Miransky:2015ava}, that is,
	\begin{eqnarray}
	{\cal A}_1&=&-\int{d^4p\over{(2\pi)^4}}{e^{-(p_x^2+p_y^2)[1/(q_{\rm u}B)+1/(|q_{\rm d}B|)]}\over \left[p_4^2+E_{\rm u1}^2\right]\left[p_4^2+E_{\rm d0}^2\right]}{\rm Tr}\Bigg\{(m-p_4\gamma_4-p_3\gamma^3)\left[(1+i\gamma^1\gamma^2)L_1\left({2(p_x^2+p_y^2)\over q_{\rm u}B}\right)-(1-i\gamma^1\gamma^2)\right]\nonumber\\
	&&(-m-p_4\gamma_4-p_3\gamma^3)(1-i\gamma^1\gamma^2)\Bigg\}-\int{d^4p\over{(2\pi)^4}}{e^{-(p_x^2+p_y^2)[1/(q_{\rm u}B)+1/(|q_{\rm d}B|)]}\over \left[p_4^2+E_{\rm u0}^2\right]\left[p_4^2+E_{\rm d1}^2\right]}{\rm Tr}\Bigg\{(m-p_4\gamma_4-p_3\gamma^3)(1+i\gamma^1\gamma^2)\nonumber\\
	&&(-m-p_4\gamma_4-p_3\gamma^3)\left[(1-i\gamma^1\gamma^2)L_1\left({2(p_x^2+p_y^2)\over |q_{\rm d}B|}\right)-(1+i\gamma^1\gamma^2)\right]\Bigg\}\nonumber\\
	&=&-8N_c\int{d^4p\over{(2\pi)^4}}e^{-(p_x^2+p_y^2)[1/(q_{\rm u}B)+1/(|q_{\rm d}B|)]}\left[{1\over p_4^2\!+\!E_{\rm u1}^2}\!+\!{1\over p_4^2\!+\!E_{\rm d1}^2}\right]\nonumber\\
	&=&{N_c\over 2\pi}\left(\tilde{q}_{\rm u}\tilde{q}_{\rm d}eB\right)\int_{-\infty}^{\infty}{dp_3\over(2\pi)}\left[{1\over E_{\rm d1}}\!+\!{1\over E_{\rm u1}}\right]\label{A0M}
	\end{eqnarray}
	and
	\begin{eqnarray}
	{\cal A}_2&=&\int{d^4p\over{(2\pi)^4}}{e^{-(p_x^2+p_y^2)[1/(q_{\rm u}B)+1/(|q_{\rm d}B|)]}\over \left[p_4^2+E_{\rm u1}^2\right]\left[p_4^2+E_{\rm d1}^2\right]}{\rm Tr}\Bigg\{(m-p_4\gamma_4-p_3\gamma^3)\left[(1+i\gamma^1\gamma^2)L_1\left({2(p_x^2+p_y^2)\over q_{\rm u}B}\right)-(1-i\gamma^1\gamma^2)\right]\nonumber\\
	&&+4(p_x\gamma^1+p_y\gamma^2)\Bigg\}\Bigg\{(-m-p_4\gamma_4-p_3\gamma^3)\left[(1-i\gamma^1\gamma^2)L_1\left({2(p_x^2+p_y^2)\over |q_{\rm d}B|}\right)-(1+i\gamma^1\gamma^2)\right]+4(p_x\gamma^1+p_y\gamma^2)\Bigg\}\nonumber\\
	&=&8\int{d^4p\over{(2\pi)^4}}{e^{-(p_x^2+p_y^2)[1/(q_{\rm u}B)+1/(|q_{\rm d}B|)]}\over \left[p_4^2+E_{\rm u1}^2\right]\left[p_4^2+E_{\rm d1}^2\right]}\left\{(p_4^2+p_3^2+m^2)\left[L_1\left({2(p_x^2\!+\!p_y^2)\over q_{\rm u}B}\right)+L_1\left({2(p_x^2\!+\!p_y^2)\over |q_{\rm d}B|}\right)\right]-8(p_x^2\!+\!p_y^2)\right\}\nonumber\\
	&=&-{16N_c\over \pi}\left(\tilde{q}_{\rm u}\tilde{q}_{\rm d}eB\right)^2\int_{-\infty}^{\infty}{dp_4\over2\pi}\int_{-\infty}^{\infty}{dp_3\over2\pi}{1\over \left[p_4^2+E_{\rm u1}^2\right]\left[p_4^2+E_{\rm d1}^2\right]}.\label{A1M}
	\end{eqnarray}
	Thus, the $\Omega\rightarrow0$ limit of Eq.\eqref{AFLM} and \eqref{AFLC} is indeed consistent with the results in pure magnetic field.

\section{The equality between proper-time and Landau-level presentations}\label{equality}
As illustrated in Appendix.\ref{sec:check}, the effective quark propagators can be decomposed as a sum of all Landau level Green's functions~\cite{Miransky:2015ava}. In energy-momentum space, we have
\bea
S_{\rm f}(k)&=&-i~e^{-{{\bf k}_\bot^2\over|q_{\rm f}B|}}\sum_{\rm n=0}^\infty(-1)^n{D_n(q_{\rm f} B,k)\over k_4^2\!+\!k_3^2\!+\!m^2+2n|q_{\rm f} B|},\\
D_n(q_{\rm f} B,k)&=&(m-\slashed k_4-\slashed k_3)\left[{\cal P}_+^{\rm f}L_n\left({2{\bf k}_\bot^2\over|q_{\rm f}B|}\right)-{\cal P}_-^{\rm f}L_{\rm n-1}\left({2{\bf k}_{\rm f}^2\over|q_{\rm f}B|}\right)\right]+4(\slashed k_1+\slashed k_2)L_{\rm n-1}^1\left({2{\bf k}_\bot^2\over|q_{\rm f}B|}\right),
\eea
where ${\cal P}_\pm^{\rm f}=1\pm{\cal S}(q_{\rm f} B)i\gamma^1\gamma^2$ is the spin up/down projector and $L_{\rm n}^\alpha(x)$	are the generalized Laguerre polynomials with $L_{\rm n}(x)\equiv L_{\rm n}^0(x)$ and $L_{-1}^\alpha(x)=0$.
Then, the polarization loop for $\bar{\rho}^+_1$ meson with zero three-momentum can be evaluated as the following:
\bea\label{Pirho}
\Pi_{\bar{\rho}^+_1\bar{\rho}^+_1}(B,p_4)
&=&-32N_c\sum_{\rm n=0}^\infty\sum_{\rm n'=0}^\infty\int{\di^4k\over (2\pi)^4}e^{-{{\bf k}_\bot^2\over|q_{\rm u}B|}-{{\bf k}_\bot^2\over|q_{\rm d}B|}}{(m^2+k_3^2+(k_4+p_4)k_4)L_n\left({2{\bf k}_\bot^2\over|q_{\rm u}B|}\right)L_{\rm n'}\left({2{\bf k}_\bot^2\over|q_{\rm d}B|}\right)\over ((k_4+p_4)^2+{E_{\rm un}}^2)(k_4^2+{E_{\rm dn'}}^2)}\nonumber\\
&=&-4N_c\sum_{\rm n=0}^\infty\sum_{\rm n'=0}^\infty{eB\over\pi}\int{\di {\bf k}_3\over (2\pi)}\left[{(m^2+E_{\rm un}E_{\rm dn'}+k_3^2)G_{\rm nn'}\over p_4^2+(E_{\rm un}+E_{\rm dn'})^2}\left({1\over E_{\rm un}}+{1\over E_{\rm dn'}}\right)\right],
\eea
where the dimensionless $G$ function is defined as
\bea
{{G}}_{\rm nn'}&\equiv&\int_0^\infty\di x~e^{-\left({1\over{|\tilde{q}_{\rm u}|}}+{1\over{|\tilde{q}_{\rm d}|}}\right)x}L_n\left({2x\over{|\tilde{q}_{\rm u}|}}\right)L_{\rm n'}\left({2x\over{|\tilde{q}_{\rm d}|}}\right)\nonumber\\
&=&{1\over4}\sum_{k=0}^n\sum_{k'=0}^{n'}\left(\begin{array}{c}
		n\\n-k
\end{array}\right)\left(\begin{array}{c}
		n'\\n'-k'
\end{array}\right)\left(\begin{array}{c}
		k+k'\\k
\end{array}\right)(-2{|\tilde{q}_{\rm d}|})^{k+1}(-2{|\tilde{q}_{\rm u}|})^{k'+1}.
	\eea
\end{widetext}
As ${{G}}_{\rm nn'}$ is independent of the magnetic field and energy-momentum, the matrix can be evaluated to very large $n$ and $n'$ numerically and then reserved as a special function for further manipulations. 

\begin{figure}[!htb]
	\begin{center}
		\includegraphics[width=8cm]{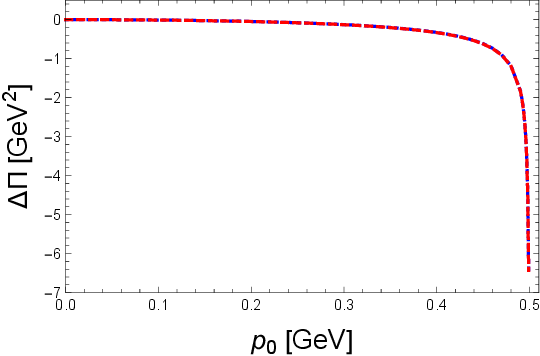}
		\caption{The comparison of $\Delta\Pi$ between proper-time (blue dotted line) and Landau-level (red dashed line) presentations for the chosen quark mass $m=0.25~{\rm GeV}$ and magnetic fields: $eB_1=1~{\rm GeV}^2$ and $eB_2=2~{\rm GeV}^2$. The plots are from Ref.~\cite{Cao:2019res}..}\label{LLPT}
	\end{center}
\end{figure}

The bare polarization function Eq.\eqref{Pirho} is ultraviolet divergent that requires for further regularization. However, we are not going to introduce any artificial cutoff at this stage for the purpose of demonstrating the equality between proper-time and Landau-level presentations. Rather, the following formally convergent term will be evaluated:
\bea
\!\!\!\!\Delta\Pi\equiv[\Pi_{\bar{\rho}^+_1\bar{\rho}^+_1}(B_2,i\,p_0)\!-\!\Pi_{\bar{\rho}^+_1\bar{\rho}^+_1}(B_2,0)]\!-\!(B_2\!\rightarrow\!B_1)
\eea
with $B_1$ and $B_2$ different magnetic fields. The comparison is illustrated in Fig.~\ref{LLPT}, where they are found to be precisely consistent with each other up to the unstable point $p_0=2m$. In principle, the equality should continue in the unstable region $p_0>2m$, but $\Delta\Pi$ is not suitable for such exploration as it diverges in the imaginary proper-time presentation.

\section{Invalidity of NJL model to explore the magnetic field effect to physical $\rho$ meson}\label{invalidity}
\begin{figure}[!htb]
	\begin{center}
		\includegraphics[width=8cm]{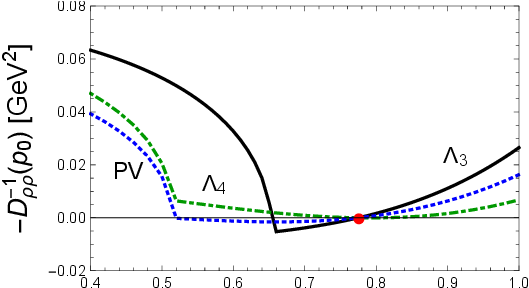}
		\includegraphics[width=8cm]{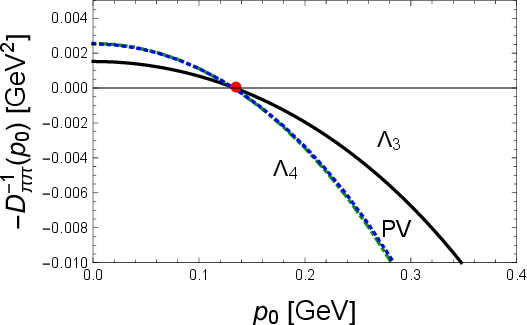}
		\caption{The effective inverse propagators of $\rho$ (upper panel) and $\pi^0$ (lower panel) mesons in vacuum with respect to different regularization schemes: three-momentum cutoff (black solid line), four-momentum cutoff (green dot-dashed line) and Pauli-Villars (blue dotted line). The red bullets are the physical $\rho$ and $\pi^0$ meson masses. In the upper panel, nonanalytic features are developed at twice the dynamical masses: $0.313~{\rm GeV}$ for $\Lambda_3$ and $\sim 0.256~{\rm GeV}$ for $\Lambda_4$ and PV regularizations~\cite{Klevansky:1992qe}. The plots are from Ref.~\cite{Cao:2019res}.}\label{prp_rhopi}
	\end{center}
\end{figure}

To explore the properties of $\rho$ meson in certain circumstances, we should choose adequate regularization schemes in NJL model.
Here, we compare three schemes with the parameters listed in Ref.~\cite{Klevansky:1992qe}: three-momentum cutoff ($\Lambda_3$), four-momentum cutoff ($\Lambda_4$) and Pauli-Villars (PV). In order to show the pion spectrum more obviously, the current quark mass $m_0=5~{\rm MeV}$ is adopted alternatively. Firstly, we study the spatial component $\rho_i$ of rho meson for $eB=0$ and illuminate the inverse propagators in the upper panel of Fig.~\ref{prp_rhopi}. 
As can be seen, there are two zero points for all the regularization schemes, of which the other one is lighter than the physical mass in the $\Lambda_3$ and PV schemes but slightly heavier in the $\Lambda_4$ scheme. Recalling the basic form of vector boson propagator in quantum field theory (QFT)~\cite{Peskin1995}:
\bea
D_{\rm VV}^{\mu\nu}(p)=-{g^{\mu\nu}-\hat{p}^\mu\hat{p}^\nu\over p^2-m^2},
\eea
we expect $-{D}^{-1}_{\bar{\rho}^+_1\bar{\rho}^+_1}>0~(<0)$ for $p_0<2m~(>2m)$. So, the signs of the $\rho$ meson propagators are wrong around the physical zero point in the $\Lambda_3$ and PV schemes, and only $\Lambda_4$ scheme is suitable to study $\rho$ meson. 
For comparison, the inverse propagators of $\pi$ meson are Si in the lower panel of Fig.~\ref{prp_rhopi}. They share the same signs around their zero points and are consistent with the form of scalar boson propagator in QFT~\cite{Peskin1995}:
\bea
D_{\rm SS}(p)={1\over p^2-m^2}.
\eea

\begin{figure}[!htb]
	\begin{center}
		\includegraphics[width=8cm]{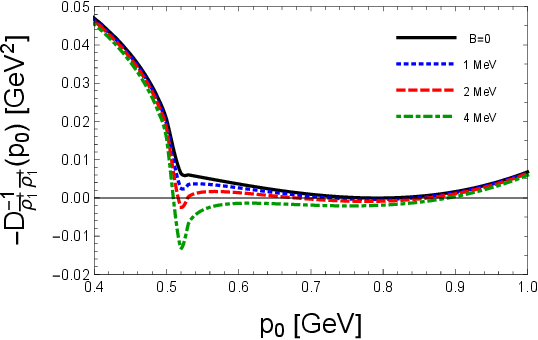}
		\caption{The effective inverse propagators of $\rho^+_1$ meson for different magnetic fields: $B=0$ (black solid line), $10^{-3}~{\rm GeV}^2$ (blue dotted line), $2*10^{-3}~{\rm GeV}^2$ (red dashed line) and $4*10^{-3}~{\rm GeV}^2$ (green dot-dashed line). There are dips around the two quark unstable point $p_0\sim2m$. The plots are from Ref.~\cite{Cao:2019res}.}\label{prho_B}
	\end{center}
\end{figure}

Secondly, we explore the magnetic effect to $\bar{\rho}^+_1$ meson by adopting the most optimistic $\Lambda_4$ scheme in  Eq.\eqref{RPFLLr}. 
The regularized terms can be transformed with the help of Feynman parameter as~\cite{Klimt:1989pm}:
\bea
&&-8N_c\int^{\Lambda_4}{\di^4k\over (2\pi)^4}{m^2\!+\!k_4(k_4+p_4)\!+\!k_3^2\over (k_4^2\!+\!E_{\bf k}^2)[(k_4+p_4)^2\!+\!E_{\bf k}^2]}\nonumber\\
&=&{N_c\over 6\pi^2}\left[-2\left(\Lambda^2-m^2\log\left(1+{\Lambda^2\over m^2}\right)\right)+(-p_4^2+2m^2)\right.\nonumber\\
&&\ \ \ \ \ \left.\int_0^1\di x\left({\Lambda^2F(x,\Lambda)}+\log\left(1-{\Lambda^2F(x,\Lambda)}\right)\right)\right],\\
&&-8N_c\int^{\Lambda_4}{\di^4k\over (2\pi)^4}{[m^2\!+\!k_4(k_4+p_4)\!+\!k_3^2]~eB\over (k_4^2\!+\!E_{\bf k}^2)^2[(k_4+p_4)^2\!+\!E_{\bf k}^2]}\nonumber\\
&=&-{N_ceB\over 4\pi^2}\!\left[\log\left(\!1\!+\!{\Lambda^2\over m^2}\!\right)\!+\!{p_4^2\over2}\!\!\int_0^1\!\!\di x\left(F(x,\Lambda)\!-\!F(x,0)\right)\right]\nonumber\\
\eea
with the auxiliary function $F(x,y)=[y^2-p_4^2(x^2-x)+m^2]^{-1}$. The numerical results are demonstrated in Fig.\ref{prho_B}.
In contrary to the point particle results or LQCD simulations, so strong dips are developed around $p_0\sim2m$ that $m_{\bar{\rho}^+_1}$ changes very quickly and even discontinuously with $B$. The instability around $2m$ indicates the decay of $\rho$ meson into quark-antiquark pair thus must be an artifact due to the absence of confinement in NJL model. Even adding the potential of Polyakov loop can not help the situation because it only induces confinement effect to the system at finite temperature~\cite{Ferreira:2013tba}. Thus, we can conclude that NJL (or PNJL) model is not suitable for the study of magnetic effect to mesonic resonances with masses $>2m$, such as vectors $\rho$ and $\omega$, and pseudoscalar $\eta'$.


\begin{thebibliography}{462}
\bibitem{baidu} Most of the images are selected from the search engine "Baidu", interested readers please check the initial sources there.
\end{thebibliography}
\end{document}